\documentclass{llncs}[13pt]
\usepackage{mathrsfs}
\usepackage[utf8]{inputenc}
\usepackage[english]{babel}
\usepackage{multicol}
\usepackage{pdfpages}
\usepackage{wrapfig}
\usepackage{color}
\usepackage{tabularx}
\usepackage{caption}
\usepackage{bm}
\usepackage{mathrsfs}
\usepackage{amssymb}
\usepackage{enumitem}
\usepackage{pdfpages}
\usepackage{float}
\usepackage{graphicx}
\usepackage{csquotes}
\usepackage{hyperref}
\usepackage{amsmath}
\usepackage[margin=1.30in]{geometry}
\usepackage{ulem}

\newtheorem{theo}{Theorem}
\newtheorem{cor}{Corollary}
\newtheorem{lem}{Lemma}
\newtheorem{prop}{Proposition}

\usepackage{ulem}
\usepackage{mathrsfs}
\usepackage[utf8]{inputenc}
\usepackage[english]{babel}
\usepackage{multicol}
\usepackage{pdfpages}
\usepackage{wrapfig}
\usepackage{color}
\usepackage{tabularx}
\usepackage{caption}
\usepackage{bm}
\usepackage{mathrsfs}
\usepackage{amssymb}
\usepackage{enumitem}
\usepackage{pdfpages}
\usepackage{float}
\usepackage{graphicx}
\usepackage{csquotes}
\usepackage{hyperref}
\usepackage{amsmath}
\usepackage{ulem}
\usepackage{lipsum}

\numberwithin{equation}{section}
\numberwithin{theorem}{section}
\numberwithin{cor}{section}
\numberwithin{lemma}{section}
\numberwithin{proposition}{section}
\numberwithin{cor}{section}
\numberwithin{eg}{section}
\numberwithin{examp}{section}

\newcommand{\R}{\mathbf{R}}
\newcommand{\T}{\mathbf{T}}

\newcommand{\bb}{\mathbf{b}}
\newcommand{\bbeta}{\bm{\beta}}
\newcommand{\bSigma}{\boldsymbol{\Sigma}}

\begin{document}

\title{Perturbation Bootstrap in Adaptive Lasso}
\titlerunning{Perturbation Bootstrap in ALASSO}  
%
\author{Debraj Das\inst{1}, Karl Gregory \inst{2} \and
S. N. Lahiri\inst{3}
}
\authorrunning{Debraj Das} 


\institute{University of Wisconsin-Madison\\
\email{ddas25@wisc.edu}
\and
University of South Carolina\\
\email{GREGORKB@stat.sc.edu}
\and
North Carolina State University\\
\email{snlahiri@ncsu.edu}
}

\maketitle              

\begin{abstract}
The Adaptive Lasso(Alasso) was proposed by Zou [\textit{J. Amer. Statist. Assoc. \textbf{101} (2006) 1418-1429}] as a modification of the Lasso for the purpose of simultaneous variable selection and estimation of the  parameters in a linear regression model. Zou (2006) established that the Alasso estimator is variable-selection consistent as well as asymptotically Normal in the indices corresponding to the nonzero regression coefficients in certain fixed-dimensional settings. In an influential paper, Minnier, Tian and Cai [\textit{J. Amer. Statist. Assoc. \textbf{106} (2011) 1371-1382}] proposed a perturbation bootstrap method and established its distributional consistency for the Alasso estimator in the fixed-dimensional setting. In this paper, however, we show that this (naive) perturbation bootstrap fails to achieve second order correctness in approximating the distribution of the Alasso estimator. We propose a modification to the perturbation bootstrap objective function and show that a suitably studentized version of our modified perturbation bootstrap Alasso estimator achieves second-order correctness even when the dimension of the model is allowed to grow to infinity with the sample size. As a consequence, inferences based on the modified perturbation bootstrap will be more accurate than the inferences based on the oracle Normal approximation. We give simulation studies demonstrating good finite-sample properties of our modified perturbation bootstrap method as well as an illustration of our method on a real data set.
\end{abstract}
\tableofcontents

\section{Introduction}
\large
Consider the multiple linear regression model
\begin{equation}\label{eqn:model}
y_{i} = \bm{x}'_{i}\bm{\beta} + \epsilon_{i}, \; \;\;\;\;     i = 1,\dots,n,
\end{equation}
where $y_1,\ldots,y_n$ are responses, $\epsilon_1,\ldots,\epsilon_n$ are independent and identically distributed (iid) random variables, $\bm{x}_1,\ldots,\bm{x}_n$ are known non-random design vectors, and $\bm{\beta}=(\beta_1,\ldots, \beta_p)$ is the $p$-dimensional vector of regression parameters. When the dimension $p$ is large, it is common to approach regression model (\ref{eqn:model}) with the assumption that the vector $\bm{\beta}$ is sparse, that is that the set $\mathcal{A}= \{j:\beta_j\neq 0\}$ has cardinality $p_0 = |\mathcal{A}|$ much smaller than $p$, meaning that only a few of the covariates are ``active''. The Lasso estimator introduced by Tibshirani (1996) is well suited to the sparse setting because of its property that it sets some regression coefficients exactly equal to 0. One disadvantage of the Lasso, however, is that it produces non-trivial asymptotic bias for the non-zero regression parameters, primarily because it shrinks all estimators toward zero [cf. Knight and Fu (2000)].

Building on the Lasso, Zou (2006) proposed the Adaptive Lasso [hereafter referred to as Alasso] estimator $\bm{\hat{\beta}}_n$ of $\bm{\beta}$ in the regression problem (\ref{eqn:model}) as
\begin{equation}\label{eqn:modelalasso}
\bm{\hat{\beta}}_n = \operatorname*{arg\,min}_{\bm{t}}\Bigg[\sum_{i=1}^{n}(y_i - \bm{x}'_i \bm{t})^2
+\lambda_n\sum_{j=1}^{p}|\tilde{\beta}_{j,n}|^{-\gamma}|t_{j}|\Bigg],
\end{equation}
where $\tilde{\beta}_{j,n}$ is the $j$th component of a root-$n$-consistent estimator $\tilde{\bm{\beta}}_n$ of $\bm{\beta}$, such as the ordinary least squares (OLS) estimator when $p\leq n$ or the Lasso or Ridge regression estimator when $p>n$, $\lambda_n>0$ is the penalty parameter, and $\gamma>0$ is a constant governing the influence of the preliminary estimator $\tilde{\bm{\beta}}_n$ on the Alasso fit.
Zou (2006) showed in the fixed-$p$ setting that under some regularity conditions and with the right choice of $\lambda_n$, the Alasso estimator enjoys the so-called oracle property [cf. Fan and Li (2001)]; that is, it is variable-selection consistent and it estimates the non-zero regression parameters with the same precision as the OLS estimator which one would compute if the set of active covariates were known. 
\par
In an important recent work, Minnier, Tian and Cai (2011) introduced the perturbation bootstrap in the Alasso setup. To state their main results, let $\bm{\beta}_{n}^{*N} = ( \beta_{1,n}^{*N},\dots,\beta_{p,n}^{*N} )'$ be the naive perturbation bootstrap Alasso estimator prescribed by Minnier, Tian and Cai (2011) and define $\hat {\mathcal{A}}_n=\{j: \hat \beta_{j,n}\neq 0\}$ and $\mathcal{A}_n^{*N}=\{j: \beta_{j,n}^{*N}\neq 0\}$.
These authors showed that under some regularity conditions and with $p$ fixed as $n\rightarrow \infty$
\begin{align*}
\mathbf{P}_*(\mathcal{A}_n^{*N}=\hat{\mathcal{A}}_n)\rightarrow 1 \;\;\text{and}\;\; \sqrt{n}(\bm{\beta}_{n}^{*N(1)}-\hat{\bm{\beta}}_{n}^{(1)})| \bm{\varepsilon} \asymp_d \sqrt{n}(\hat{\bm{\beta}}_{n}^{(1)}-\bm{\beta}^{(1)}),
\end{align*}
where $\bm{\varepsilon}_n=(\epsilon_1,\ldots,\epsilon_n)$, $\bm{z}^{(1)}$ denotes the sub-vector of $\bm{z}\in \mathcal{R}^p$ corresponding to the co-ordinates in $\mathcal{A}= \{j:\beta_{j}\neq 0\}$, ``$\asymp_d$'' denotes asymptotic equivalence in distribution, and $\mathbf{P}_*$ denotes bootstrap probability conditional on the data. 
Thus Minnier, Tian and Cai (2011) [hereafter referred to as MTC(11)] showed that, in the fixed-$p$ setting and conditionally on the data, the naive perturbation bootstrap version of the Alasso estimator is variable-selection consistent in the sense that it recovers the support of the Alasso estimator with probability tending to one and that its distribution conditional on the data converges at the same time to that of the Alasso estimator for the non-zero regression parameters. But the accuracy of inference for non-zero regression parameters relies on the rate of convergence of the bootstrap distribution of $\sqrt{n}(\bm{\beta}_{n}^{*N(1)}-\hat{\bm{\beta}}_{n}^{(1)})| \bm{\varepsilon}$ to the distribution of $\sqrt{n}(\hat{\bm{\beta}}_{n}^{(1)}-\bm{\beta}^{(1)})$ after proper studentization. Furthermore, Chatterjee and Lahiri (2013) showed that the convergence of the Alasso estimators of the nonzero regression coefficients to their oracle Normal distribution is quite slow, owing to the bias induced by the penalty term in (\ref{eqn:modelalasso}). Thus, it would be important for the accuracy of inference if second-order correctness can be achieved in approximating the distribution of the Alasso estimator by the perturbation bootstrap. Second-order correctness implies that the distributional approximation has a uniform error rate of $o_p(n^{-1/2})$. We show in this paper, however, that the distribution of the naive perturbation bootstrap version of the Alasso estimator, as defined by MTC(11), cannot be second order correct even in fixed dimension. For more details, see Section \ref{sec:naiveinconsistency}.
\par
We introduce a modified perturbation bootstrap for the Alasso estimator for which second order correctness does hold, even when the number of regression parameters $p=p_n$ is allowed to increase with the sample size $n$.
We also show in Proposition \ref{prop:compute} that the modified perturbation bootstrap version of the Alasso estimator (defined in Section \ref{sec:mpb}) can be computed by minimizing simple criterion functions.  This makes our bootstrap procedure computationally simple and inexpensive. 

 In this paper, we consider some pivotal quantities based on Alasso estimators and establish that the modified perturbation bootstrap estimates the distribution of these pivotal quantities up to second order, i.e. with an error that is of much smaller magnitude than what we would obtain by using the Normal approximation under the knowledge of the true active set of covariates.  We will refer to the Normal approximation which uses knowledge of the true set of active covariates as the oracle Normal approximation.  Our main results show that the modified perturbation bootstrap method enables, for example, the construction of confidence intervals for the nonzero regression coefficients with smaller coverage error than those based on the oracle Normal approximation.
\par
More precisely, we consider pivots which are studentizations of the quantities
\[
\sqrt{n}\bm{D}_n(\hat{\bm{\beta}}_n - \bm{\beta}) \quad \text{ and } \quad \sqrt{n}\bm{D}_n(\hat{\bm{\beta}}_n - \bm{\beta}) + \breve{\bm{b}}_n,
\]
where $\bm{D}_n$ is a $q\times p$ matrix ($q$ fixed) producing $q$ linear combinations of interest of $\hat{\bm{\beta}}_n - \bm{\beta}$ and where $\breve{\bm{b}}_n$ is a bias correction term which we will define in section \ref{sec:mainresults}.  We find that in the $p\leq n$ case, the modified perturbation bootstrap can estimate the distribution of the first pivot with an error of order $o_p(n^{-1/2})$ (see Theorem \ref{thm:RstarloD}). This is much smaller than the error of the oracle Normal approximation, which was shown in Theorem 3.1 of Chatterjee and Lahiri (2013) to be of the order $O_p(n^{-1/2}+||\bm{b}_n||+c_n)$, where $\bm{b}_n$ is the bias targeted by $\breve{\bm{b}}_n$ and $c_n>0$ is determined by the initial estimator $\tilde{\bm{\beta}}_n$ and the tuning parameters $\lambda_n$ and $\gamma$; both $||\bm{b}_n||$ and $c_n$ are typically greater in magnitude than $n^{-1/2}$  and hence determine the rate of the oracle Normal approximation. We also discover that the bias correction in the second pivot improves the error rate so that the modified perturbation bootstrap estimator achieves the rate $O_p(n^{-1})$ (see Theorem \ref{thm:RcheckstarloD}), which is a significant improvement over the best possible rate of oracle Normal approximation, namely $O(n^{-1/2})$.  In the $p>n$ case, we find that the modified perturbation bootstrap estimates the distributions of studentized versions of both the bias-corrected and un-bias-corrected pivots with the rate $o_p(n^{-1/2})$ (see Theorems \ref{thm:hiD1}, \ref{thm:hiD2} and \ref{thm:hiD3}), establishing the second-order correctness of our modified perturbation bootstrap in the high-dimensional setting. We have explored the cases when the dimension $p$ is increasing polynomially with $n$ and when $p$ is increasing exponentially with $n$. Our adding to the pivot a bias correction term may bring to mind the desparsified Lasso introduced independently by Zhang and Zhang (2014) and van de Geer et al. (2014); these authors construct a nonsparse estimator of $\bm{\beta}$ by adding a Lasso-based bias correction to a biased, nonsparse estimator of $\bm{\beta}$ which is linear in the response values $y_1,\dots,y_n$.  They consider first-order properties of this nonsparse estimator, establishing asymptotic normality under sparsity conditions.  In contrast, we consider the sparse Alasso estimator of $\bm{\beta}$ and correct the bias of a pivot based on the form of the Alasso estimator. We establish second order results of our proposed perturbation bootstrap method in both before and after bias correction. The main motivation behind the bias correction is to achieve the error rate $O_p(n^{-1})$. 
\par
We show that the naive perturbation bootstrap of MTC(11) is not second-order correct (see Theorem \ref{thm:naiveinconsistent}) by investigating the Karush-Kuhn-Tucker (KKT) condition [cf. Boyd and Lieven (2004)] corresponding to their minimization problem. It is shown that second order correctness is not attainable by the naive version of the perturbation bootstrap, primarily due to lack of proper centering of the naive bootstrapped Alasso criterion function. We derive the form of the centering constant by analyzing the corresponding approximation errors using the theory of Edgeworth expansion. To accommodate the centering correction, we modify the perturbation bootstrap criterion function for the Alasso; see Section \ref{sec:mpb} for details. In addition, we also find out that it is beneficial, from both theoretical and computational perspectives, to modify the perturbation bootstrap version of the initial estimators in a similar way. To prove second order correctness of the modified perturbation bootstrap Alasso, the key steps are to find an Edgeworth expansion of the bootstrap pivotal quantities based on the modified criterion function and to compare it with the Edgeworth expansion of the sample pivots. We want to mention that the dimension $p$ of the regression parameter vector can grow polynomially in the sample size $n$ at a rate depending on the number of finite polynomial moments of the error distribution. Extension to the case in which $p$ grows exponentially with $n$ is possible under the assumption of finiteness of moment generating function of the regression errors. In this regime, we have explored separately two important special cases, namely when the errors are Sub-Gaussian and Sub-Exponential.
\par
We conclude this section with a brief literature review. The perturbation bootstrap was introduced by Jin, Ying, and Wei (2001) as a resampling procedure where the objective function has a U-process structure. Work on the perturbation bootstrap in the linear regression setup is limited. Some work has been carried out by Chatterjee and Bose (2005), MTC(11), Zhou, Song and Thompson (2012), and Das and Lahiri (2016). As a variable selection procedure, Tibshirani (1996) introduced the Lasso. Zou (2006) proposed the Alasso as an improvement over the Lasso. For the Alasso and related popular penalized estimation and variable selection procedures, the residual bootstrap has been investigated by Knight and Fu (2000), Hall, Lee and Park (2009), Chatterjee and Lahiri (2010, 2011, 2013), Wang and Song (2011), MTC(11),  Van De Geer et al. (2014), and Camponovo (2015), among others.
\par
The rest of the paper is organized as follows. The modified perturbation bootstrap for the Alasso is introduced and discussed in Section \ref{sec:mpb}. Assumptions and explanations of those are presented in Section \ref{sec:assum}. Negative results on the naive perturbation bootstrap approximation proposed by MTC(11) are discussed in \ref{sec:naiveinconsistency}. Main results concerning the estimation properties of the studentized modified perturbation bootstrap pivotal quantities as well as intuitions and explanations behind the modification of the modified perturbation bootstrap are given in Section \ref{sec:mainresults}. Section \ref{sec:simulation} presents simulation results exploring the finite-sample performance of the modified perturbation bootstrap in comparison with other methods for constructing confidence intervals based on Alasso estimators. Proofs are presented in Section \ref{sec:proofs}. Section \ref{sec:1.777} states concluding remarks.

\section{The modified perturbation bootstrap for the ALASSO}
\label{sec:mpb}
Let $G_1^*,\ldots, G_n^*$ be $n$ independent copies of a non-degenerate random variable $G^* \in [0,\infty)$ having expectation $\mu_{G^*}$. These quantities will serve as perturbation quantities in the construction of the perturbation bootstrap Alasso estimator. 
We define our bootstrap version of the Alasso estimator as the minimizer of a carefully constructed penalized objective function which involves the Alasso predicted values $\hat{y}_i = \bm{x}'_i\bm{\hat{\beta}}_n$, $i=1,\dots,n$ as well as the observed values $y_i,\dots,y_n$.
These sets of values appear in the objective function in two perturbed least-squares criteria. 
Similar modification is also needed in defining the bootstrap versions of the Alasso initial estimators, see  (\ref{eqn:mpbi}). The motivation behind this construction is detailed in Section \ref{sec:naiveinconsistency}. We point out in Section \ref{sec:mainresults} why the naive perturbation bootstrap formulation of MTC(11) fails to achieve second order correctness.

We formally define the modified perturbation bootstrap version $\bm{\hat{\beta}}_n^*$ of the Alasso estimator $\bm{\hat{\beta}}_n$  as
\begin{align}\label{eqn:mpb}
\bm{\hat{\beta}_n^*} = \operatorname*{arg\,min}_{\bm{t}^*}&\Bigg[\sum_{i=1}^{n}(y_i - \bm{x}'_i \bm{t}^*)^2(G^*_i-\mu_{G^*}) \nonumber\\
&+\sum_{i=1}^{n}(\hat{y}_i-\bm{x}'_i\bm{t}^*)^2(2\mu_{G^*}-G_i^*)+\mu_{G^*}\lambda_n\sum_{j=1}^{p}|\tilde{\beta}_{j,n}^*|^{-\gamma}|t_{j}^*|\Bigg],
\end{align}
where $\tilde{\beta}_{j,n}^*$ is the $j$th component of $\tilde{\bm{\beta}}_{n}^*$, the modified perturbation bootstrap version of the Alasso initial estimator $\tilde{\bm{\beta}}_n$. We construct $\tilde{\bm{\beta}}^*_n$ as
\begin{align}\label{eqn:mpbi}
\tilde{\bm{\beta}}_n^* = \operatorname*{arg\,min}_{\bm{t}^*}&\Bigg[\sum_{i=1}^{n}(y_i - \bm{x}'_i \bm{t}^*)^2(G^*_i-\mu_{G^*}) \nonumber\\
&+\sum_{i=1}^{n}(\hat{y}_i-\bm{x}'_i\bm{t}^*)^2(2\mu_{G^*}-G_i^*)+\mu_{G^*}\tilde{\lambda}_n\sum_{j=1}^{p}|t_{j}^*|^l\Bigg],
\end{align}
where $\tilde{\lambda}_n=0$ when $\tilde{\bm{\beta}}_n$ is taken as the OLS, which we use when $p\leq n$, and $l=1$ or $2$ according as the initial estimator $\tilde{\bm{\beta}}_n$ is taken as the Lasso or Ridge regression estimator when $p>n$. Note that $\tilde{\lambda}_n$ may be different from $\lambda_n$.



\par
We point out that the modified perturbation bootstrap estimators can be computed using existing algorithms. 
Define $\bm{L}_1(\bm{t}) = \sum_{i=1}^{n}(y_i - \bm{x}'_i \bm{t})^2(G^*_i-\mu_{G^*})+\sum_{i=1}^{n}(\hat{y}_i-\bm{x}'_i\bm{t})^2(2\mu_{G^*}-G_i^*)+\mu_{G^*}\tilde{\lambda}_n\sum_{j=1}^{p}c_j|t_{j}|^l$ 
for some non-negative constants $c_j$, $j =1,\cdots,p$. Now set $z_i=\hat{y}_i+\hat{\epsilon}_i\mu_{G^*}^{-1}(G_i^*-\mu_{G^*})$, where $\hat \epsilon_i = y_i - \hat{y}_i$ for $i=1,\dots,n$ and let $\bm{L}_2(\bm{t})=\sum_{i=1}^{n}\big(z_i-\bm{x}'_i\bm{t}\big)^2+\tilde{\lambda}_n\sum_{j=1}^{p}c_j|t_{j}|^l$. Then we have the following proposition.

\begin{prop}\label{prop:compute}
$\operatorname*{arg\,min}_{\bm{t}}\bm{L}_1(\bm{t})=\operatorname*{arg\,min}_{\bm{t}}\bm{L}_2(\bm{t})$.
\end{prop}

This proposition allows us to compute $\tilde{\bm{\beta}}^*_n$ as well as $\hat{\bm{\beta}}^*_n$ by minimizing standard objective functions on some pseudo-values. Note that the modified perturbation bootstrap versions of the Alasso estimator as well as of the Alasso initial estimator can be obtained simply by properly perturbing the Alasso residuals in the decomposition $y_i=\hat{y}_i + \hat{\epsilon}_i$, $i = 1,\dots,n$.

\section{Assumptions}\label{sec:assum}

We first introduce some notations required for stating our assumptions and useful for the proofs later. We denote the true parameter vector as $\bm{\beta}_n = (\beta_{1,n},\dots,\beta_{p,n} )'$, where the subscript $n$ emphasizes that the dimension $p:=p_n$ may grow with the sample size $n$.
Set $\mathcal{A}_n=\{j: \beta_{j,n}\neq 0\}$ and  $p_0:=p_{0,n}=|\mathcal{A}_n|$. For simplicity, we shall suppress the subscript $n$ in the notations $p_n$ and $p_{0n}$. Without loss of generality, we shall assume that $\mathcal{A}_n=\{1,\dots,p_0\}$. Let $\bm{C}_n=n^{-1}\sum_{i=1}^{n}\bm{x}_i\bm{x}'_i$ and partition it according to $\mathcal{A}_n = \{1,\dots,p_0\}$ as 
\begin{equation*}
\bm{C}_{n} = \begin{bmatrix}
\bm{C}_{11,n} \;\;\;\bm{C}_{12,n}\\
\bm{C}_{21,n}\;\;\; \bm{C}_{22,n}
\end{bmatrix},
\end{equation*}

where $\bm{C}_{11,n}$ is of dimension $p_0\times p_0$. Define $\tilde{\bm{x}}_i=\bm{C}_n^{-1}\bm{x}_i$ (when $p\leq n$) and $sgn(x) =-1, 0 ,1$ according as $x<0$, $x=0$, $x>0$, respectively. Suppose $\bm{D}_n$ is a known $q\times p$ matrix with $\text{tr}(\bm{D}_n\bm{D}'_n)=O(1)$ and $q$ is not dependent on $n$. Let $\bm{D}_n^{(1)}$ contain the first $p_0$ columns of $\bm{D}_n$.

Define
\begin{align*}
\bm{S}_{n} = \begin{bmatrix}
\bm{D}_n^{(1)}\bm{C}_{11,n}^{-1}\bm{D}_n^{(1)'}.\sigma^2 \;\;\;\bm{D}_n^{(1)}\bm{C}_{11,n}^{-1}\bar{\bm{x}}^{(1)}_n.\mu_{3}\\
\bar{\bm{x}}^{(1)'}_n\bm{C}_{11,n}^{-1}\bm{D}_n^{(1)'}.\mu_3\;\;\;\;\;\;\;\;\; (\mu_4-\sigma^4)
\end{bmatrix},
\end{align*}
where $\bar{\bm{x}}_n=n^{-1}\sum_{i=1}^{n}\bm{x}_i=(\bar{\bm{x}}^{(1)\prime}_n,\bar{\bm{x}}^{(2)\prime}_n)^\prime$, $\sigma^2=\mathbf{Var}(\epsilon_1)=\mathbf{E}(\epsilon_1^2)$, and where $\mu_3$ and $\mu_4$ are, respectively, the third and fourth central moments of $\epsilon_1$. Define in addition the $q\times p_0$ matrix $\check{\bm{D}}_n^{(1)}=\bm{D}_n^{(1)}\bm{C}_{11,n}^{-1/2}$ and the $p_0 \times 1$ vector $\check{\bm{x}}_i^{(1)}=\bm{C}_{11,n}^{-1/2}\bm{x}_i^{(1)}$. Let $K$ be a positive constant and $r$ be a positive integer $\geq 3$ unless otherwise specified. $||\cdot||$ and $||\cdot||_{\infty}$ respectively denote the Euclidean norm and the Sup norm. $c\wedge d$ denotes $\min\{c, d\}$ for two real numbers $c$ and $d$. By $\mathbf{P_*}$ and $\mathbf{E_*}$ we denote, respectively, probability and expectation with respect to the distribution of $G^{*}$ conditional upon the observed data. 




\par
We now introduce our assumptions. 
\begin{enumerate}[label=(A.\arabic*)]
\item Let $\eta_{11,n}$ denote the smallest eigenvalue of the matrix $\bm{C}_{11,n}$.
\begin{enumerate}[label=(\roman*)]
 \item $\eta_{11,n}>Kn^{-a}$ for some $a\in [0,1)$.
  \item $\max\{n^{-1}\sum_{i=1}^{n}|x_{i,j}|^{2r}:1\leq j \leq p\}+\{n^{-1}\sum_{i=1}^{n}\big|(\bm{C}_{11,n}^{-1})_{j.}\bm{x}_{i}^{(1)}\big|^{2r}:1\leq j \leq p_0\} = O(1)$.
  \item $\max\{n^{-1}\sum_{i=1}^{n}|\tilde{x}_{i,j}|^{2r}:1\leq j \leq p\} = O(1)$, where $\tilde{x}_{i,j}$ is the $j$th element of $\tilde{\bm{x}}_i$. (when $p\leq n$)
	\item[(iii)$'$]$\max\{c_{11,n}^{j,j}: 1\leq j\leq p_0\}=O(1)$, where $c_{11,n}^{j,j}$ is the $(j,j)$th element of $C_{11,n}^{-1}$. (when $p > n$)	
\end{enumerate}
\item There exists a $\delta \in (0,1)$ such that for all $n>\delta^{-1}$,
\begin{enumerate}[label=(\roman*)]
  \item sup$\{\bm{x}'\check{\bm{D}}_n^{(1)}\check{\bm{D}}_n^{(1)'}\bm{x}:\bm{x}\in \mathcal{R}^{q}, ||\bm{x}||=1\}<\delta^{-1}$.
  \item $n^{-1}\sum_{i=1}^{n}||\check{\bm{D}}_n^{(1)}\check{\bm{x}}_i^{(1)}\check{\bm{x}}_i^{(1)'}\check{\bm{D}}_n^{(1)'}||^r = O(1)$. 
  \item inf$\{\bm{x}'\bm{S}_n\bm{x}:\bm{x}\in \mathcal{R}^{q+1}, ||\bm{x}||=1\}>\delta$.
\end{enumerate}
\item $\max\{|\beta_{j,n}|:j\in \mathcal{A}_n\}=O(1)$ and min$\{|\beta_{j,n}|:j\in \mathcal{A}_n\}\geq Kn^{-b}$ for some $b\geq 0$ such that $4b< 1$ and $a+2b\leq 1$, where $a$ is defined as in (A.1)(i).
\item \begin{enumerate}[label=(\roman*)]
  \item $\mathbf{E}|\epsilon_1|^{r}< \infty$. $\mathbf{E}\epsilon_1=0$.
  \item $(\epsilon_1, \epsilon_1^{2})$ satisfies Cramer's condition:\\ 
\hspace*{5mm}$\limsup_{||(t_1,t_2)||\rightarrow \infty}\mathbf{E}(exp(i(t_1\epsilon_1+t_2\epsilon_1^{2})))<1$.
\end{enumerate}
\item \begin{enumerate}[label=(\roman*)]
  \item $\mathbf{E_*}(G_{1}^{*})^{r} < \infty$. 
$\mathbf{Var}(G_1^*) =\sigma^2_{G^*}= \mu_{G^*}^2$, $\mathbf{E_*}(G_1^* - \mu_{G^*})^3 = \mu_{G^*}^3$.
  \item $G_{i}^{*}$ and $\epsilon_i$ are independent for all $1\leq i\leq n$.

\item $((G^*_1-\mu_{G^*}), (G_1^* - \mu_{G^*})^{2})$ satisfies Cramer's condition:\\ 
\hspace*{9mm}$\limsup_{||(t_1,t_2)||\rightarrow \infty}\mathbf{E_*}(exp(i(t_1(G_1^*-\mu_{G^*})+t_2(G_1^* - \mu_{G^*})^{2})))<1$
\end{enumerate}
\item There exists $\delta_1 \in (0,1)$ such that for all $n>\delta_1^{-1}$,

\begin{enumerate} [label=(\roman*)]
\item
$\dfrac{\lambda_n}{\sqrt{n}}\leq \delta_1^{-1}n^{-\delta_1} \text{min}\Big{\{}\dfrac{n^{-b\gamma}}{p_0},\dfrac{n^{-b\gamma-a/2}}{\sqrt{p_0}}\Big{\}}$.
\item $\dfrac{\lambda_n}{\sqrt{n}} n^{\gamma/2} \geq \delta_1 n^{\delta_1} p_0$
\item $p_{0}=o\big(n^{1/2}(\log n)^{-3/2}\big)$.

\end{enumerate}
\item[(A.7)] There exists $C\in(0,\infty)$ and $\delta_2\in(0,\gamma^{-1}\delta_1)$, $\delta_1$ being defined in the assumption (A.6), such that
\begin{align*}
&\mathbf{P}\Big(\max\{\big|\sqrt{n}(\tilde{\beta}_{j,n}-\beta_{j,n})\big|:1\leq j\leq p\}>C.n^{\delta_2}\Big)=o(n^{-1/2})\\
& \mathbf{P_*}\Big(\max\{\big|\sqrt{n}(\tilde{\beta}_{j,n}^*-\hat{\beta}_{j,n})\big|:1\leq j\leq p\}>C.n^{\delta_2}\Big)=o_p(n^{-1/2})
\end{align*}

\end{enumerate}
\par

Now we explain the assumptions briefly.  Assumption (A.1) describes the regularity conditions needed on the growth of the design vectors. Assumption (A.1)(i) is a restriction on the smallest eigenvalue of $\bm{C}_{11,n}$. Assumption (A.1)(i) is a weaker condition than assuming that $\bm{C}_{11,n}$ converges to a positive definite matrix. (A.1)(ii) and (iii) are needed to bound the weighted sums of types $\big[\sum_{i=1}^{n}\bm{x}_i\epsilon_i\big]$, $\big[\sum_{i=1}^{n}\tilde{\bm{x}}_i\epsilon_i\big]$,  $\big[\bm{C}_{11,n}^{-1}\sum_{i=1}^{n}\bm{x}_i^{(1)}\epsilon_i\big]$ (second one only when $p\leq n$). For $r=2$ (A.1)(iii) is equivalent to the condition that the diagonal elements of the matrix $\bm{C}_n^{-1}$ are uniformly bounded. Also for general value of $r$, (A.1)(ii) and (iii) are much weaker than conditioning on $l_r$-norms of the design vectors. Here the value of $r$ is specified by the underlying Edgeworth expansion. Assumption (A.1)(iii) requires $p\leq n$ and hence is not defined when $p>n$. Note that the condition (A.1)(iii)$'$ needs $p_0\leq n$ which is true in our setup due to assumption (A.6)(iii).

\par
Assumptions (A.2)(i) bounds the eigenvalues of the matrix $\bm{D}_n^{(1)}\bm{C}_{11,n}^{-1}\bm{D}_n^{(1)'}$ away from infinity. It is necessary to obtain bounds needed in the studentized setup. Assumption (A.2)(ii) is a condition similar to the conditions in (A.1)(ii) and (iii); but involving the $q\times p$ matrix $\bm{D}_n$. This condition is needed for showing necessary closeness of the covariance matrix estimators $\breve{\bm{\Sigma}}_n, \tilde{\bm{\Sigma}}_n$ [defined in Section \ref{sec:mainresults}] to their population counterparts (for details see Lemma \ref{lem:Sigma}).  
Assumption (A.2)(iii) bounds the minimum eigen value of the matrix $S_n$ away from $0$. This condition along with the Cramer conditions given in (A.4) and (A.5) enable certain Edgeworth expansions.
\par
Assumption (A.3) separates the relevant covariates from the non-relevant ones. The condition on the minimum is needed to ensure that the non-zero regression coefficients cannot converge to zero faster than the error rate, that is not faster than $O(n^{-1/2})$. We mention that one can assume $b<1/2$ instead of assuming $b<1/4$, but with the price of putting another restriction on the penalty parameter $\lambda_n$. We do not consider such a setting here. {We also want to point  out that it is not possible to relax this minimal signal condition by the bias correction, considered in Section \ref{sec:mainresults}.  With further relaxation, the bias of the Alasso estimator will be larger than the estimation error which is of order $O_p(n^{-1/2})$ and hence second-order correctness cannot be achieved by perturbation bootstrap in more relaxed minimal signal condition.}
\par
Assumption (A.4)(i) is a moment condition on the error term needed for valid Edgeworth expansion. Assumption (A.4)(ii) is Cramer's condition on the errors, which is very common in the literature of Edgeworth expansions; it is satisfied when the distribution of $(\epsilon_1, \epsilon_1^{2})$ has a non-degenerate component which is absolutely continuous with respect to the Lebesgue measure [cf. Hall (1992)]. Assumption (A.4)(ii) is only needed to get a valid Edgeworth expansion for the original Alasso estimator in the studentized setup. Assumptions (A.5)(i) and (iii) are the analogous conditions that are needed on the perturbing random quantities to get a valid Edgeworth expansion in the bootstrap setting. Assumption (A.5)(ii) is natural, since the $\epsilon_i$ are present already in the data generating process, whereas $G_i^*$ are introduced by the user. One can look for Generalized Beta and Generalized Gamma families for suitable choices of the distribution of $G^*$. The pdf of Generalized Beta family of distributions is
\begin{equation*}
  GB(y;f,g,h,\omega, \rho)=\left\{
  \begin{array}{@{}ll@{}}
    \dfrac{|f|y^{f\omega-1}\Big(1-(1-c)(y/g)^f\Big)^{\rho-1}}{g^{f\omega}B(\omega,\rho)\Big(1+c(y/g)^f\Big)^{\omega+\rho}}\;\;\;& \text{for}\;\; 0<y^f<\dfrac{g^f}{1-h}\\
     0\;\; &  \text{otherwise}
  \end{array}\right.
\end{equation*} 
where $0\leq h\leq 1$ and other parameters are all positive. We interpret $1/0$ as $\infty$. The function $B(\omega,\rho)$ is the beta function. Choices of the distribution of $G^*$ can be obtained by finding solution of $(f,g,h,\omega,\rho)$ from the following two equations
\begin{align*}
\dfrac{B(\omega+2/f,\rho)}{B(\omega,\rho)}& {}_2F_1\big[\omega+2/f,2/f;h;\omega+\rho+2/f\big]\\
&=2\bigg[\dfrac{B(\omega+1/f,\rho)}{B(\omega,\rho)}{}_2F_1\big[\omega+1/f,1/f;h;\omega+\rho+1/f\big]\bigg]^2\\
\text{and}\;\; \dfrac{B(\omega+3/f,\rho)}{B(\omega,\rho)}{}_2&F_1\big[\omega+3/f,3/f;h;\omega+\rho+3/f\big]\\
&=5\bigg[\dfrac{B(\omega+1/f,\rho)}{B(\omega,\rho)}{}_2F_1\big[\omega+1/f,1/f;h;\omega+\rho+1/f\big]\bigg]^3
\end{align*}
where ${}_2F_1$ denotes hypergeometric series. The pdf of Generalized Gamma family of distributions is given by
\begin{equation*}
  GG(y;\omega, \rho, \nu)=\left\{
  \begin{array}{@{}ll@{}}
    \dfrac{(\nu/\omega^\rho)y^{\rho-1}e^{(y/\omega)^{\nu}}}{\Gamma(\rho/\nu)}\;\;\;& \text{for}\;\; y>0\\
     0\;\; &  \text{otherwise}
  \end{array}\right.
\end{equation*} 
where all the parameters are positive and $\Gamma(\cdot)$ denotes the gamma function. For this family, the suitable choices of the distribution of $G^*$ can be obtained by considering any positive value of the parameter $\omega$ and solving the following two equations for $(\rho,\nu)$,
\begin{align*}
&\Big[\Gamma((\rho+2)/\nu)\Big]* \Gamma(\rho/\nu) = 2\Big[\Gamma((\rho+1)/\nu)\Big]^2\\
\text{and}\;\; &\Big[\Gamma((\rho+3)/\nu)\Big]* \Big[\Gamma(\rho/\nu)\Big]^2 = 5\Big[\Gamma((\rho+1)/\nu)\Big]^3.
\end{align*}
One immediate choice of the distribution of $G^*$ from Generalized Beta family is the Beta$(\alpha, \beta)$ distribution with $3\alpha=\beta=3/2$. We have utilized this distribution as the distribution of the perturbing quantities $G_i^*$'s in our simulations, presented in Section \ref{sec:simulation}. Outside these two generalized family of distributions, one possible choice is the distribution of $(M_1+M_2)$ where $M_1$ and $M_2$ are independent and $M_1$ is a Gamma random variable with shape and scale parameters $0.008652$ and $2$ respectively and $M_2$ is a Beta random variable with both the parameters $0.036490$. Another possible choice is the distribution of $(M_3+M_4)$ where $M_3$ and $M_4$ are independent and $M_3$ is an Exponential random variable with mean $\big{(}79-15\sqrt{33}\big{)}/16$ and $M_4$ is an Inverse Gamma random variable with both shape and scale parameters $\big{(}4 + \sqrt{11/3}\big{)}$.

Assumptions (A.6)(i) and (ii) can be compared with the condition (c) $\lambda_n/\sqrt{n}$ $\rightarrow 0$ and $n^{\gamma/2}\lambda_n/\sqrt{n}\rightarrow \infty$ [cf. Zou (2006), Caner and Fan (2010)]. Whereas (c) is ensuring the oracle normal approximation, (A.6)(i) and (ii) are required for obtaining Edgeworth expansions. Lastly, (A.6)(iii) limits how quickly the number of non-zero regression coefficients may grow. Though it would seem that $p_0=O(n)$ with $p_0\leq n$ should be a sufficient restriction on the growth rate of $p_0$ for approximating the distribution of the Alasso estimator, a careful analysis reveals that further reduction in the growth rate of $p_0$ is necessary for accommodating the studentization. Clearly it is difficult to comprehend what possible choices of $p_0,\lambda_n, \gamma, a, b$ would satisfy the assumptions presented in (A.6). Thus it is better to present some possible choices of those parameters.

First consider $a=0$ and $b=0$, that is assume that the smallest eigenvalue of $\bm{C}_{11,n}$ and the smallest non-zero regression coefficients are bounded away from 0. In that case it is easy to check that one set of possible choices are $p_0 = O(n^{\gamma/5})$ and $\lambda_n = C.n^{1/2-\gamma/4}$ for some constant $C> 0$, provided $\gamma \in (0,2)$. In particular if $\gamma =1$ then the choices of $p_0$ and $\lambda_n$ maybe respectively $p_0=O(n^{1/5})$ and $\lambda_n = C. n^{1/4}$ when $a=b=0$. Again $p_0$ can grow with $n$ at the rate $o(n^{1/2}(\log n)^{-3/2})$, when $\gamma>2$ and $\lambda_n = C. n^{(2-\gamma)/6}$ for some constant $C>0$ whenever $a=b=0$.

In general if $a\in [0,1/2)$ and $b<1/4$, then it can be shown that the possible choices of $\gamma$, $p_0$ and $\lambda_n$ are respectively $4a/(1-2b)< \gamma < 2/(1+2b)$, $p_0 = O(n^{[(1-2b)\gamma]/5})$ and $\lambda_n = C.n^{1/2-\gamma/4-b\gamma/2}$ for some constant $C> 0$. On the other hand if $a \in [1/2,1)$ and $a+2b<1$, one set of possible choices would be $\gamma\geq 2$, $p_0 = O(n^{2/3-(a+2b\gamma+4c)/3})$ and $\lambda_n = C.n^{1/6-(a+2b\gamma+c)/3}$ for some constants $c, C >0$. With $a=1/2$ and $b =0$, clearly the choices of $p_0$ and $\lambda_n$ reduce to $p_0=O(n^{1/2-\delta})$ and $\lambda_n=C.n^{-\delta/4}$ for some $\delta, C>0$.
\par
Assumption (A.7) places deviation bounds on both the sample and bootstrap initial estimators which are needed to get valid Edgeworth expansions. These conditions are satisfied by OLS estimator in $p\leq n$ case [cf. Lemma \ref{lem:W}]. Note that non-bootstrap part of (A.7) is satisfied if there exists a linear approximation of the type $\sum_{i=1}^{n}a_{i,j}\epsilon_i$ of $\sqrt{n}(\tilde{\beta}_{j,n}-\beta_{j,n})$, where $\max\Big{\{}\sum_{i=1}^{n}|a_{i,j}|^r: 1\leq j \leq p\Big{\}}=o\big{(}p^{-1}n^{-1/2+r\delta_2}\big{)}$ and $\mathbf{E}(|\epsilon_1|^r)< \infty$ for some $r\geq 3$. The bootstrap deviation bound corresponding to (A.7) holds provided similar approximation exits with $(G_1^*-\mu_{G^*})$ in place of $\epsilon_1$. More precisely, for the Ridge estimator and for its perturbation bootstrap version defined in Section $\ref{sec:mpb}$, if for some $r\geq 4$, the conditions
\begin{enumerate}[label=(\alph*)]
\item $\mathbf{E}|\epsilon_1|^r+\mathbf{E}_{*}(G_1^{*})^r < \infty$.
\item $\max\{n^{-1}\sum_{i=1}^{n}(|\bm{x}_i|^{2r} + |\breve{\bm{x}}_i|^{2r}):1\leq j \leq p\}=O(n^{{\delta_2}/2})$ for all $i \in \{1,\cdots, n\}$.
\item $\max\big{\{}\bm{e}'_j(\bm{C}_n+\tilde{\lambda}_n n^{-1}\bm{I}_p)^{-1}\bm{\beta}_n:1\leq j \leq p\big{\}} = O(n^{(1+\delta_2)/2}\tilde{\lambda}_n^{-1})$.
\item $\sup\big{\{}\bm{e}'_j(\bm{C}_n+\tilde{\lambda}_n n^{-1}\bm{I}_p)^{-1}\bm{z}_n:||\bm{z}_n||\leq 1 \big{\}} = O(n^{(1+\delta_2)/2}\tilde{\lambda}_n^{-1})$ for all $j \in\{1,\cdots, p\}$.
\end{enumerate}
are satisfied, then the assumption (A.7) holds. Here $\{\bm{e}_1,\cdots,\bm{e}_p\}$ is the standard basis of $\mathcal{R}^p$, $\breve{\bm{x}}_i=(\bm{C}_n+\tilde{\lambda}_n n^{-1}\bm{I}_p)^{-1}\bm{x}_i$ and $\tilde{\lambda}_n$ is the penalty parameter corresponding to the Ridge estimator [cf. Sec $\ref{sec:mpb}$]. This follows analogously to proposition 8.4 of Chatterjee and Lahiri (2013) after applying Lemma $\ref{lem:concentration}$, stated in Section \ref{sec:proofs}.

\section{ Impossibility of Second-order correctness of the naive perturbation bootstrap}
\label{sec:naiveinconsistency}
In this section we describe the naive perturbation bootstrap as defined by MTC(11) for the Alasso and show that second-order correctness can not be achievable by their naive perturbation bootstrap method.  When the objective function is the usual least squares criterion function the naive perturbation bootstrap Alasso estimator $\bm{\beta}_n^{*N}$ is defined in MTC(11) as
\begin{align}\label{eqn:naivebetastar}
\bm{\beta}_n^{*N}= \operatorname*{arg\,min}_{\bm{v}_n^*}&\Bigg[\sum_{i=1}^{n}(y_i - \bm{x}'_i \bm{v}_n^*)^2G^*_i +\lambda_n^*\sum_{j=1}^{p}|\tilde{\beta}_{j,n}^{*N}|^{-\gamma}|v_{j,n}^*|\Bigg],
\end{align}
where 
\begin{enumerate}[label=(\roman*)]
\item $\lambda_n^*>0$ is such that $\lambda_n^*n^{-1/2}\rightarrow 0$ and $\lambda_n^*\rightarrow \infty$ as $n\rightarrow \infty$.
\item the initial naive bootstrap estimator is defined as 
\begin{equation*}
\tilde{\bm{\beta}}_{n}^{*N}= \operatorname*{arg\,min}_{\bm{v}_n^*}\Big[\sum_{i=1}^{n}(y_i - \bm{x}'_i \bm{v}_n^*)^2G^*_i\Big]
\end{equation*}
and $\tilde{\beta}_{j,n}^{*N}$ is the $j$th component of $\tilde{\bm{\beta}}_{n}^{*N}$.
\item $\{G_1^*,\ldots, G_n^*\}$ is a set of iid non-negative random quantities with mean and variance both equal to 1.
\end{enumerate}
\par
Note that the initial estimator $\tilde{\bm{\beta}}_{n}^{*N}$ is unique only when $p$ is less than or equal to $n$.  We now consider the quantity $\bm{u}_n^{*N}=\sqrt{n}(\bm{\beta}_n^{*N}-\bm{\hat{\beta}}_n)$, which we can show from (\ref{eqn:naivebetastar}) to be the minimizer \begin{align}\label{eqn:ustarnaive}
\bm{u}_n^{*N}= \operatorname*{arg\,min}_{\bm{w}_n^*}\Bigg[\bm{w}_n^{*'}\bm{C}_n^{*}\bm{w}_n^{*} -2\bm{w}_n^{*}\bm{W}_n^{*}+\lambda_n^*\sum_{j=1}^{p}|\tilde{\beta}_{j,n}^{*N}|^{-\gamma}\Big(|\hat{\beta}_{j,n}+\dfrac{w_{j,n}^*}{\sqrt{n}}|-|\hat{\beta}_{j,n}|\Big)\Bigg],
\end{align}
where $\hat{\beta}_{j,n}$ is the $j$th component of the Alasso estimator $\bm{\hat{\beta}}_n$, $\bm{C}_n^*=n^{-1}\sum_{i=1}^{n}\bm{x}_i\bm{x}'_iG_i^*$, and $\bm{W}_n^{*}=n^{-1/2}\sum_{i=1}^{n}\hat{\epsilon}_i\bm{x}_iG_i^*$. To describe the solution of MTC(11), assume $\mathcal{A}=\{j:\beta_j\neq 0\}=\{1,\dots,p_0\}$. MTC(11) claimed that when $\gamma=1$ and $p$ is fixed, $\left((\bm{u}_{n1}^{*N})',\bm{0}\right)'$ is a solution of (\ref{eqn:ustarnaive}) for sufficiently large $n$, where 
\begin{equation*}
\bm{u}_{n1}^{*N}=\bm{C}_{11,n}^{-1}n^{-1/2}\sum_{i=1}^{n}\epsilon_i\bm{x}_i^{(1)}(G_i^*-1) \; \text{and}\; ||\bm{u}_n^{*N}-\left((\bm{u}_{n1}^{*N})',\bm{0}\right)'||_{\infty}=o_{p_*}(1).
\end{equation*}
However, to achieve second order correctness, we need to obtain a solution $\left((\bm{u}_{n2}^{*N})',\bm{0}\right)'$ of (\ref{eqn:ustarnaive}) such that $||\bm{u}_n^{*N}-\left((\bm{u}_{n2}^{*N})',\bm{0}\right)'||_{\infty}=o_{p_*}(n^{-1/2})$. We show that such an $\bm{u}_{n2}^{*N}$ has the form
\begin{equation*}
\bm{u}_{n2}^{*N}=\bm{C}_{11,n}^{*-1}\Big{[}\bm{W}_n^{*(1)}-\dfrac{\lambda_n^*}{\sqrt{n}}\tilde{\bm{s}}_n^{*(1)}\Big{]}
\end{equation*}
for sufficiently large $n$, where $\bm{W}_n^{*(1)}$ is the first $p_0$ components of $\bm{W}_n^*$ and the $j$th component of $\tilde{\bm{s}}_n^{*(1)}$ equals to $\text{sgn}(\hat{\beta}_{j,n})||\tilde{\beta}_{jn}^{*N}|^{-\gamma}$, $j\in \mathcal{A}$ (Here we drop the subscript $n$ from the notations of true parameter values since we are considering $p$ to be fixed in this section). We establish this fact by exploring the KKT condition corresponding to (\ref{eqn:ustarnaive}), which is given by
\begin{align}\label{eqn:KKT}
2\bm{C}_n^*\bm{w}_n^*-2\bm{W}_n^{*}+\dfrac{\lambda_n^*}{\sqrt{n}}\bm{\Gamma}_n^*\bm{l}_n=\bm{0},
\end{align}
for some $\bm{l}_n=(l_{1n},\ldots, l_{pn})'$ with $l_{j,n}\in [-1,1]$ for $j=1,\ldots,p$ and $\bm{\Gamma}_n^*=\text{diag}\big(|\tilde{\beta}_{1n}^{*N}|^{-\gamma},$ $\ldots, |\tilde{\beta}_{pn}^{*N}|^{-\gamma}\big)$. 
Since $\bm{C}_n^*$ is a non-negative definite matrix, (\ref{eqn:ustarnaive}) is a convex optimization problem; hence (\ref{eqn:KKT}) is both necessary and sufficient in solving (\ref{eqn:ustarnaive}). 
\par
Note that $\bm{W}_n^*$ is not centered and hence we need to adjust the solution $\left((\bm{u}_{n2}^{*N})',\bm{0}\right)'$ for centering before investigating if the naive perturbation bootstrap can asymptotically correct the distribution of Alasso up to second order. Clearly, the centering adjustment term is $\bm{Ad}_n^{*}=\big{(}\bm{Ad}_n^{*(1)\prime},\bm{0}^{\prime}\big{)}^{\prime}$ where $\bm{Ad}_n^{*(1)}=\bm{C}_{11,n}^{*-1}n^{-1/2}\sum_{i=1}^{n}\hat{\epsilon}_i\bm{x}_i^{(1)}$. It follows from the steps of the proofs of the results of Section \ref{sec:mainresults} that we need $||\bm{Ad}_n^*||=o_{p_*}(n^{-1/2})$ to achieve second-order correctness. We show that this is indeed not the case even in the fixed $p$ setting.

\par
More precisely, we negate the second-order correctness of the naive perturbation bootstrap of MTC(11) by first showing that 
$\left((\bm{u}_{n2}^{*N})',\bm{0}'\right)'$ satisfies the KKT condition (\ref{eqn:KKT}) exactly with bootstrap probability converging to 1. Then we show that 
$\sqrt{n}||\bm{Ad}_n^*||$ diverges in bootstrap probability to $\infty$, which in turn implies that the conditional cdf of $\bm{F}_n^{*N}=\sqrt{n}\big{(}\bm{\beta}_n^{*N}-\bm{\hat{\beta}}_n\big{)}$ can not approximate the cdf of $\bm{F}_n=\sqrt{n}\big{(}\bm{\hat{\beta}}_n-\bm{\beta}\big{)}$ with the uniform  accuracy $O_p(n^{-1/2})$, needed for the validity of second-order correctness. 
We formalize these arguments in the following theorem.
\par

\begin{theo} \label{thm:naiveinconsistent}
Let $p$ be fixed and $\bm{C}_n\rightarrow \bm{C}$, a positive definite matrix. Define $Z_n^{*-1}=\sqrt{n}||\bm{Ad}_n^*||$. Suppose, $(\log n/n)^{1/2}.\max\{\lambda_n, \lambda_n^*\}\rightarrow 0$ and $(\log n)^{-(\gamma+1)/2}.\min\{\lambda_n, \lambda_n^{*}\}$ $.\min\{1,$ $n^{(\gamma-1)/2}\}\rightarrow \infty$ as $n\rightarrow \infty$. Also assume that (A.1)(i), (ii) and (A.4)(i) hold with $r=4$. Then there exists a sequence of borel sets $\{\bm{A}_n\}_{n\geq 1}$ with $\mathbf{P}(\bm{\varepsilon}_n\in \bm{A}_n)\rightarrow 1$ and given $\bm{\varepsilon}_n =(\epsilon_1,\ldots,\epsilon_n)'\in \bm{A}_n$, the following conclusions hold.
\begin{itemize}
\item[(a)]

$\mathbf{P_*}\bigg(\bm{u}_{n}^{*N}=\left((\bm{u}_{n2}^{*N})',\bm{0}'\right)'\bigg)=1-o(n^{-1/2})$.

\item[(b)]

$\mathbf{P_*}\Big{(}Z_n^*> \epsilon\Big{)}=o(n^{-1/2})$ for any $\epsilon > 0$.

\item[(c)]

$\sup\limits_{\bm{x}\in \mathcal{R}^p}\Big{|}\mathbf{P_*}\big{(}\bm{F}_n^{*N}\leq \bm{x}\big{)}-\mathbf{P}\big{(}\bm{F}_n \leq \bm{x}\big{)}\Big{|} \geq K. \dfrac{\lambda_n}{\sqrt{n}}$ for some $K>0$. 
\end{itemize}
\end{theo}

\begin{remark}Theorem \ref{thm:naiveinconsistent} (a), (b) state that the naive perturbation bootstrap is incompetent in approximating the distribution of Alasso up to second order. The fundamental reason behind second order incorrectness is the inadequate centering in the form of $\sqrt{n}(\bm{\beta}_n^{*N}-\hat{\bm{\beta}}_n)$. Although the adjustment term necessary for centering is $o_{p_*}(1)$, which essentially helps to establish distributional consistency in MTC(11), the term is coarser than $n^{-1/2}$, leading to second order incorrectness. Additionally, it is worth mentioning that studentization will also not help in achieving second-order correctness by naive perturbation bootstrap of MTC(11), since the necessary centering correction cannot be accomplished by any sort of studentization. Part (c) conveys uniformly how far the naive bootstrap cdf is from the original cdf.  
\end{remark}

\section{Modified Perturbation Bootstrap and its Higher Order Properties}
\label{sec:mainresults}
This section is divided into two sub-sections. The first one describes briefly the motivation behind considering the perturbation bootstrap modification in Alasso. The second sub-section describes higher order asymptotic properties of our modified perturbation bootstrap method.
\subsection{Motivation for the modified perturbation bootstrap}

Theorem \ref{thm:naiveinconsistent} establishes that the naive perturbation bootstrap of MTC(11) does not provide a solution for approximating the distribution of $\sqrt{n}(\bm{\hat{\beta}}_n-\bm{\beta}_n)$ up to second order. As it is mentioned earlier, the problem occurs because $\bm{W}_n^*$ is not centered. Let $\breve{\bm{W}}_n^*$ denotes the centered version of $\bm{W}_n^*$, that is $\breve{\bm{W}}_n^*=n^{-1/2}\sum_{i=1}^{n}\hat{\epsilon}_i\bm{x}_i(G_i^*-\mu_{G^*})$, and consider the vector equation
\begin{align}\label{eqn:mKKT}
2\bm{C}_n^*\bm{w}_n^*-2\breve{\bm{W}}_n^{*}+\dfrac{\lambda_n^*}{\sqrt{n}}\bm{\Gamma}_n^*\bm{l}_n=\bm{0},
\end{align}
which is same as (\ref{eqn:KKT}) after replacing $\bm{W}_n^*$ with $\breve{\bm{W}}_n^*$. Note that the solution to (\ref{eqn:mKKT}) is of the form $((\bm{u}_{n3}^{*(1)})',\bm{0}')'$, where $\bm{u}_{n3}^{*(1)}=\bm{C}_{11,n}^{*-1}\Big{[}\breve{\bm{W}}_n^{*(1)}-\dfrac{\lambda_n^*}{\sqrt{n}}\tilde{\bm{s}}_n^{*(1)}\Big{]}$. Although this form is adequate for achieving second-order correctness in fixed dimension, there are some computational and higher-dimensional issues that we now address.

\par
Note that $\bm{C}_{11,n}^*$ is a matrix involving random quantities $\{G_1^*,\dots,G_n^*\}$. Thus $\bm{C}_{11,n}^*$ will not remain same for each bootstrap iteration and hence each bootstrap iteration will require computing the inverse of $\bm{C}_{11,n}^*$ afresh. This is computationally expensive and the expense increases as the number of non-zero regression parameters increases. Therefore it will be computationally advantageous if we can replace $\bm{C}_{11,n}^*$ by $\bm{C}_{11,n}$ in the form of $\bm{u}_{n3}^{*(1)}$.

\par
Now define, $\bm{u}_{n4}^{*(1)}=\bm{C}_{11,n}^{-1}\Big{[}\breve{\bm{W}}_n^{*(1)}-\dfrac{\lambda_n^*}{\sqrt{n}}\tilde{\bm{s}}_n^{*(1)}\Big{]}$. If we look closely at the bias term $-\dfrac{\lambda_n^*}{\sqrt{n}}\bm{C}_{11,n}^{-1}$ $\tilde{\bm{s}}_n^{*(1)}$, then it is clear that the primary contribution of the bias towards $\bm{u}_{n4}^{*(1)}$ is $-\dfrac{\lambda_n^*}{\sqrt{n}}\bm{C}_{11,n}^{-1}\tilde{\bm{s}}_n^{(1)}$, where $j$th component of $\tilde{\bm{s}}_n^{(1)}$ is equal to $\text{sgn}(\hat{\beta}_{j,n})||\tilde{\beta}_{jn}|^{-\gamma}$, $j\in \mathcal{A}$, where $\tilde{\beta}_{j,n}$ is the $j$th component of the OLS estimator $\tilde{\bm{\beta}}_n$. By Taylor's expansion, $\big{(}\tilde{\bm{s}}_n^{*(1)}-\tilde{\bm{s}}_n^{(1)}\big{)}$ depends on the OLS residuals. The OLS residuals again depend on all $p$ estimated regression parameters, unlike Alasso residuals which depend only on the estimates of the $p_0$ non-zero components.  Since it is needed to bound $||\tilde{\bm{s}}_n^{*(1)}-\tilde{\bm{s}}_n^{(1)}||_{\infty}$ for achieving valid edgeworth expansion, we will come up with an implicit bound on the dimension $p$, which we do not want to impose. On the other hand, if the difference depends on Alasso residuals instead of OLS ones, then the implicit condition will be on $p_0$ and this is reasonable as $p_0$ can be much smaller than $p$. Additionally, $\tilde{\bm{\beta}}_n^{*N}$ involves inversion of the random matrix $\bm{C}_n^*$ and hence it is computationally expensive. Thus if $\bm{C}_n^*$ can be replaced by some fixed matrix, say $\bm{C}_n$, then the bootstrap will be computationally advantageous.

However, if we implement the modification described in Section \ref{sec:mpb}, then both the theoretical and computational shortcomings of the perturbation bootstrap method become resolved and the second-order correctness is achieved even in increasing dimension under some mild regularity conditions. Additionally, we also have the nice structure due to the modification, which enables us to employ existing computational algorithms, as pointed out in Proposition \ref{prop:compute}.

\subsection{Higher Order Results}
Define, $\bm{T}_n=\sqrt{n}\bm{D}_n(\bm{\hat{\beta}}_n-\bm{\beta}_n)$. Without loss of generality we assume that $\mathcal{A}_n=\{j: \beta_{j,n}\neq 0\}=\{1,\dots,p_0\}$. Hence, by Taylor's expansion it is immediate from the form of Alasso estimator that $\bm{\Sigma}_n=n^{-1}\sum_{i=1}^{n}\big(\bm{\xi}_i^{(0)}+\bm{\eta}_i^{(0)}\big)\big(\bm{\xi}_i^{(0)}+\bm{\eta}_i^{(0)}\big)'$ or $\bar{\bm{\Sigma}}_n=n^{-1}\sum_{i=1}^{n}\bm{\xi}_i^{(0)}\bm{\xi}_i^{\prime(0)}$ can be considered as the asymptotic variance of $\bm{T}_n/\sigma$ at sample size $n$. Here $\bm{\xi_i}^{(0)}=\bm{D}_n^{(1)}\bm{C}_{11,n}^{-1}\bm{x}_i^{(1)}$, $\bm{\eta}_i^{(0)}=\bm{D}_n^{(1)}\bm{C}_{11,n}^{-1}\bm{\eta}_i$. For each $i\in\{1,\ldots,n\}$, $\bm{\eta}_i$ is a $p_0\times 1$ vector with $j$th element
$\Big(\dfrac{\lambda_n}{2n}\tilde{x}_{i,j}\dfrac{\gamma}{|\beta_{j,n}|^{\gamma+1}}sgn(\beta_{j,n})\Big)$ where $\tilde{\bm{x}}_i=\bm{C}_n^{-1}\bm{x}_i$ (when $p\leq n$) and $sgn(x) =-1, 0 ,1$ according as $x<0$, $x=0$, $x>0$, respectively, as defined earlier.  The bias corresponding to $\bm{T}_n$ is $-\bm{b}_n=-\bm{D}_n^{(1)}\bm{C}_{11,n}^{-1}$ $\bm{s}_n^{(1)}\dfrac{\lambda_n}{2\sqrt{n}}$, where $\bm{D}_n^{(1)}$ and $\bm{C}_{11,n}$ are as defined earlier and $\bm{s}_n^{(1)}$ is a $p_0\times 1$ vector with $j$th element $\text{sgn}(\beta_{j,n})|\beta_{j,n}|^{-\gamma}$. Although $\bar{\bm{\Sigma}}_n$ is defined for all $p$, $\bm{\Sigma}_n$ is only defined when $p\leq n$. $\bar{\bm{\Sigma}}_n$ is also the asymptotic variance of $[\T_n + \bb_n]/\sigma$.

Define the set $\hat{\mathcal{A}}_n=\{j: \hat \beta_{j,n}\neq 0\}$ and  $\hat p_{0,n}=|\hat{\mathcal{A}}_n|$, supposing, without loss of generality, that $\hat{\mathcal{A}}_n=\{1,\ldots,\hat{p}_{0,n}\}$. We then partition the matrix $\bm{C}_n=n^{-1}\sum_{i=1}^{n}\bm{x}_i\bm{x}'_i$ as 
\begin{equation*}
\bm{C}_{n} = \begin{bmatrix}
\hat{\bm{C}}_{11,n} \;\;\;  \hat{\bm{C}}_{12,n}\\
\hat{\bm{C}}_{21,n}\;\;\; \hat{\bm{C}}_{22,n}
\end{bmatrix},
\end{equation*}
where $\hat{\bm{C}}_{11,n}$ is of dimension $\hat{p}_{0,n}\times \hat{p}_{0,n}$. Similarly, we define $\hat{\bm{D}}_n^{(1)}$ as the matrix containing the first $\hat{p}_{0,n}$ columns of $\bm{D}_n$ and we define $\hat{\bm{x}}_i^{(1)}$ as the vector containing the first $\hat{p}_{0,n}$ entries of $\bm{x}_i$. Hence, the bias-correction term $\breve \bb_n$ corresponding to $\bm{T}_n$ can be defined as
\[
\breve{\bm{b}}_n=\hat{\bm{D}}_n^{(1)}\hat{\bm{C}}_{11,n}^{-1}\hat{\bm{s}}_n^{(1)}\dfrac{\lambda_n}{2\sqrt{n}},
\]
where $\hat{\bm{s}}_n^{(1)}$ is the $\hat p_{0,n}\times 1$ vector with $j$th entry equal to $\text{sgn}(\hat{\beta}_{j,n})|\tilde \beta_{j,n}|^{-\gamma}$, $j\in \hat{\mathcal{A}_n}$.

\par
Therefore, the studentized pivots can be constructed as
\[
\R_n  = \left\{ \begin{array}{ll} \hat \sigma_n^{-1} \hat \bSigma_n^{-1/2} \T_n  & \text{for $p \leq n$} \\\hat \sigma_n^{-1} \check \bSigma_n^{-1/2} \T_n & \text{for $p>n$} \end{array}  \right. \quad \text{ and } \quad
\check \R_n =  \check \sigma_n^{-1} \check \bSigma_n^{-1/2}  [  \T_n + \breve \bb_n  ],
\]
where the matrices $\hat \bSigma_n$ and $\check \bSigma_n$ have the form
\begin{equation} \label{eqn:Sigmas}
\hat{\bm{\Sigma}}_n =n^{-1}\sum_{i=1}^{n}\big(\hat{\bm{\xi}}_i^{(0)}+\hat{\bm{\eta}}_i^{(0)}\big)\big(\hat{\bm{\xi}}_i^{(0)}+\hat{\bm{\eta}}_i^{(0)}\big)' \quad \text{ and} \quad  \check{\bm{\Sigma}}_n = n^{-1}\sum_{i=1}^{n}\hat{\bm{\xi}}_i^{(0)}\hat{\bm{\xi}}_i^{(0)'},
\end{equation}
and
\[
\hat \sigma_n^2 = n^{-1}\sum_{i=1}^n \hat \epsilon_i^2 \quad \text{ and} \quad \check \sigma_n^2 = n^{-1}\sum_{i=1}^n \tilde \epsilon_i^2,
\]
where $\hat \epsilon_i =  y_i  -  \bm{x}_i^\prime \hat \bbeta_{n} $, $\tilde \epsilon_i = y_i  - \sum_{j \in \hat{\mathcal{A}}_n} x_{ij}\tilde \beta_{j,n}$, $\hat{\bm{\xi}_i}^{(0)}=\hat{\bm{D}}_n^{(1)}\hat{\bm{C}}_{11,n}^{-1}\hat{\bm{x}}_i^{(1)}$ and  $\hat{\bm{\eta}}_i^{(0)}=\hat{\bm{D}}_n^{(1)}\hat{\bm{C}}_{11,n}^{-1}$ $\hat{\bm{\eta}}_i$, with
\[
\hat{\bm{\eta}}_i = \Big(\dfrac{\lambda_n}{2n}\tilde{x}_{i,j}\dfrac{\gamma}{|\hat{\beta}_{j,n}|^{\gamma+1}} \text{sgn}(\hat{\beta}_{j,n})\Big)_{j \in \hat{\mathcal{A}_n}}.
\]
\par
We construct perturbation bootstrap versions $\bm{R}_n^*$ and $\check{\bm{R}}_n^*$ of $\bm{R}_n$ and $\check{\bm{R}}_n$ first by replacing $\bm{T}_n$ with $\bm{T}_n^* =\sqrt{n}\bm{D}_n(\bm{\hat{\beta}}_n^*-\bm{\hat{\beta}}_n)$. We then replace $\hat{\bm{\Sigma}}_n$ and $\check{\bm{\Sigma}}_n$ with $\breve{\bm{\Sigma}}_n$ and $\tilde{\bm{\Sigma}}_n$, respectively, which we define by replacing $\hat{\bm{\xi}}_i^{(0)}$  with $\breve{\bm{\xi}}_i^{(0)} =  \hat{\bm{\xi}}_i^{(0)}\hat\epsilon_i$ and $\hat{\bm{\eta}}_i^{(0)}$ with $\breve{\bm{\eta}}_i^{(0)} = \hat{\bm{\eta}}_i^{(0)}\hat\epsilon_i$ in (\ref{eqn:Sigmas}). We replace $\breve{\bm{b}}_n$ with $\breve{\bm{b}}^*_n= \hat{\bm{D}}_n^{*(1)}\hat{\bm{C}}_{11,n}^{*-1}\hat{\bm{s}}_n^{*(1)}\lambda_n/(2\sqrt{n})$, where $\hat{\bm{s}}_n^{*(1)}$ is the $|\hat{\mathcal{A}^*_n}|\times 1$ vector with $j$th entry equal to $\text{sgn}(\hat{\beta}^*_{j,n})|\tilde \beta^*_{j,n}|^{-\gamma}$, $j\in \hat{\mathcal{A}^*_n}=\{j : \hat{\bm{\beta}}_{j,n}^* \neq 0\}$. The matrix $\hat{\bm{C}}_{11,n}^*$ is the $|\hat{\mathcal{A}}^*_n|\times|\hat{\mathcal{A}}^*_n|$ sub-matrix of $\bm{C}_n$ with rows and columns in $\hat{\mathcal{A}}^*_n$ and $\hat{\bm{D}}^{*(1)}_n$ is the  $q \times|\hat{\mathcal{A}}^*_n|$ sub-matrix of $\bm{D}_n$ with columns in $\hat{\mathcal{A}}^*_n$.  Lastly, we need
\[
\hat \sigma_n^{*2} = n^{-1}\mu_{G^*}^{-2}\sum_{i=1}^n \hat \epsilon_i^{*2} \left(G_i^* - \mu_{G^*}\right)^2 \quad \text{ and} \quad \check \sigma_n^{*2} = n^{-1}\mu_{G^*}^{-2}\sum_{i=1}^n \tilde \epsilon_i^{*2}\left(G_i^* - \mu_{G^*}\right)^2,
\]
where $\hat{\epsilon}_i^*=y_i-\bm{x}'_i\bm{\hat{\beta}}_n^*$, $\tilde{\epsilon}_i^*=y_i-\sum_{j \in  \hat{\mathcal{A}}_n^*} x_{ij}\tilde \beta_{j,n}^*$. With these we construct $\R_n^*$ and $\check \R_n^*$ as
\[
 \R_n^*  = \left\{ \begin{array}{ll} \hat{\sigma}_n^{*-1}\hat \sigma_n \breve \bSigma_n^{-1/2} \T_n^*  & \text{for $p \leq n$} \\ \hat{\sigma}_n^{*-1} \hat \sigma_n \tilde \bSigma_n^{-1/2} \T_n^* & \text{for $p>n$} \end{array}  \right. \quad \text{ and } \quad 
\check \R_n^* = \check \sigma_n^{*-1} \check \sigma_n \tilde \bSigma_n^{-1/2}[\T_n^* + \breve \bb_n^*].
\]
\par

We are motivated to look at these studentized or pivot quantities by the fact that studentization improves the rate of convergence of bootstrap estimators in many settings [cf. Hall (1992)].  

\subsubsection{Results for $p\leq n$.}

\begin{theo}\label{thm:RstarloD}
Let (A.1)--(A.6) hold with $r=6$. Then
\begin{align*}
\sup\limits_{B \in  \mathcal{C}_q}\big|\mathbf{P_*}\big(\mathbf{R}_n^*\in B\big) - \mathbf{P}\big(\mathbf{R}_n\in B\big)\big|=o_p(n^{-1/2})
\end{align*}
\end{theo}

Theorem \ref{thm:RstarloD} shows that after proper studentization, the modified perturbation bootstrap approximation of the distribution of the Alasso estimator is second-order correct. The error rate reduces to $o_p(n^{-1/2})$ from $O(n^{-1/2})$, the best possible rate obtained by the oracle Normal approximation. This is a significant improvement from the perspective of inference. As a consequence, the precision of the percentile confidence intervals based on $\mathbf{R}_n^*$ will be greater than that of confidence intervals based on the oracle Normal approximation.
\par
We point out that the error rate in Theorem \ref{thm:RstarloD} cannot be reduced to the optimal rate of $O_p(n^{-1})$, unlike in the fixed-dimension case. To achieve this optimal rate by our modified bootstrap method, we now consider a bias corrected pivot $\check{\mathbf{R}}_n$ and its modified perturbation bootstrap version $\check{\mathbf{R}}_n^*$.  The following theorem states that it achieves the optimal rate. 
\begin{theo}\label{thm:RcheckstarloD}
Let (A.1)--(A.6) hold with $r=8$. Then
\begin{align*}
\sup\limits_{B \in  \mathcal{C}_q}\big|\mathbf{P_*}\big(\check{\mathbf{R}}_n^*\in B\big) - \mathbf{P}\big(\check{\mathbf{R}}_n\in B\big)\big|=O_p(n^{-1})
\end{align*}
\end{theo}

Theorem \ref{thm:RcheckstarloD} suggests that the modified perturbation bootstrap achieves notable improvement in the error rate over the oracle Normal approximation irrespective of the order of the bias term. Thus Theorem \ref{thm:RcheckstarloD} establishes the perturbation bootstrap method as an effective method for approximating the distribution of the Alasso estimator when $p\leq n$.

\subsubsection{Results for $p>n$}

We now present results for the quality of perturbation bootstrap approximation when the dimension $p$ of the regression parameter can be much larger than the sample size $n$. We consider the initial estimator $\tilde{\bm{\beta}}_n$ to be some bridge estimator, for example Lasso or Ridge estimator, in defining the Alasso estimator by (\ref{eqn:modelalasso}). The bootstrap version of Lasso or Ridge is defined by (\ref{eqn:mpbi}). Higher order results are presented separately for two cases based on growth of $p$ with sample size $n$. First we consider the case when $p$ can grow polynomially and then we move to the situation when $p$ can grow exponentially.
\paragraph{$p$ grows polynomially}
\begin{theo}\label{thm:hiD1}
Let (A.1)(i), (ii), (iii)$'$ and (A.2)--(A.6) and (A.7) hold and $p=O(n^{(r-3)/2})$ for some positive integer $r\geq 3$. Now if $b=0$ [cf. condition (A.3) in Section \ref{sec:assum}] and $r\geq 8$, then we have
\begin{align*}
&\sup\limits_{B \in  \mathcal{C}_q}\big|\mathbf{P_*}\big(\mathbf{R}_n^*\in B\big) - \mathbf{P}\big(\mathbf{R}_n\in B\big)\big|=o_p(n^{-1/2})\\
&\sup\limits_{B \in  \mathcal{C}_q}\big|\mathbf{P_*}\big(\check{\mathbf{R}}_n^*\in B\big) - \mathbf{P}\big(\check{\mathbf{R}}_n\in B\big)\big|=o_p(n^{-1/2}).
\end{align*}
\end{theo}
Theorem \ref{thm:hiD1} states that our proposed modified perturbation bootstrap approximation is second-order correct, even when $p$ grows polynomially with $n$. The error rate obtained by our proposed method is significantly better than $O(n^{-1/2})$, which is the best-attainable rate of the oracle Normal approximation. When $p$ can grow at a polynomial rate with $n$, the validity of our method depends on the existence of some polynomial moment of the error distribution.
To see why, note that it is essential to have
\begin{align}\label{eqn:reqcon}
&\mathbf{P}\Big(\max_{1\leq j \leq p}|\breve{W}_{j,n}|>K.\sqrt{\log n}\Big)= o(n^{-1/2})\;\;\;\;\;\;\;\;\text{and}\nonumber\\
&\mathbf{P_*}\Big(\max_{1\leq j \leq p}|\breve{W}_{j,n}^*|>K.\sqrt{\log n}\Big)= o_p(n^{-1/2})
\end{align}
to obtain second-order correctness, as presented in Theorem \ref{thm:hiD1}. Here $K\in (1, \infty)$ is a constant, $\breve{W}_{j,n}=n^{-1/2}\sum_{i=1}^{n}\epsilon_ix_{i,j}$ and $\breve{W}_{j,n}^*=n^{-1/2}\sum_{i=1}^{n}\hat{\epsilon}_ix_{i,j}(G_i^*-\mu_{G^*})$. In view of Lemma \ref{lem:concentration}, the following bound is needed to conclude (\ref{eqn:reqcon})
\begin{align*}
p.\Big(\max_{1\leq j \leq p}\big[\sum_{i=1}^{n}|x_{i,j}|^{2r}\big]\Big)\Big(\mathbf{E}|\epsilon_1|^r\Big)^2=o\Big(n^{(r-1)/2}(\log n)^{r/2}\Big)
\end{align*}
Clearly under the assumption $\max\{n^{-1}\sum_{i=1}^{n}|x_{i,j}|^{2r}:1\leq j \leq p\}=O(1)$ [cf. condition (A.1) (ii)], we must have $p = o\big(n^{(r-3)/2}(\log n)^{r/2}\big)$ provided $\mathbf{E}|\epsilon_1|^r<\infty$. Therefore in view of condition (A.1) (ii), $p$ can grow like $\Big(a_n.n^l.(\log n)^{l+3/2}\Big)$ where $a_n\rightarrow 0$ as $n\rightarrow \infty$, provided $\mathbf{E}|\epsilon_1|^{2l+3} <\infty$. This implies that $p$ can grow polynomially with $n$  under the assumption that some polynomial moment of the error distribution exists.

\paragraph{$p$ grows exponentially}

When $p$ grows exponentially with some fractional power of $n$, existence of polynomial moment of some order of regression errors $\epsilon_i$'s [cf. condition (A.4) (i)] is not enough to achieve higher order accuracy. Indeed, we need to have some control over the moment generating function of the error variable. Following two important cases are considered in this setting.
\vspace*{5mm}

\underline{\textbf{Errors are Sub-Gaussian}}:
Suppose error $\epsilon_1$ is sub-gaussian. This means that there exists $d > 0$ such that 
\begin{align}\label{eqn:subgaussianor}
\mathbf{E}[e^{\kappa \epsilon_1}]\leq e^{\kappa^2 d^2/2}\;\; \text{for all}\; \kappa \in \mathcal{R}.
\end{align}

When the regression errors have sub-gaussian tails, we need to choose the perturbing quantities $G_i^*$'s effectively to have sub-gaussian tails, that is there exists $d^*>0$ such that
\begin{align}\label{eqn:subgaussianboot}
\mathbf{E_*}[e^{\kappa (G_1^*-\mu_{G^*})}]\leq e^{\kappa^2 d^{*2}/2}\;\; \text{for all}\; \kappa \in \mathcal{R}.
\end{align}
\begin{theo}\label{thm:hiD2}
Let (A.1)(i), (ii), (iii)$'$ and (A.2)--(A.6) and (A.7) hold with $r=8$ and $b=0$. Also assume that (\ref{eqn:subgaussianor}) \& (\ref{eqn:subgaussianboot}) hold and $p=O\Big(\exp\big({n^{(\delta_1-\gamma\delta_2)}}\big)\Big)$ where $\delta_1$ and $\delta_2$ are defined in assumptions (A.6) and (A.7) in Section \ref{sec:assum}. Then the conclusions of Theorem \ref{thm:hiD1} hold.
\end{theo}

\vspace*{5mm}

\underline{\textbf{Errors are Sub-Exponential}}:
Consider the regression errors to be sub-exponential, that is there exist positive parameters $d ,h$ such that 
\begin{align}\label{eqn:subexpoor}
\mathbf{E}[e^{\kappa \epsilon_1}]\leq e^{\kappa^2 d^2/2}\;\; \text{for all}\; |\kappa| < 1/h.
\end{align}
Similar to sub-gaussian case, we need to choose the perturbing quantities $G_i^*$'s to be sub-exponential besides the errors being sub-exponential, that is there exist positive parameters $d^* ,h^*$ such that 
\begin{align}\label{eqn:subexpoboot}
\mathbf{E_*}[e^{\kappa (G_1^*-\mu_{G^*})}]\leq e^{\kappa^2 d^{*2}/2}\;\; \text{for all}\; |\kappa| < 1/h^*.
\end{align}
\begin{theo}\label{thm:hiD3}
Let (A.1)(i), (ii), (iii)$'$ and (A.2)--(A.5), (A.6)(i), (ii) and (A.7) hold with $r=8$ and $b=0$. Also assume that (\ref{eqn:subexpoor}) and (\ref{eqn:subexpoboot}) hold.
\begin{itemize}
\item[\emph{(a)}] If $p=O\Big(\exp\big({n^{(\delta_1-\gamma\delta_2)}}\big)\Big)$ and $p_0 = O\Big(n^{(1-\delta_1+\gamma\delta_2)/2}\Big)$ are satisfied where $\delta_1$ and $\delta_2$ are defined in assumptions (A.6) and (A.7) in Section \ref{sec:assum}, then the conclusions of Theorem \ref{thm:hiD1} hold.
\item[\emph{(b)}] If $p=O\Big(\exp\big(n\big)\Big)$, $ n^{(-\delta_1+\gamma\delta_2)} = o\big(p_0^2/n\big)$ and $p_0/\sqrt{n} = o\big((\log n)^{-3/2}\big)$ are satisfied where $\delta_1$ and $\delta_2$ are defined in assumptions (A.6) and (A.7) in Section \ref{sec:assum}, then the conclusions of Theorem \ref{thm:hiD1} hold.
\end{itemize}
\end{theo}

Theorem \ref{thm:hiD2} and \ref{thm:hiD3} show that our perturbation bootstrap method remains valid as a second order correct method even when the dimension $p$ grows exponentially with some fractional power of $n$. Moreover, we can achieve exponential growth of $p$ in some situations when errors are sub-exponential, as stated in part (b) of Theorem \ref{thm:hiD3}. To obtain higher order results stated in Theorem \ref{thm:hiD2} and  Theorem \ref{thm:hiD3}, we need to relax (\ref{eqn:reqcon}) a bit for $j=p_0+1,\dots,p$. It follows from the proofs and condition (A.6)(ii) that we can relax (\ref{eqn:reqcon}) for $j=p_0+1,\dots,p$, to the following
\begin{align}\label{eqn:reqcon1}
&\mathbf{P}\Big(\max_{p_0+1\leq j \leq p}|\breve{W}_{j,n}|>K.n^{(\delta_1-\gamma\delta_2)}.p_0\Big)= o(n^{-1/2})\;\;\;\;\;\;\;\;\text{and}\nonumber\\
&\mathbf{P_*}\Big(\max_{p_0+1\leq j \leq p}|\breve{W}_{j,n}^*|>K.n^{(\delta_1-\gamma\delta_2)}.p_0\Big)= o_p(n^{-1/2}),
\end{align}
keeping higher order results valid. Now consider using Hoeffding's inequality in sub-gaussian case and Bernstein's inequality in sub-exponential case. As a result, the following two bounds are needed respectively in sub-gaussian and sub-exponential case to conclude (\ref{eqn:reqcon1})
\begin{align*}
&p.\exp\Bigg(-\dfrac{C_1.n^{1+2(\delta_1-\gamma\delta_2)}.p_0^2}{2.\max_{1\leq j \leq p}\Big[\sum_{i=1}^{n}\big(|x_{i,j}|^{2}+|x_{i,j}|^{4}\big)\Big]}\Bigg)=o\Big(n^{-1/2}\Big)\;\;\;\;\;\; \text{and}\\
& p.\exp\Bigg(-\dfrac{C_2.n^{1+2(\delta_1-\gamma\delta_2)}.p_0^2}{2\Big(\max_{1\leq j \leq p}\Big[\sum_{i=1}^{n}\big(|x_{i,j}|^{2}+|x_{i,j}|^{4}\big)\Big]+ C_3. n^{1/2+(\delta_1-\gamma\delta_2)}.p_0\Big)}\Bigg)=o\Big(n^{-1/2}\Big).
\end{align*}
$C_1, C_2, C_3$ are some positive constants.
In view of the assumption $\max\{n^{-1}\sum_{i=1}^{n}|x_{i,j}|^{2r}:1\leq j \leq p\}=O(1)$ [cf. condition (A.1) (ii)], the first bound is implied by $p = o\Big(\exp\big(C.n^{2(\delta_1-\gamma\delta_2)}.p_0^2\big)$ $.n^{-1/2}\Big)$, whereas $p = o\bigg(\exp\Big(\dfrac{C.n^{2(\delta_1-\gamma\delta_2)}.p_0^2}{1+p_0.n^{-1/2+(\delta_1-\gamma\delta_2)}}\Big).n^{-1/2}\bigg)$ is required to obtain the second bound. Here $C$ is some positive constant. These requirements on the growth of $p$ are implying the growth conditions stated in Theorem \ref{thm:hiD2} and Theorem \ref{thm:hiD3}.

\begin{remark}
Note that the matrices $\breve{\bm{\Sigma}}_n$ and $\tilde{\bm{\Sigma}}_n$ used in defining the bootstrap pivots do not depend on $G_1^*,\dots,G_n^*$. Hence it is not required to compute the negative square roots of these matrices for each Monte Carlo bootstrap iteration; these must only be computed once. This is a notable feature of our modified perturbation bootstrap method from the perspective of computational complexity. 
\end{remark}

\begin{remark}
When the dimension $p$ is increasing exponentially, then it is important to choose the distribution of $G_i^*$'s appropriately depending on whether the regression errors are sub-gaussian or sub-exponential. Note that if a random variable $W_1$ has distribution $Beta(a_1,b_1)$, then by Hoeffding's inequality,
\begin{align*}
\mathbf{E}\Big[e^{\kappa (W_1-\mathbf{E}W_1)}\Big]\leq e^{\kappa^2 /8}\;\; \text{for all}\; \kappa \in \mathcal{R}
\end{align*}
and hence $W_1$ is sub-gaussian with parameter value $1/4$, for any choice of $(a_1,b_1)$. On the other hand, if $W_2$ has Gamma distribution with shape parameter $a_2$ and scale parameter $b_2$ then 
\begin{align*}
\log\mathbf{E}\Big[e^{\kappa (W_2-\mathbf{E}W_2)}\Big]&=-a_2b_2\kappa-a_2\log (1-b_2\kappa), \;\; \text{for}\;\; |\kappa|<1/b_2\\
&\leq \dfrac{a_2b_2^2\kappa^2}{{2(1-b_2\kappa)}},\;\; \text{for}\;\; |\kappa| <1/b_2\\
& \leq a_2b_2^2\kappa^2,\;\; \text{for}\;\; |\kappa| <1/2b_2
\end{align*}
where the first inequality follows from the fact that $-\log (1-u)\leq u+\dfrac{u^2}{2(1-u)}$ for $0\leq u<1$. Therefore $W_2$ is sub-exponential with parameters $(b_2\sqrt{2a_2}, 2b_2)$ and hence $W_1+W_2$ is also sub-exponential with parameters $(\sqrt{1/4+2a_2b_2^2}, 2b_2)$ when $W_1$ and $W_2$ are independent.
These observations imply that $Beta(1/2,$ $3/2)$ is an appropriate choice for the distribution of $G_i^*$'s when the errors are sub-gaussian and the distribution of $(M_1+M_2)$ is an appropriate choice for the distribution of $G_i^*$'s when the errors are sub-exponential where $M_1$ and $M_2$ are independent and $M_1$ is a Gamma random variable with shape and scale parameters $0.008652$ and $2$ respectively and $M_2$ is a Beta random variable with both the parameters $0.036490$.  
\end{remark}

\begin{remark}
Let us consider the problem of simultaneous inference. Suppose we want to make inference simultaneously for the regression parameters $\beta_{j,n}$ for all $j$ in the index set $\mathcal{J}_n$.

First suppose that $|\mathcal{J}_n|$, the cardinality of $\mathcal{J}_n$, is fixed. Then assuming without loss of generality that $|\mathcal{J}_n|=\{1,\dots,l\}$ and taking $\bm{D}_n = (\bm{I}_l, \bm{0})$, we can use Theorems \ref{thm:RstarloD}, \ref{thm:RcheckstarloD}, \ref{thm:hiD1}, \ref{thm:hiD2}, \ref{thm:hiD3} to make simultaneous inference. Obviously we need to utilize the fact that the perturbation bootstrap approximation holds uniformly over all convex sets of $\mathcal{R}^l$.

Now suppose that $|\mathcal{J}_n|$ is increasing with $n$. In this scenario simultaneous inference is not possible with a mere choice of the matrix $\bm{D}_n$. There are two possible ways out. One way out is to establish the validity of the bootstrap in approximating the distribution of $\max\{\sqrt{n}|\hat{\beta}_{j,n}-\beta_{j,n}|:j\in\mathcal{J}_n\}$. The Edgeworth expansion theory used in this paper is a well-developed technique in fixed dimensional settings; however, its validity in increasing dimension, more precisely how the error rate depends on the dimension \sout{the dependence of the error rate on dimension}, is still unknown, and hence future investigation is necessary. Instead of using Edgeworth expansions, one can also explore the utility of the techniques developed in Chernozhukov et al. (2013) to establish the validity of bootstrap in approximating the distribution of $\max\{\sqrt{n}|\hat{\beta}_{j,n}-\beta_{j,n}|:j\in\mathcal{J}_n\}$ based upon the component-wise asymptotic normality of $\sqrt{n}(\hat{\bm{\beta}}_{n}-\bm{\beta}_{n})$ and of $\sqrt{n}(\hat{\bm{\beta}}^*_{n}-\hat{\bm{\beta}}_{n})|\bm{\varepsilon}$, where $\bm{\varepsilon} = (\epsilon_1,\dots,\epsilon_n)'$. The second way out is to use component-wise bootstrap approximation dictated by Theorems \ref{thm:RstarloD}, \ref{thm:RcheckstarloD}, \ref{thm:hiD1}, \ref{thm:hiD2}, \ref{thm:hiD3} and then combine them using the well-known Bonferroni correction procedure. For example, suppose we want to construct a $100(1-\alpha)\%$ confidence region for $(\beta_{1,n},\dots,\beta_{l_n,n})$, where $|\mathcal{J}_n|=l_n$ is increasing with $n$ and $\alpha$ is the family wise error rate (FWER) of the region. Define $R_{j,n}=\mathbf{R}_n$ and $\check{R}_{j,n}=\check{\mathbf{R}}_n$, corresponding to defining $\bm{D}_n$ as a unit row vector in $\mathcal{R}^p$ with $j$th component equal to 1. Define $\hat{u}_\Omega^j$ as the $(1-\Omega)$th quantile of the bootstrap distribution of $|\check{R}_{j,n}^*|$ for $j\in\{1,\dots,l_n\}$ for $\Omega\in (0,1)$. Then one can have the following corollary:
\begin{cor}\label{cor:FWER}
Suppose $p_0(p_0\wedge l_n) = o(n)$ when $p\leq n$ and $p_0(p_0\wedge l_n) = o(n^{1/2})$ when $p > n$. Then if $\Omega < \alpha/l_n$, $\{(\beta_{1,n},\dots,\beta_{l_n,n}): |\check{R}_{j,n}|\leq \hat{u}_\Omega^j, j=1,\dots, l_n\}$ is a confidence region for $(\beta_{1,n},\dots,\beta_{l_n,n})$ with FWER $\leq \alpha$ for sufficiently large n.
\end{cor}
Proof: It is enough to show $\mathbf{P}(|\check{R}_{j,n}|>\hat{u}_\Omega^j$ for at least one $j)\leq \alpha$. Without loss of generality assume that $\mathcal{A}_n=\{j:\beta_{j,n}\neq 0\}=\{1,\dots, p_0\}$. Hence note that for sufficiently large n,
\begin{align*}
\mathbf{P}(|\check{R}_{j,n}|>\hat{u}_\Omega^j\; \text{for}\;& \text{at least }\; \text{one}\; j) \leq \sum_{j=1}^{l_n} \mathbf{P}(|\check{R}_{j,n}|>\hat{u}_\Omega^j)\\
&\leq \sum_{j=1}^{l_n} \mathbf{P}_*(|\check{R}^*_{j,n}|>\hat{u}_\Omega^j) + \sum_{j=1}^{l_n} |\mathbf{P}_*(|\check{R}^*_{j,n}|>\hat{u}_\Omega^j)-\mathbf{P}(|\check{R}_{j,n}|>\hat{u}_\Omega^j)|\\
&= \sum_{j=1}^{l_n} \mathbf{P}_*(|\check{R}^*_{j,n}|>\hat{u}_\Omega^j) + \sum_{j=1}^{p_0 \wedge l_n} |\mathbf{P}_*(|\check{R}^*_{j,n}|\leq \hat{u}_\Omega^j)-\mathbf{P}(|\check{R}_{j,n}|\leq \hat{u}_\Omega^j)|\\
&\leq \alpha,
\end{align*}
where the first inequality follows from Boole's inequality. The third equality follows from the fact that $\check{R}_{j,n}=0$ and $\check{R}^*_{j,n}=0$ for sufficiently large n (cf. proof of Lemma 6 in the section \ref{sec:proofs}). The fourth inequality is a consequence of 
\begin{align}\label{eqn:all}
%
%
\max_{j=1,\dots, p_0}\sup\limits_{B \in  \mathcal{C}_q}\big|\mathbf{P_*}\big(\check{\mathbf{R}}_n^*\in B\big) - \mathbf{P}\big(\check{\mathbf{R}}_n\in B\big)\big|=\left\{ \begin{array}{ll} O_p\Big{(}p_0/n\Big{)},\; \text{when $p\leq n$}\\ o_p\Big{(}p_0/\sqrt{n}\Big{)},\; \text{when $p> n$} \end{array} \right.
\end{align}
and due to the assumption that $p_0(p_0\wedge l_n) = o(n)$ when $p\leq n$ and $p_0(p_0\wedge l_n) = o(n^{1/2})$ when $p > n$. Equation (\ref{eqn:all}) is a direct consequence of Theorems \ref{thm:RcheckstarloD}, \ref{thm:hiD1}, \ref{thm:hiD2}, \ref{thm:hiD3}. We want to point out that since $\Omega$ represents bootstrap probability, it should be identified with a random variable which takes the value $\Omega$ with probability 1 and hence Corollary \ref{cor:FWER} holds only on a set $\bm{Q}_n$ with $\mathbf{P}(\bm{Q}_n)\rightarrow 1$ as $n\rightarrow \infty$. But we have omitted those subtleties to keep the corollary simple and easy to understand. Also note that the confidence region of Corollary \ref{cor:FWER} can be utilized for testing $\beta_{j,n}=0$ simultaneously for $j\in\{1,\dots,l_n\}$. Construction of confidence regions and multiple testing can similarly be carried out with $R_{j,n}$ and $R_{j,n}^*$ instead of $\check{R}_{j,n}$ and $\check{R}_{j,n}^*$ for $j\in\{1,\dots,l_n\}$.

\end{remark}

\section{Simulation results}
\label{sec:simulation}

We study through simulation the coverage of one-sided and two-sided $95\%$ confidence intervals for individual nonzero regression coefficients constructed via the pivot quantities $\bm{R}_n$ and $\check{\bm{R}}_n$ as well as via their modified perturbation bootstrap versions $\R^*_n$ and $\check{\bm{R}}^*_n$.  To make further comparisons, we also construct confidence intervals based on a Normal approximation to the distribution of a local quadratic approximation pivot $\bm{R}_n^{\text{LQA}}$, which uses the estimator of $\text{Cov}((\beta_j,j\in\hat A_n)^\prime)$ proposed in the original Alasso paper by Zou (2006).  We also consider the confidence interval from the oracle Normal approximation, which is based on the closeness in distribution of $\bm{T}_n$ to a Normal$(0,\sigma^2 \bm{D}^{(1)}\bm{C}_{11,n}^{-1}\bm{D}^{(1)} )$ random variable, where we use the true active set of covariates $\mathcal{A}_n$ to compute $\bm{C}_{11,n}^{-1}$. We denote this by $\bm{R}_n^{\text{oracle}}$. For the sake of comparison, we also consider the confidence intervals based on the naive perturbation bootstrap from MTC(11) which in that paper are denoted by $CN^{*Q}$ and $CN^{*N}$. 

\begin{figure}[ht]
\begin{center}
\includegraphics[width=\textwidth]{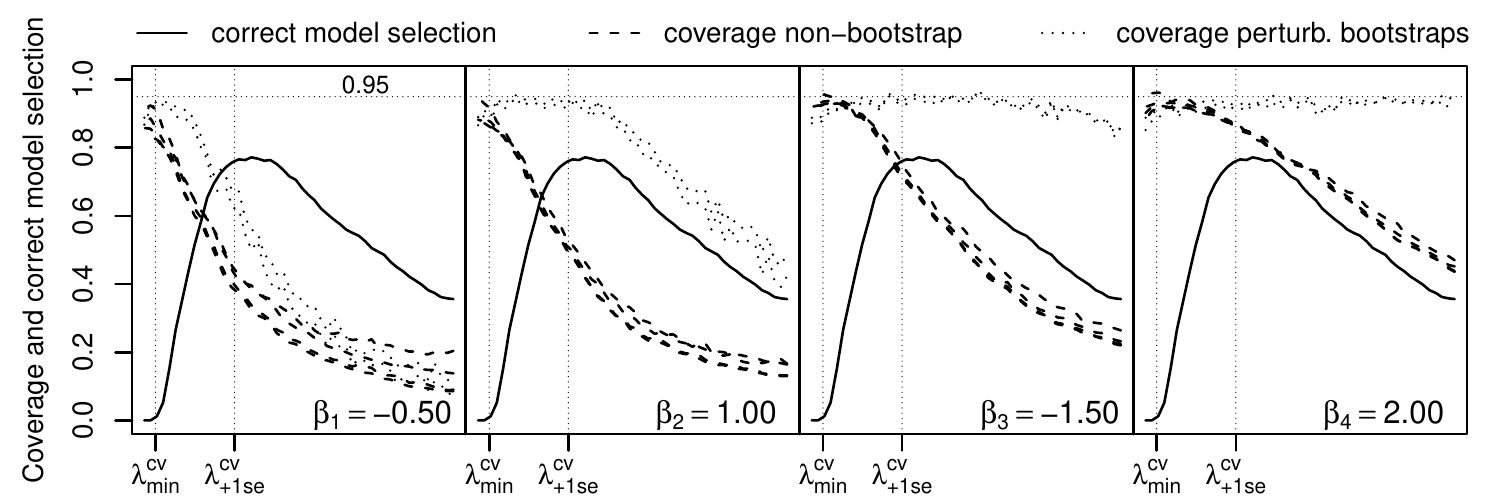}
\end{center}
	\caption{Coverage of $\beta_1$, $\beta_2$, $\beta_3$, and $\beta_4$ over $500$ simulation runs of the confidence intervals based on $\R_n^{\text{LQA}}$, $\R_n^{\text{oracle}}$, $\R_n$,  $\check \R_n$ (dashed curves),  $\R_n^{*}$, and  $\check \R_n^{*}$ (dotted curves) along with the frequency of correct model selection (solid curve) over a grid of fifty $\lambda_n$ values in the $(n,p,p_0)=(200,80,4)$ case.  Vertical lines show median choices of $\lambda_n$ over $500$ simulation runs when selected by minimizing the crossvalidation estimate of prediction error ($\lambda^{\operatorname{cv}}_{\operatorname{min}}$) or under the $1$-standard error rule ($\lambda^{\operatorname{cv}}_{\operatorname{+1se}}$). }
\label{fig:lambdagrid_n200p80s4}
\end{figure}

\par
Under the settings 
\[
(n,p,p_0) \in \left\{(200,80,4), (150,250,6),(200,500,8) \right\},
\]
we generate $n$ independent copies $(X_1,Y_1),\dots,(X_n,Y_n)$ of $(X,Y) \in \mathbb{R}^p \times \mathbb{R}$ from the model $Y = X^\prime\bbeta + \epsilon$,
where $\epsilon$ is a standard normal random variable, $X = (X_1,\dots,X_p)^\prime$ is a mean-zero multivariate normal random vector such that
\[
\text{Cov}(X_j,X_k) = \mathbf{1}(j=k) + 0.3^{|j-k|}\mathbf{1}(j\leq p_0)\mathbf{1}(k \leq p_0)\mathbf{1}(j\neq k)
\] 
for $1 \leq j,k \leq p$, and $\bbeta = (\beta_1,\dots,\beta_p)^\prime$ with $\beta_j$ defined as $\beta_j = (1/2)j(-1)^{j}\mathbf{1}(j \leq p_0)$ for $j=1,\dots,p$. 

\par
We compute the empirical coverage over $500$ simulated data sets of one- and two-sided confidence intervals for each nonzero regression coefficient under crossvalidation-selected values of $\tilde \lambda_n$ and $\lambda_n$, where $\tilde \lambda_n$ is the value of the tuning parameter used to obtain the preliminary Lasso estimate $\tilde \bbeta_n$ and $\lambda_n$ is the value of the tuning parameter used to obtain the Alasso estimate $\hat \bbeta_n$. We use $\gamma=1$ throughout.  For each of the $500$ simulated data sets, $1000$ Monte Carlo draws of the independent random variables $G_1^*,\dots,G_n^* \sim \text{Beta}(1/2, 3/2)$ were drawn in order to create $1000$ Monte Carlo draws of the bootstrap pivots.

\begin{table}[ht]
\centering
\addtolength{\tabcolsep}{4pt} 
\caption{Empirical coverage of 95\% confidence intervals for nonzero regression coefficients by Alasso under $(n,p,p_0)=(200,80,4)$  using $\tilde \lambda_n=0$ and crossvalidation choice of $\lambda_n$. The median $\lambda_n$ choice was $0.987 \cdot n^{1/4}$. One-sided intervals are bounded in the $sgn(\beta_j)$ direction.} 
\begin{tabular}{r|cc|cc|cc|cc}
 \multicolumn{9}{l}{Coverage and \textit{(avg.~width)} of two-sided 95\% CIs:   $(n,p,p_0)=(200,80,4)$  }\\ \hline\hline 
&\multicolumn{2}{l|}{}&\multicolumn{2}{l|}{}&\multicolumn{2}{l|}{}&\multicolumn{2}{l}{}\\[-2ex]
$\beta_j$ & $\R_n^{\text{LQA}}$ & $\R_n^{\text{oracle}}$ & $CN^{*Q}$ & $CN^{*N}$ & $\R_n$ & $\check \R_n$ & $\R_n^{*}$   & $\check \R_n^{*}$  \\ \hline\hline
-0.50 & 0.42 & 0.31 & 0.10 & 0.45 & 0.31 & 0.42 & 0.61 & 0.68 \\ 
   & \textit{(0.44)} & \textit{(0.30)} & \textit{(0.28)} & \textit{(0.31)} & \textit{(0.30)} & \textit{(0.29)} & \textit{(0.26)} & \textit{(0.31)} \\ 
  1.00 & 0.54 & 0.49 & 0.16 & 0.77 & 0.49 & 0.57 & 0.95 & 0.96 \\ 
   & \textit{(0.37)} & \textit{(0.31)} & \textit{(0.47)} & \textit{(0.48)} & \textit{(0.31)} & \textit{(0.30)} & \textit{(0.39)} & \textit{(0.44)}  \\ 
  -1.50 & 0.75 & 0.73 & 0.36 & 0.89 & 0.74 & 0.76 & 0.93 & 0.93 \\ 
   & \textit{(0.34)} & \textit{(0.31)} & \textit{(0.46)} & \textit{(0.47)} & \textit{(0.32)} & \textit{(0.30)} & \textit{(0.37)} & \textit{(0.41)} \\ 
  2.00 & 0.86 & 0.86 & 0.59 & 0.93 & 0.86 & 0.87 & 0.92 & 0.92 \\ 
   & \textit{(0.32)} & \textit{(0.30)} & \textit{(0.39)} & \textit{(0.39)} & \textit{(0.31)} & \textit{(0.29)} & \textit{(0.31)} & \textit{(0.34)}  \\ \hline\hline
 \multicolumn{9}{l}{ Coverage of one-sided 95\% CIs}\\
\hline\hline
-0.50 & 0.29 & 0.23 & 0.08 & 0.36 & 0.23 & 0.36 & 0.63 & 0.70 \\ 
  1.00 & 0.44 & 0.41 & 0.12 & 0.65 & 0.41 & 0.50 & 0.96 & 0.98 \\ 
  -1.50 & 0.64 & 0.61 & 0.29 & 0.82 & 0.62 & 0.68 & 0.95 & 0.96 \\ 
  2.00 & 0.79 & 0.78 & 0.48 & 0.88 & 0.78 & 0.80 & 0.95 & 0.96 \\   \end{tabular}
\label{tab:p80n200cvlambda}
\end{table}

\begin{table}[ht]
\centering
\addtolength{\tabcolsep}{4pt} 
\caption{Empirical coverage of 95\% confidence intervals for nonzero regression coefficients by Alasso under $(n,p,p_0)=(150,250,6)$ using crossvalidation choices of $\tilde \lambda_n$ and $\lambda_n$. The median $\tilde \lambda_n$ and $\lambda_n$ choices were $0.014 \cdot n^{1/2}$ and $0.119 \cdot n^{1/4}$. One-sided intervals are bounded in the $sgn(\beta_j)$ direction.} 
\begin{tabular}{r|cc|cc|cc|cc}
 \multicolumn{9}{l}{Coverage and \textit{(avg.~width)} of two-sided 95\% CIs:   $(n,p,p_0)=(150,250,6)$  }\\ \hline\hline 
&\multicolumn{2}{l|}{}&\multicolumn{2}{l|}{}&\multicolumn{2}{l|}{}&\multicolumn{2}{l}{}\\[-2ex]
$\beta_j$ & $\R_n^{\text{LQA}}$ & $\R_n^{\text{oracle}}$ & $CN^{*Q}$ & $CN^{*N}$ & $\R_n$ & $\check \R_n$ & $\R_n^{*}$   & $\check \R_n^{*}$  \\ \hline\hline
-0.50 & 0.76 & 0.67 & 0.52 & 0.78 & 0.67 & 0.80 & 0.81 & 0.83\\ 
   & \textit{(0.71)} & \textit{(0.31)} & \textit{(0.41)} & \textit{(0.45)} & \textit{(0.32)} & \textit{(0.38)} & \textit{(0.39)} & \textit{(0.50)} \\ 
  1.00 & 0.82 & 0.75 & 0.62 & 0.87 & 0.76 & 0.86 & 0.90 & 0.93 \\ 
   & \textit{(0.52)} & \textit{(0.32)} & \textit{(0.42)} & \textit{(0.43)} & \textit{(0.33)} & \textit{(0.40)} & \textit{(0.42)} & \textit{(0.56)} \\ 
  -1.50 & 0.89 & 0.86 & 0.78 & 0.90 & 0.87 & 0.92 & 0.91 & 0.95 \\ 
   & \textit{(0.53)} & \textit{(0.32)} & \textit{(0.40)} & \textit{(0.40)} & \textit{(0.33)} & \textit{(0.40)} & \textit{(0.40)} & \textit{(0.53)} \\ 
  2.00 & 0.88 & 0.84 & 0.82 & 0.87 & 0.85 & 0.91 & 0.87 & 0.94 \\ 
   & \textit{(0.46)} & \textit{(0.33)} & \textit{(0.38)} & \textit{(0.39)} & \textit{(0.33)} & \textit{(0.40)} & \textit{(0.38)} & \textit{(0.50)} \\ 
  -2.50 & 0.90 & 0.86 & 0.85 & 0.88 & 0.87 & 0.91 & 0.88 & 0.94 \\ 
   & \textit{(0.42)} & \textit{(0.32)} & \textit{(0.37)} & \textit{(0.37)} & \textit{(0.33)} & \textit{(0.40)} & \textit{(0.36)} & \textit{(0.48)} \\ 
  3.00 & 0.89 & 0.85 & 0.85 & 0.87 & 0.86 & 0.92 & 0.87 & 0.93 \\ 
   & \textit{(0.45)} & \textit{(0.31)} & \textit{(0.34)} & \textit{(0.34)} & \textit{(0.32)} & \textit{(0.38)} & \textit{(0.33)} & \textit{(0.43)} \\ 
\hline\hline
 \multicolumn{9}{l}{ Coverage of one-sided 95\% CIs}\\
\hline\hline
-0.50 & 0.71 & 0.63 & 0.45 & 0.75 & 0.64 & 0.73 & 0.84 & 0.88 \\ 
  1.00 & 0.76 & 0.69 & 0.53 & 0.81 & 0.69 & 0.81 & 0.91 & 0.95 \\ 
  -1.50 & 0.84 & 0.81 & 0.72 & 0.86 & 0.82 & 0.88 & 0.90 & 0.94 \\ 
  2.00 & 0.85 & 0.82 & 0.76 & 0.84 & 0.83 & 0.86 & 0.89 & 0.92 \\ 
  -2.50 & 0.87 & 0.83 & 0.82 & 0.86 & 0.84 & 0.88 & 0.89 & 0.92 \\ 
  3.00 & 0.86 & 0.83 & 0.82 & 0.84 & 0.84 & 0.88 & 0.87 & 0.92 \\ 
\end{tabular}
\label{tab:p250n150cvlambda}
\end{table}

\par
When $p\leq n$ we set $\tilde \lambda_n=0$, whereby we use the ordinary least squares estimate for the preliminary estimator $\tilde \bbeta_n$.  When $p>n$, the value of  $\tilde \lambda_n$ is chosen via $10$-fold crossvalidation and $\tilde \bbeta_n$ is computed under the selected value of $\tilde \lambda_n$. Once $\tilde \bbeta_n$ is obtained, $10$-fold crossvalidation is used to select $\lambda_n$. The values $\tilde \lambda_n$ and $\lambda_n$ are thereafter held fixed for all bootstrap computations on the same dataset. In each crossvalidation procedure, the largest value of the tuning parameter for which the crossvalidation prediction error lies within one standard error of its minimum is used so that greater penalization is preferred; see Friedman et al. (2010). 
\par

\begin{table}[ht]
\centering
\addtolength{\tabcolsep}{4pt} 
\caption{Empirical coverage of 95\% confidence intervals for nonzero regression coefficients by Alasso under $(n,p,p_0)=(200,500,8)$ using crossvalidation choices of $\tilde \lambda_n$ and $\lambda_n$. The median $\tilde \lambda_n$ and $\lambda_n$ choices were $0.01 \cdot n^{1/2}$ and $0.30 \cdot n^{1/4}$. One-sided intervals are bounded in the $sgn(\beta_j)$ direction.} 
\begin{tabular}{r|cc|cc|cc|cc}
 \multicolumn{9}{l}{Coverage and \textit{(avg.~width)} of two-sided 95\% CIs:   $(n,p,p_0)=(200,500,8)$  }\\ \hline\hline 
&\multicolumn{2}{l|}{}&\multicolumn{2}{l|}{}&\multicolumn{2}{l|}{}&\multicolumn{2}{l}{}\\[-2ex]
$\beta_j$ & $\R_n^{\text{LQA}}$ & $\R_n^{\text{oracle}}$ & $CN^{*Q}$ & $CN^{*N}$ & $\R_n$ & $\check \R_n$ & $\R_n^{*}$   & $\check \R_n^{*}$  \\ \hline\hline
-0.50 & 0.79 & 0.68 & 0.56 & 0.86 & 0.70 & 0.81 & 0.87 & 0.92 \\ 
   & \textit{(0.69)} & \textit{(0.26)} & \textit{(0.38)} & \textit{(0.42)} & \textit{(0.27)} & \textit{(0.33)} & \textit{(0.36)} & \textit{(0.46)} \\ 
  1.00 & 0.84 & 0.75 & 0.65 & 0.86 & 0.77 & 0.87 & 0.88 & 0.94 \\ 
   & \textit{(0.54)} & \textit{(0.27)} & \textit{(0.34)} & \textit{(0.35)} & \textit{(0.28)} & \textit{(0.35)} & \textit{(0.34)} & \textit{(0.44)} \\ 
  -1.50 & 0.90 & 0.85 & 0.83 & 0.88 & 0.85 & 0.91 & 0.86 & 0.94 \\ 
   & \textit{(0.45)} & \textit{(0.27)} & \textit{(0.31)} & \textit{(0.31)} & \textit{(0.28)} & \textit{(0.35)} & \textit{(0.31)} & \textit{(0.41)} \\ 
  2.00 & 0.89 & 0.84 & 0.86 & 0.85 & 0.85 & 0.92 & 0.86 & 0.95 \\ 
   & \textit{(0.44)} & \textit{(0.28)} & \textit{(0.30)} & \textit{(0.30)} & \textit{(0.28)} & \textit{(0.35)} & \textit{(0.30)} & \textit{(0.40)} \\ 
  -2.50 & 0.93 & 0.89 & 0.87 & 0.88 & 0.89 & 0.91 & 0.89 & 0.93 \\ 
   & \textit{(0.46)} & \textit{(0.28)} & \textit{(0.30)} & \textit{(0.30)} & \textit{(0.28)} & \textit{(0.35)} & \textit{(0.30)} & \textit{(0.39)} \\ 
  3.00 & 0.91 & 0.85 & 0.87 & 0.86 & 0.86 & 0.91 & 0.86 & 0.92 \\ 
   & \textit{(0.46)} & \textit{(0.27)} & \textit{(0.30)} & \textit{(0.30)} & \textit{(0.28)} & \textit{(0.35)} & \textit{(0.29)} & \textit{(0.39)} \\ 
  -3.50 & 0.91 & 0.86 & 0.87 & 0.87 & 0.87 & 0.92 & 0.87 & 0.95 \\ 
   & \textit{(0.48)} & \textit{(0.27)} & \textit{(0.30)} & \textit{(0.30)} & \textit{(0.28)} & \textit{(0.35)} & \textit{(0.29)} & \textit{(0.39)} \\ 
  4.00 & 0.89 & 0.86 & 0.87 & 0.87 & 0.86 & 0.90 & 0.84 & 0.92 \\ 
   & \textit{(0.45)} & \textit{(0.26)} & \textit{(0.28)} & \textit{(0.28)} & \textit{(0.27)} & \textit{(0.33)} & \textit{(0.28)} & \textit{(0.36)} \\ 
\hline\hline
 \multicolumn{9}{l}{ Coverage of one-sided 95\% CIs}\\
\hline\hline
-0.50 & 0.72 & 0.62 & 0.48 & 0.82 & 0.63 & 0.75 & 0.89 & 0.94 \\ 
  1.00 & 0.79 & 0.70 & 0.59 & 0.80 & 0.71 & 0.81 & 0.87 & 0.94 \\ 
  -1.50 & 0.87 & 0.79 & 0.79 & 0.82 & 0.80 & 0.87 & 0.87 & 0.92 \\ 
  2.00 & 0.85 & 0.80 & 0.81 & 0.82 & 0.82 & 0.86 & 0.85 & 0.91 \\ 
  -2.50 & 0.89 & 0.84 & 0.84 & 0.86 & 0.85 & 0.88 & 0.86 & 0.91 \\ 
  3.00 & 0.86 & 0.79 & 0.83 & 0.82 & 0.81 & 0.86 & 0.84 & 0.90 \\ 
  -3.50 & 0.88 & 0.82 & 0.85 & 0.85 & 0.83 & 0.88 & 0.85 & 0.91 \\ 
  4.00 & 0.89 & 0.83 & 0.84 & 0.84 & 0.85 & 0.87 & 0.84 & 0.90 \\ 
\end{tabular}
\label{tab:p500n200cvlambda}
\end{table}

We begin our discussion of the simulation results with Figure \ref{fig:lambdagrid_n200p80s4}, which presents for the case $(n,p,p_0)=(200,80,4)$ a study of how the coverages of the confidence intervals based on the various pivots are affected by the choice of $\lambda_n$ and by the magnitude of the regression coefficients.  Each panel of Figure \ref{fig:lambdagrid_n200p80s4} corresponds to one of the $p_0=4$ non-zero regression coefficients, where the magnitude of the coefficients increases from left to right.  Each panel shows the coverage over $500$ simulated data sets of the confidence intervals based on the pivots $\R_n^{\text{LQA}}$, $\R_n^{\text{oracle}}$, $\R_n$,  $\check \R_n$ (dashed curves),  $\R_n^{*}$, and  $\check \R_n^{*}$ (dotted curves) plotted against $50$ choices of the tuning parameter $\lambda_n$, increasing from left to right. Also appearing in each panel is a solid curve tracing the proportion of times the true model was selected by the Alasso estimator.  The two vertical lines in each panel are positioned at the median choices of $\lambda_n$ when it is selected as the minimizer of the crossvalidation estimate of the prediction error and when the one-standard-error rule is used.  We do not show curves for the  $CN^{*Q}$ and $CN^{*N}$ intervals in Figure \ref{fig:lambdagrid_n200p80s4}, as they exhibited poorer performance and gave the plots a cluttered appearance.
\par
We see that for small values of $\lambda_n$ the confidence intervals based on all the pivots achieve close-to-nominal coverage.  For such small values of $\lambda_n$, however, model selection scarcely occurs.  As larger values of $\lambda_n$ are chosen, the coverage of the confidence intervals tends to drop, the drop being more gradual the larger in magnitude the regression coefficient.  The confidence intervals based on the perturbation bootstrap pivots $\R_n^{*}$ and  $\check \R_n^{*}$, however, are able to sustain nominal coverage for much larger values of $\lambda_n$ than the others, such that they are able to achieve close-to-nominal coverage for the model-selection-optimal choice of $\lambda_n$ for all but the smallest regression coefficient. 


\par
Table \ref{tab:p80n200cvlambda} displays the coverage results for the $n > p$ case $(n,p,p_0)  = (200,80,4)$ under the crossvalidation choice of $\lambda_n$ using the one-standard-error rule and Tables \ref{tab:p250n150cvlambda} and \ref{tab:p500n200cvlambda} for the $n \leq p $ cases $(n,p,p_0) \in \{(150,250,6),(200,500,8)\}$ under crossvalidation choices of $\tilde \lambda_n$ and $\lambda_n$, where both are chosen using the one-standard-error rule.  The median values of the crossvalidation selections of $\tilde \lambda_n$ and $\lambda_n$ under each setting are provided in the table captions in the forms $c_1 \cdot n^{1/2}$ and $c_2 \cdot n^{1/4}$ where $c_1$ and $c_2$ are constants. These correspond to the forms of the theoretical choices of $\tilde \lambda_n$ and $\lambda_n$ under the choice of $\gamma=1$.

In Table \ref{tab:p80n200cvlambda}, we see that under $(n,p,p_0)=(200,80,4)$ the modified perturbation bootstrap intervals based on $\R_n^*$ and $\check \R_n^*$ achieve the closest-to-nominal coverage.  The two-sided $\check \R_n^*$ interval achieves sub-nominal coverage for the smallest regression coefficient $\beta_j=-0.50$, as this coefficient was occasionally estimated to be zero, but achieves close-to-nominal coverage for the larger regression coefficients.  The coverage of the other intervals is much more dramatically effected by the magnitude of the regression coefficient $\beta_j$, a phenomenon which is even more pronounced in the one-sided coverages; for example, the coverage of the $\check \R_n$ interval rises from $0.36$ for $\beta_1 = -0.50$ to $0.80$ for $\beta_4 = 2.00$.  Given that the modified perturbation bootstrap distributions of $\R_n^*$ and $\check \R_n^*$ result in much closer-to-nominal coverages than the Normal approximations to the distributions of $\R_n$ and $\check \R_n$, we may conclude that the sample size is too small for the asymptotically-Normal pivots to have sufficiently approached their limiting distribution; the second-order correctness of the modified perturbation bootstrap is thus apparent.

\par
In the $p > n$ settings, the modified perturbation bootstrap interval based on $\check \R_n^*$ continues to perform well.  Under the $(n,p,p_0)=(150,250,6)$ setting, for which Table \ref{tab:p250n150cvlambda} shows the results, the $\check \R_n^*$ interval achieves the nominal coverage across all regression coefficients except for the smallest in magnitude for both two- and one-sided intervals.  Here also we see a difference between the performance of the confidence intervals based on $\R_n^*$ and $\check \R_n^*$, owing to the bias correction; the coverage of the $\R_n^*$ interval tends to be sub-nominal for both one- and two-sided intervals.   The confidence intervals based on the asymptotic normality of the respective pivot all have sub-nominal coverage for most of the regression coefficients, and their coverages are dramatically affected by the magnitude of the true regression coefficient.

\par
The results are similar for the $(n,p,p_0)=(200,500,8)$ case, for which Table \ref{tab:p500n200cvlambda} shows the results. The only confidence interval which reliably achieves close-to-nominal coverage is the modified perturbation bootstrap interval based on $\check \R_n^*$. We note that the width of the $\check \R_n^*$ interval seems to adapt more to the magnitude of the regression coefficient than the widths of the Normal-based confidence intervals, which remain, with the exception of the $\R_n^{\text{LQA}}$ interval, fairly constant across all magnitudes of $\beta_j$, resulting in poorer coverage for smaller regression coefficients.  In contrast, the $\check \R_n^*$ interval is able to achieve nominal coverage even for the smallest values of $\beta_j$ by producing suitably wider confidence intervals.

\par

We see that the modified perturbation bootstrap is able to produce reliable confidence intervals for regression coefficients in the high-dimensional setting under data-based choices of the tuning parameter, and, importantly, under levels of penalization large enough for model selection to occur.

\section{Data analysis}
\label{sec:dataanalysis}

\begin{figure}[ht]
\begin{center}
\includegraphics[width=.75\textwidth]{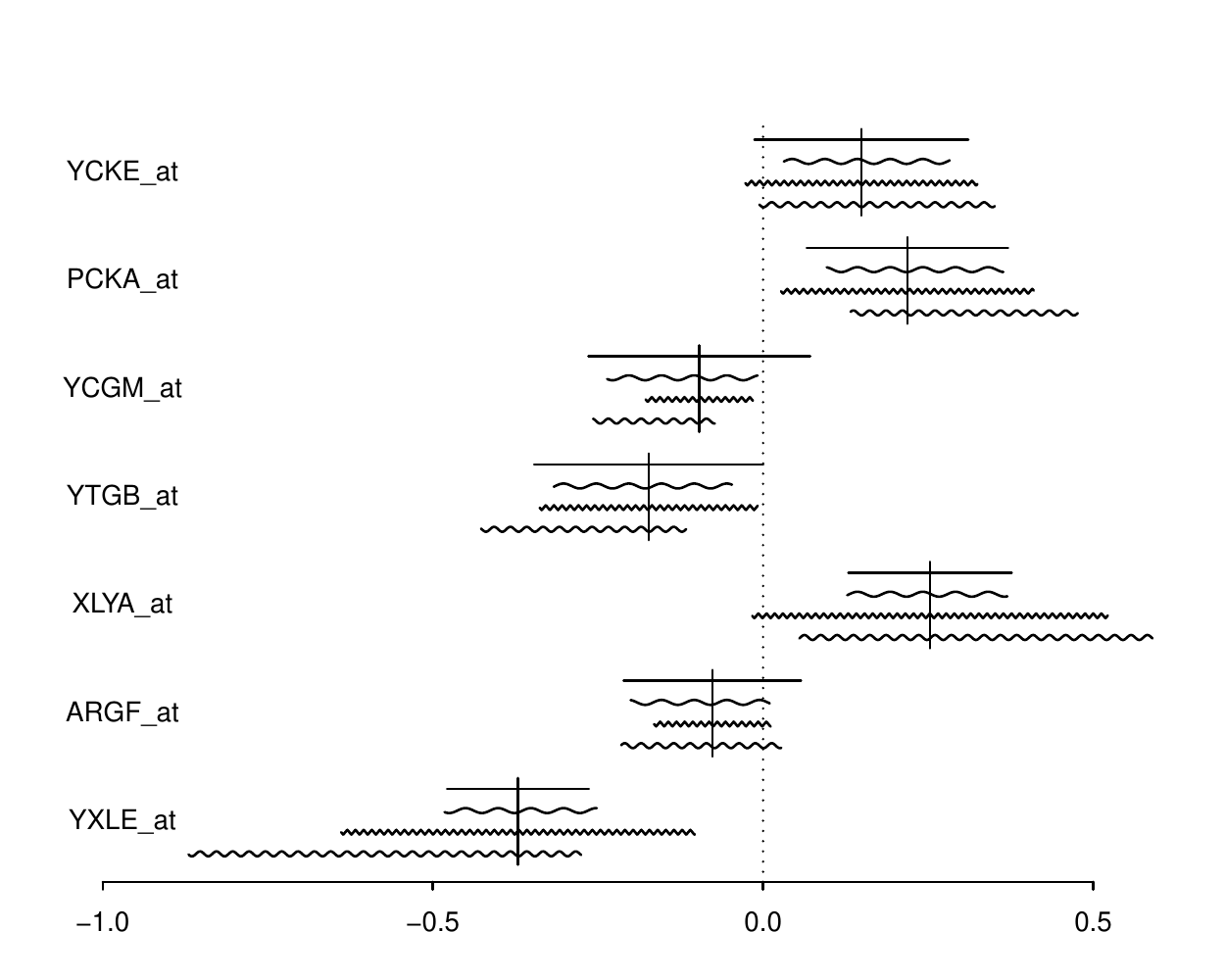}
\end{center}
\caption{Confidence intervals based on $\R_n^{\text{LQA}}$ (straight), $\check \R_n$ (wavy), $CN^{*N}$ (jagged), and $\check \R_n^*$ (wiggly) for each of the Alasso selected genes from the \texttt{riboflavin} data set.}
\label{fig:riboflavin1}
\end{figure}

To illustrate the construction of confidence intervals for regression coefficients in the high-dimensional linear regression model using the modified perturbation bootstrap, we present an analysis of the \texttt{riboflavin} data set considered in B\"{u}hlmann et al. (2014), which those authors make publicly available in their supplementary material.  The data contains $n=71$ independent records consisting of a response variable which is the logarithm of the riboflavin production rate and of $4088$ gene expression levels in batches of \textit{Bacillis subtilis} bacteria.  Of the $4088$, we pre-select $200$ genes by sorting them in order of decreasing empirical variance and keeping the first $200$.  We then fit the linear regression model to the data set with $n=71$ and $p=200$ and compute confidence intervals for the regression coefficients selected by the Alasso procedure.  The variables selected by our methods were different from those discovered in B\"{u}hlmann et al.~(2014).  We choose $\tilde \lambda_n$ and $\lambda_n$ using $10$-fold crossvalidation.  Figure \ref{fig:riboflavin1} displays the confidence intervals for the Alasso-selected covariates obtained from the $\R_n^{\text{LQA}}$, $\check \R_n$, $CN^{*N}$, and $\check \R_n^*$ pivots, where $1000$ bootstrap replicates were used for the bootstrap-based intervals.

\par
The interval based on the $\R_n^{\text{LQA}}$ pivot (straight line) and the $CN^{*N}$ interval (jagged), are symmetric around the estimated value of the regression coefficient (the $CN^{*N}$ interval is formed by adding and substracting an upper quantile of a Normal distribution with a bootstrap-estimated variance). The intervals based on $\check \R_n$ are asymmetric owing to the bias correction (which is quite small in this example) and, in the case of the $\check \R_n^*$ interval, owing to the bias correction and to the asymmetry of the bootstrap distribution of $\check \R_n^*$.  For some of the coefficients, the $\check \R_n^*$ interval is highly asymmetric, suggesting that the distribution of the pivot $\check \R_n$ may still be far from Normal.
\par


\section{Proofs}\label{sec:proofs}
\par
\subsection{Notations}
We denote the true parameter vector as $\bm{\beta}_n = (\beta_{1,n},\dots,\beta_{p,n} )'$, where the subscript $n$ emphasizes that the dimension $p:=p_n$ may grow with the sample size $n$.
Set $\mathcal{A}_n=\{j: \beta_{j,n}\neq 0\}$ and  $p_0:=p_{0,n}=|\mathcal{A}_n|$. For simplicity, we shall suppress the subscript $n$ in the notations $p_n$ and $p_{0n}$. Without loss of generality, we shall assume that $\mathcal{A}_n=\{1,\dots,p_0\}$. Let $\bm{C}_n=n^{-1}\sum_{i=1}^{n}\bm{x}_i\bm{x}'_i$ and partition it according to $\mathcal{A}_n = \{1,\dots,p_0\}$ as 
\begin{equation*}
\bm{C}_{n} = \begin{bmatrix}
\bm{C}_{11,n} \;\;\;\bm{C}_{12,n}\\
\bm{C}_{21,n}\;\;\; \bm{C}_{22,n}
\end{bmatrix},
\end{equation*}
where $\bm{C}_{11,n}$ is of dimension $p_0\times p_0$. Define $\tilde{\bm{x}}_i=\bm{C}_n^{-1}\bm{x}_i$ (when $p\leq n$) and $sgn(x) =-1, 0 ,1$ according as $x<0$, $x=0$, $x>0$, respectively. Suppose $\bm{D}_n$ is a known $q\times p$ matrix with $\text{tr}(\bm{D}_n\bm{D}'_n)=O(1)$ and $q$ is not dependent on $n$. Let $\bm{D}_n^{(1)}$ contains the first $p_0$ columns of $\bm{D}_n$.
Define
\begin{align*}
\bm{S}_{n} = \begin{bmatrix}
\bm{D}_n^{(1)}\bm{C}_{11,n}^{-1}\bm{D}_n^{(1)'}.\sigma^2 \;\;\;\bm{D}_n^{(1)}\bm{C}_{11,n}^{-1}\bar{\bm{x}}^{(1)}_n.\mu_{3}\\
\bar{\bm{x}}^{(1)'}_n\bm{C}_{11,n}^{-1}\bm{D}_n^{(1)'}.\mu_3\;\;\;\;\;\;\;\;\; (\mu_4-\sigma^4)
\end{bmatrix},
\end{align*}
where $\bar{\bm{x}}_n=n^{-1}\sum_{i=1}^{n}\bm{x}_i=(\bar{\bm{x}}^{(1)\prime}_n,\bar{\bm{x}}^{(2)\prime}_n)^\prime$, $\sigma^2=\mathbf{Var}(\epsilon_1)=\mathbf{E}(\epsilon_1^2)$, and where $\mu_3$ and $\mu_4$ are, respectively, the third and fourth central moments of $\epsilon_1$. Let $K$ be a positive constant and $r$ be a positive integer $\geq 3$ unless otherwise specified. By $\mathbf{P_*}$ and $\mathbf{E_*}$ we denote, respectively, probability and expectation with respect to the distribution of $G^{*}$ conditional upon the observed data.

\par
Define $\breve{\bm{W}}_n=n^{-1/2}\sum_{i=1}^{n}\epsilon_i\bm{x}_i$ and $\breve{\bm{W}}_n^*=n^{-1/2}\sum_{i=1}^{n}\hat{\epsilon}_i\bm{x}_i(G_i^*-\mu_{G^*})$. 
Write $\breve{\bm{W}}_n^{(0)}=\breve{\bm{W}}_n$, $\breve{\bm{W}}_n^{*(0)}=\breve{\bm{W}}_n^*$, $p^{(0)}=p$, $p^{(1)}=p_0$, and $p^{(2)}=p-p_0$. Define, $\tilde{b}_n=\sigma^{-1}\bm{\Sigma}_n^{-1/2}\bm{b}_n$ when $p\leq n$ and $\tilde{b}_n=\sigma^{-1}\bar{\bm{\Sigma}}_n^{-1/2}\bm{b}_n$ when $p> n$. Recall that $\bm{b}_n=\bm{D}_n^{(1)}\bm{C}_{11,n}^{-1}\bm{s}_n^{(1)}\dfrac{\lambda_n}{2\sqrt{n}}$, where $\bm{D}_n^{(1)}$ and $\bm{C}_{11,n}$ are as defined earlier and $\bm{s}_n^{(1)}$ is a $p_0\times 1$ vector with $j$th element $\text{sgn}(\beta_{j,n})|\beta_{j,n}|^{-\gamma}$.
Note that under the conditions (A.2)(i), (A.3), and (A.6)(i), $||\bm{\Sigma}_n||=O(1)$, $||\bar{\bm{\Sigma}}_n||=O(1)$, $||\bm{D}_n^{(1)}\bm{C}_{11,n}^{-1/2}||=O(1)$ and $||\bm{s}_n^{(1)}||\leq K\sqrt{p_0}.n^{b\gamma}$. Hence, $||\tilde{\bm{b}}_n|| =O(n^{-\delta_1})$. 
Define $\check{\bm{\xi}}_i^{(0)}=\breve{\bm{\Sigma}}_n^{-1/2}\hat{\bm{\xi}}_i^{(0)}$ and $\check{\bm{\eta}}_i^{(0)}=\breve{\bm{\Sigma}}_n^{-1/2}\hat{\bm{\eta}}_i^{(0)}$ or $\check{\bm{\xi}}_i^{(0)}=\tilde{\bm{\Sigma}}_n^{-1/2}\hat{\bm{\xi}}_i^{(0)}$ and $\check{\bm{\eta}}_i^{(0)}=\tilde{\bm{\Sigma}}_n^{-1/2}\hat{\bm{\eta}}_i^{(0)}$, $i =1,\ldots, n$, according as $p\leq n$ or $p>n$. Here $\breve{\bm{\Sigma}}_n$, $\tilde{\bm{\Sigma}}_n$, $\hat{\bm{\xi}}_i^{(0)}$ and $\hat{\bm{\eta}}_i^{(0)}$ are as defined in the next section. Also define ${\check{\bm{b}}_n}=\breve{\bm{\Sigma}}_n^{-1/2}\bm{D}_n^{(1)} \bm{C}_{11,n}^{-1}\check{\bm{s}}_n^{(1)}\dfrac{\lambda_n}{2\sqrt{n}}$ when $p\leq n$ and ${\check{\bm{b}}_n}=\tilde{\bm{\Sigma}}_n^{-1/2}\bm{D}_n^{(1)} \bm{C}_{11,n}^{-1}\check{\bm{s}}_n^{(1)}\dfrac{\lambda_n}{2\sqrt{n}}$ when $p> n$, where $\check{\bm{s}}_n^{(1)}=(\check{s}_{1n},\ldots, \check{s}_{p_0n})'$ and $\check{s}_{j,n}=sgn(\hat{\beta}_{j,n})|\hat{\beta}_{j,n}|^{-\gamma}$.

\par
We denote by $||\cdot||$ and $||\cdot||_{\infty}$, respectively, the $L^2$ and $L^{\infty}$ norm. 
For a non-negative integer-valued vector $\bm{\alpha} = (\alpha_1, \alpha_2,\ldots,\alpha_l)'$ and a function $f = (f_1,f_2,\ldots,f_l):\ \mathscr{R}^l\ \rightarrow \ \mathscr{R}^l$, $l\geq 1$, write $|\bm{\alpha}| = \alpha_1 +\ldots+ \alpha_l$, $\bm{\alpha}! = \alpha_1!\ldots \alpha_l!$, $f^{\bm{\alpha}} = (f_1^{\alpha_1})\ldots(f_l^{\alpha_l})$, and $D^{\bm{\alpha}}f_1 = D_1^{\alpha_1}\ldots D_l^{\alpha_l}f_1$, where $D_jf_1$ denotes the partial derivative of $f_1$ with respect to the $j$th component of the argument, $1\leq j \leq l$. 
For $\bm{t} =(t_1,\ldots t_l)'\in \mathscr{R}^l$ and $\bm{\alpha}$ as above, define $t^{\bm{\alpha}} = t_1^{\alpha_1}\ldots t_l^{\alpha_l}$. Let $\bm{\Phi}_V$ denote the multivariate Normal distribution with mean $\mathbf{0}$ and dispersion matrix $\bm{V}$ having $j$th row $\bm{V}_{j.}$ and let $\bm{\phi}_V$ denote the density of $\bm{\Phi}_V$. We write $\bm{\Phi}_V= \bm{\Phi}$ and $\bm{\phi}_V = \bm{\phi}$ when $\bm{V}$ is the identity matrix. Define for any set $B\subseteq \mathcal{R}^p$ and any $\bm{b}\in \mathcal{R}^p$, $B+\bm{b}=\{\bm{a}+\bm{b}:\bm{a}\in B\}$.

\par
Define, $\bm{A}_{1n}=$ $\Big{\{}\big{\{}||\breve{\bm{W}}_n^{(1)}||_{\infty}\leq K\sqrt{\log n}\big{\}}\cap\big{\{}||\breve{\bm{W}}_n^{(2)}||_{\infty} \leq K\sqrt{\log n}\big{\}}\cap \big{\{} ||\sqrt{n}\big(\tilde{\bm{\beta}}-\bm{\beta}\big)||_{\infty}\leq K\sqrt{\log n}\big{\}}\Big{\}}$ for $p\leq n$ and $\bm{A}_{1n}=\Big{\{}\big{\{}||\breve{\bm{W}}_n^{(1)}||_{\infty}\leq K\sqrt{\log n}\big{\}}\cap\big{\{}||\breve{\bm{W}}_n^{(2)}||_{\infty} \leq K\sqrt{\log n}\big{\}}\cap \big{\{} ||\sqrt{n}\big(\tilde{\bm{\beta}}-\bm{\beta}\big)||_{\infty} \leq C.n^{\delta_2}\big{\}}\Big{\}}$ for $p>n$. We have assumed $\mathcal{A}_n=\{1,\ldots$ $,p_0\}$. $\breve{\bm{W}}_n^{(1)}$ and $\breve{\bm{W}}_n^{(2)}$ are respectively first $p_0$ and last $(p-p_0)$ components of $\breve{\bm{W}}_n$.
Note that, $\mathbf{P}(\bm{A}_{1n})\geq 1-O(p.n^{-(r-2)/2})$ for $p\leq n$ and $\mathbf{P}(\bm{A}_{1n})\geq 1-o(n^{-1/2})$ for $p>n$ [cf.Lemma 8.1 of Chatterjee and Lahiri (2013)].\\ 
Note that, $\check{\bm{b}}_n=O_p(n^{-\delta_1})$, by Lemma \ref{lem:betahat} and \ref{lem:Sigma}, described below. Suppose, $r_1=\min\{a\in \mathcal{N}:||\check{\bm{b}}_n||^{a+1}=o_p(n^{-1/2})\}$, $\mathcal{N}$ being the set of natural numbers. Define the conditional Lebesgue density of two-term Edgeworth expansion of $\bm{R}_{n}^*$ as
\begin{align*}
\xi_n^*(\bm{x})=&\phi(\bm{x})\Bigg[1+\sum_{k=1}^{r_1}\dfrac{1}{k!}\big{\{}\sum_{|\bm{\alpha}|=k}\check{\bm{b}}_n^{\alpha}H_{\bm{\alpha}}(\bm{x})\big{\}}+\dfrac{1}{\sqrt{n}}\bigg[ \dfrac{1}{6}\sum_{|\bm{\alpha}|=3}\bm{t}^{\bm{\alpha}}\bar{\bm{\xi}}_n^{*(1)}({\bm{\alpha}})H_{\bm{\alpha}}(\bm{x})\\ 
&-\dfrac{1}{2\hat{\sigma}_n^2}\Big{\{}\sum_{|\bm{\alpha}|=1}\bm{t}^{\bm{\alpha}}\bar{\bm{\xi}}_n^{*(3)}({\bm{\alpha}})H_{\bm{\alpha}}(\bm{x})
+\sum_{|\bm{\alpha}|=1}\sum_{|\bm{\zeta}|=2}\bm{t}^{\bm{\alpha}+\bm{\zeta}}\bar{\bm{\xi}}_n^{*(3)}({\bm{\alpha}})\bar{\bm{\xi}}_n^{*(1)}({\bm{\zeta}})H_{\bm{\alpha}+\bm{\zeta}}(\bm{x})\Big{\}}\bigg]\Bigg],
\end{align*}
where $x\in \mathcal{R}^q$, $\bar{\bm{\xi}}_{n}^{*(j)}(\bm{\alpha})=n^{-1}\sum_{i=1}^{n}\Big(\check{\bm{\xi}}_i^{(0)}\hat{\epsilon}_i^j\Big)^{\bm{\alpha}}$, $j=0,1,\ldots$ and $H_{\bm{\alpha}}(\bm{x})=(-D)^{\bm{\alpha}}\phi(\bm{x})$, where $\phi(\bm{\cdot})$ is the standard normal density on $\mathcal{R}^q$.

\subsection{Preliminary Lemmas} Lemmas necessary for the proofs of the results, are stated in this section, along with their proofs.

\begin{lem}\label{lem:concentration}
Suppose $Y_1,\dots,Y_n$ are zero mean independent r.v.s and $\mathbf{E}(|Y_i|^t)< \infty$ for $i = 1,\dots,n$ and $\sum_{i = 1}^{n}\mathbf{E}(|Y_i|^t) = \sigma_t$; $S_n = \sum_{i = 1}^{n}Y_i$. Then, for any $t\geq 2$ and $x>0$
\begin{equation*}
P[|S_n|>x]\leq C[\sigma_t x^{-t} + exp(-x^2/\sigma_2)] 
\end{equation*}
\end{lem}
\textbf{Proof of Lemma \ref{lem:concentration}}.
This inequality was proved in Fuk and Nagaev (1971).

 \begin{lem}\label{lem:W}
 Under assumptions \emph{(A.1)}, \emph{(A.3)}, \emph{(A.4)(i)} and \emph{(A.5)(i), (ii)} with $r=3$,
 \begin{enumerate}[label=(\roman*)]
 \item $\mathbf{P_*}\big(||\breve{\bm{W}}_n^{*(1)}||>K\sqrt{p_0\log n}\big)= O_p(p_0.n^{-(r-2)/2})$.
  \item $\mathbf{P_*}\big(||\breve{\bm{W}}_n^{*(l)}||_{\infty}>K\sqrt{\log n}\big)= O_p(p^{(l)}.n^{-(r-2)/2})$, for $l=0, 1, 2$.
\item  $\mathbf{P_*}\big(||\sqrt{n}\big(\tilde{\bm{\beta}}_n^*-\bm{\hat{\beta}}_n\big)||_{\infty}>K\sqrt{\log n}\big)= O_p(p.n^{-(r-2)/2})$, when $p\leq n$.
 \end{enumerate}
 \end{lem}
 
\textbf{Proof of Lemma \ref{lem:W}}.
 This lemma follows through the same line of Lemma 8.1 of Chatterjee and Lahiri (2013) and employing Lemma \ref{lem:concentration}, stated above.
 

 \begin{lem}\label{lem:betahatnaive}
 Suppose $p$ is fixed. Then under condition \emph{(A.1)(ii)} and \emph{(A.4)(i)} with r=2, 
 \begin{equation*}
\mathbf{P_*}\Big(||\tilde{\bm{\beta}}_n^{*N} - \tilde{\bm{\beta}}_n||=o\big{(}n^{-1/2}(\log n)^{1/2}\big{)}\Big) \geq 1 - o_p\big{(}n^{-1/2}\big{)}
\end{equation*} 
 \end{lem}

\textbf{Proof of Lemma \ref{lem:betahatnaive}}.
This lemma is proved in Proposition 4.1 of Das and Lahiri (2016).

 \begin{lem}\label{lem:betahat}
Suppose assumptions \emph{(A.1)-(A.3)}, \emph{(A.4)(i)}, \emph{(A.5)(i), (ii)} and \emph{(A.6)} hold with $r=4$. Then
 \begin{align*}
 ||\bm{\hat{\beta}}_n-\bm{\beta}_n||_{\infty}=O_p(n^{-1/2})\;\; \text{and on the set } \bm{A}_{1n},\;\; ||\bm{\hat{\beta}}_n^*-\hat{\bm{\beta}}_n||_{\infty}=O_{p_*}(n^{-1/2})
 \end{align*}
 \end{lem}
 
\textbf{Proof of Lemma \ref{lem:betahat}}.
This lemma follows from Markov inequality and using the condition $\{n^{-1}\sum_{i=1}^{n}\big|(\bm{C}_{11,n}^{-1})_{j.}\bm{x}_{i}^{(1)}\big|^{2r}:1\leq j \leq p_0\} = O(1)$ [stated in assumption (A.1)(ii)], after observing the form of the Alasso estimator obtained in (8.5) of Theorem 8.2 (a) of Chatterjee and Lahiri (2013) and the solution $\hat{\bm{u}}_n^*$ of the equation \ref{eqn:KKTinproofs} obtained in the proof of part (a) of Lemma \ref{lem:Rstarconverge}.
 
\begin{lem}\label{lem:Sigma}
Under the assumptions \emph{(A.1)-(A.3)}, \emph{(A.4)(i)} and \emph{(A.6)(i)} and \emph{(iii)} with $r=6$, we have
\begin{align*}
&||\hat{\bm{\Sigma}}_n-\bm{\Sigma}_n||=o_p(n^{-(1+\delta_1)/2}),\; ||\check{\bm{\Sigma}}_n-\bar{\bm{\Sigma}}_n||=o_p(n^{-1})\\
&||\breve{\bm{\Sigma}}_n-\sigma^2\bm{\Sigma}_n||,||\tilde{\bm{\Sigma}}_n-\sigma^2\bar{\bm{\Sigma}}_n||=O_p(n^{-1/2}),
\end{align*}
where $\delta_1$ is as defined in assumption \emph{(A.6)}.
\end{lem}
 
\textbf{Proof of Lemma \ref{lem:Sigma}}. 
First we show that 
 $||\hat{\Sigma}_n-\Sigma_n||=o_p(n^{-(1+\delta_1)/2})$. Note that by Lemma \ref{lem:betahat}, for $n\geq n_0$ (for some $n_0$),
 \begin{align*}
 \hat{\bm{\Sigma}}_n-\bm{\Sigma}_n=&n^{-1}\sum_{i=1}^{n}\big(\hat{\bm{\eta}}_i^{(0)}-\bm{\eta}_i^{(0)}\big)\big(\bm{\xi}_i^{(0)}+\bm{\eta}_i^{(0)}\big)'+n^{-1}\sum_{i=1}^{n}\big(\bm{\xi}_i^{(0)}+\hat{\bm{\eta}}_i^{(0)}\big)'\big(\hat{\bm{\eta}}_i^{(0)}-\bm{\eta}_i^{(0)}\big)'
 \end{align*}
 where on the set on the set $\bm{A}_{1n}$ we have
 \begin{align*}
 \sum_{i=1}^{n}||{\bm{\eta}}_i^{(0)}-\bm{\eta}_i^{(0)}||^2&\leq K^2(\gamma)||\bm{D}_n^{(1)}\bm{C}_{11,n}^{-1}||^2.\dfrac{\lambda_n^2}{n^2}\Big(\operatorname*{\max}_{1\leq j \leq p}\sum_{i=1}^{n}|\tilde{x}_{i,j}|^2\Big)||\bm{\hat{\beta}}_n^{(1)}-\bm{\beta}_n^{(1)}||^2.n^{2b(\gamma+2)}\\
 & \leq K(\gamma, \delta_1)n^{-(1+2\delta_1)}
 \end{align*}
 and 
 \begin{align*}
 &n^{-1}\sum_{i=1}^{n}||\bm{\xi}_i^{(0)}||^2 + n^{-1}\sum_{i=1}^{n}||\bm{\eta}_i^{(0)}||^2 +n^{-1}\sum_{i=1}^{n}||\hat{\bm{\eta}}_i^{(0)}||^2\\
 & \leq tr\big(\bm{D}_n^{(1)}\bm{C}_{11,n}^{-1}\bm{D}_n^{(1)}\big)+K||\bm{D}_n^{(1)}\bm{C}_{11,n}^{-1}||^2.\dfrac{\lambda_n^2}{n^2}\Big(\operatorname*{\max}_{1\leq j \leq p}n^{-1}\sum_{i=1}^{n}|\tilde{x}_{i,j}|^2\Big)||. p_0 n^{2b(\gamma+1)}\\
 & =O(1),
 \end{align*}
 since $||\bm{C}_{11,n}^{-1/2}||^2\leq K \min\{p_0, n^a\}$ and $||\bm{D}_n^{(1)}\bm{C}_{11,n}^{-1/2}||^2\leq q||\bm{D}_n^{(1)}\bm{C}_{11,n}^{-1}\bm{D}_n^{(1)'}||
 $.
 
 Therefore, by the Cauchy-Schwarz inequality, we have $||\hat{\bm{\Sigma}}_n-\bm{\Sigma}_n||=o_p(n^{-(1+\delta_1)/2})$. It follows directly from Lemma  \ref{lem:betahat} that $\check{\bm{\Sigma}}_n=\bar{\bm{\Sigma}}_n$ for sufficiently large n. Hence $||\check{\bm{\Sigma}}_n-\bar{\bm{\Sigma}}_n||=o_p(n^{-1})$.
 \par
 Now to prove the second part, note that for $n\geq n_1$ (for some $n_1$),
  \begin{align*}
 \tilde{\bm{\Sigma}}_n-\sigma^2\bar{\bm{\Sigma}}_n=& n^{-1}\sum_{i=1}^{n}\bm{\xi}_i^{(0)}\bm{\xi}_i^{(0)'}(\hat{\epsilon}_i^2-\sigma^2)
 \end{align*}
 and
 \begin{align*}
 \breve{\bm{\Sigma}}_n-\sigma^2\bm{\Sigma}_n=& n^{-1}\sum_{i=1}^{n}\bm{\xi}_i^{(0)}\bm{\xi}_i^{(0)'}(\hat{\epsilon}_i^2-\sigma^2) + n^{-1}\sum_{i=1}^{n}\big(\hat{\bm{\eta}}_i^{(0)}-\bm{\eta}_i^{(0)}\big)\bm{\xi}_i^{(0)'}\hat{\epsilon}_i^2\\ &+n^{-1}\sum_{i=1}^{n}\bm{\eta}_i^{(0)}\bm{\xi}_i^{(0)'}(\hat{\epsilon}_i^2-\sigma^2)+n^{-1}\sum_{i=1}^{n}\bm{\xi}_i^{(0)}\big(\hat{\bm{\eta}}_i^{(0)}-\bm{\eta}_i^{(0)}\big)'\hat{\epsilon}_i^2\\ &+n^{-1}\sum_{i=1}^{n}\bm{\xi}_i^{(0)}\bm{\eta}_i^{(0)'}(\hat{\epsilon}_i^2-\sigma^2) + n^{-1}\sum_{i=1}^{n}\hat{\bm{\eta}}_i^{(0)}\big(\hat{\bm{\eta}}_i^{(0)}-\bm{\eta}_i^{(0)}\big)'\hat{\epsilon}_i^2\\
 & +n^{-1}\sum_{i=1}^{n}\big(\hat{\bm{\eta}}_i^{(0)}-\bm{\eta}_i^{(0)}\big)\bm{\eta}_i^{(0)'}\hat{\epsilon}_i^2+n^{-1}\sum_{i=1}^{n}\bm{\eta}_i^{(0)}\bm{\eta}_i^{(0)'}(\hat{\epsilon}_i^2-\sigma^2).
 \end{align*}
 Now we need to find the order of the term $||n^{-1}\sum_{i=1}^{n}\bm{\xi}_i^{(0)}\bm{\xi}_i^{(0)'}(\hat{\epsilon}_i^2-\sigma^2)||$ to find the order of $||\breve{\bm{\Sigma}}_n-\sigma^2\bm{\Sigma}_n||$, since other terms can be shown to be of smaller order by using H\"older's inequality. Note that by Lemma \ref{lem:concentration}, (A.1)(ii),
 \begin{align*}
 \mathbf{P}\Big(\Big{\{}||\sum_{i=1}^{n}\bm{\xi}_i^{(0)}\bm{\xi}_i^{(0)'}(\epsilon_i^2-\sigma^2)||>K. n^{1/2}\Big{\}} \Big)\rightarrow 0\; \text{as}\; K\rightarrow \infty
 \end{align*}
  and due to Lemma \ref{lem:betahat}, (A.1)(ii) and (A.2)(i) and (ii),
  \begin{align*}
 \mathbf{P}\Big(\Big{\{}||\sum_{i=1}^{n}\bm{\xi}_i^{(0)}\bm{\xi}_i^{(0)'}(\hat{\epsilon}_i^2-\epsilon_i^2)||>K.n^{1/2}\Big{\}}\Big)\rightarrow 0\; \text{as}\; K\rightarrow \infty
  \end{align*}
Hence, the second part of Lemma \ref{lem:Sigma} follows.
 
 \begin{lem}\label{lem:Rstarconverge}
Let $p\leq n$ and suppose that \emph{(A.1)}--\emph{(A.6)} hold with $r=6$. Then on a set $A_{2n}$ with $\mathbf{P}(\bm{\varepsilon}\in \bm{A}_{2n})\rightarrow 1$, when $\bm{\varepsilon}\in \bm{A}_{2n}$, we have
\begin{itemize}
\item[\emph{(a)}] if $p\leq n$, then
\begin{equation*}
\sup\limits_{B \in  \mathcal{C}_q} \big|\mathbf{P_*}(\bm{R}_n^*\in B) - \int_B\xi^*_n(\bm{x})d\bm{x}\big| = o(n^{-1/2}),
\end{equation*} 
\item[\emph{(b)}] if $p>n$, $b=0$ and additionally conditions \emph{(A.7)} and \emph{(A.1)(iii)}$'$ (in place of \emph{(A.1)(iii)}) hold, then 
\begin{equation*}
\sup\limits_{B \in  \mathcal{C}_q} \big|\mathbf{P_*}(\bm{R}_n^*\in B) - \int_B\xi^*_n(\bm{x})d\bm{x}\big| = o(n^{-1/2}).
\end{equation*} 
\end{itemize}

 \end{lem}
\textbf{Proof of Lemma \ref{lem:Rstarconverge}}.
The modified perturbation bootstrap Alasso estimator is given by
 \begin{align*}
\bm{\hat{\beta}_n^*} = \operatorname*{arg\,min}_{\bm{t}^*}&\Bigg[\sum_{i=1}^{n}(y_i - \bm{x}'_i \bm{t}^*)^2(G^*_i-\mu_{G^*}) \nonumber\\
&+\sum_{i=1}^{n}[\bm{x}'_i(\bm{t}^*-\bm{\hat{\beta}}_n)]^2(2\mu_{G^*}-G_i^*)+\mu_{G^*}\lambda_n\sum_{j=1}^{p}|\tilde{\beta}_{j,n}^*|^{-\gamma}|t_{j,n}^*|\Bigg].
\end{align*}

Now, writing $\bm{\hat{u}}_n^*=\sqrt{n}\big(\bm{\hat{\beta}}_n^*-\bm{\hat{\beta}}_n\big)$, we have
 
 \begin{align}
\bm{\hat{u}}_n^{*}&= \operatorname*{arg\,min}_{\bm{v}^*}\Bigg[\bm{v}^{*\prime}\bm{C}_n\bm{v}^{*} -2\bm{v}^{*\prime}\mu_{G^*}^{-1}\breve{\bm{W}}_n^{*}+\lambda_n\sum_{j=1}^{p}|\tilde{\beta}_{j,n}^{*}|^{-\gamma}\Big(|\hat{\beta}_{j,n}+\dfrac{v_{j}^*}{\sqrt{n}}|-|\hat{\beta}_{j,n}|\Big)\Bigg]\nonumber\\\label{eqn:Zn}
&= \operatorname*{arg\,min}_{\bm{v}^*} \bm{Z}_n(\bm{v}^*)\;\;\;\;\; \text{(say)}.
\end{align}

Note that $\bm{Z}_n(\bm{v}^*)$ is convex in $\bm{v}^*$. Hence, the KKT condition is necessary and sufficient. The KKT condition corresponding to (\ref{eqn:Zn}) is given by
\begin{align}\label{eqn:KKTinproofs}
2\bm{C}_n\bm{v}^*-2\mu_{G^*}^{-1}\breve{\bm{W}}_n^{*}+\dfrac{\lambda_n}{\sqrt{n}}\breve{\bm{\Gamma}}_n^*\breve{\bm{l}}_n=\bm{0}
\end{align}
for some $\breve{l}_{j,n}\in [-1,1]$ for all $j\in\{1,\ldots,p\}$, where $\breve{\bm{l}}_n=(\breve{l}_{1n},\ldots, \breve{l}_{pn})'$ and $\breve{\bm{\Gamma}}_n^*=diag\big(|\tilde{\beta}_{1n}^{*}|^{-\gamma},$ $\ldots, |\tilde{\beta}_{pn}^{*}|^{-\gamma}\big)$. It is easy to show that on the set $\bm{A}_{1n}$, $\Big(\big(\bm{\hat{u}}_n^{*(1)}\big)', \bm{0}'\Big)'$, where $\bm{\hat{u}}_n^{*(1)}=\bm{C}_{11,n}^{-1}\Big[\mu_{G^*}^{-1}\breve{\bm{W}}_n^{*(1)}-\dfrac{\lambda_n}{2\sqrt{n}}\tilde{\bm{s}}_n^{*(1)}\Big]$ is the unique solution of (\ref{eqn:KKTinproofs}) and hence $\bm{\hat{u}}_n^{*}=\Big(\big(\bm{\hat{u}}_n^{*(1)}\big)', \bm{0}'\Big)'$, is the unique solution of the minimization problem (\ref{eqn:Zn}), where $\tilde{\bm{s}}_n^{*(1)}=(\tilde{s}_{1n}^*,\ldots, \tilde{s}_{p_0n}^*)$ and $\tilde{s}_{j,n}^*=sgn(\hat{\beta}_{j,n})|\tilde{\beta}_{j,n}^*|^{-\gamma}$. 

To prove part (a) note that
\begin{align}\label{eq:rrs}
\hat{\sigma}_n^*\hat \sigma_n^{-1}\bm{R}_n^*&= \breve{\bm{\Sigma}}_n^{-1/2}\bm{T}_n^*\\
&=\breve{\bm{\Sigma}}_n^{-1/2}\bm{D}_n^{(1)}\bm{\hat{u}}_n^{*(1)}\nonumber\\
&= \breve{\bm{\Sigma}}_n^{-1/2}\bm{D}_n^{(1)}\bm{C}_{11,n}^{-1}\Big[\mu_{G^*}^{-1}\breve{\bm{W}}_n^{*(1)}-\dfrac{\lambda_n}{2\sqrt{n}}\tilde{\bm{s}}_n^{*(1)}\Big]\nonumber\\
&=\mu_{G^*}^{-1}n^{-1/2}\sum_{i=1}^{n}\big(\check{\bm{\xi}}_i^{(0)}+\check{\bm{\eta}}_i^{(0)}\big)\hat{\epsilon}_i(G_i^*-\mu_{G^*})-\check{\bm{b}}_n+Q_{1n}^*\nonumber\\
&=\bm{T}_{1n}^*-\check{\bm{b}}_n+Q_{1n}^*\;\;\;\; \text(say)
\end{align}
\par
Again note that
\begin{align}
&\mu_{G^*}^2\big[\hat{\sigma}_n^{*2}-\hat{\sigma}_n^2\big]\nonumber\\
=&n^{-1}\sum_{i=1}^{n}\hat{\epsilon}_i^{*2}(G_i^*-\mu_{G^*})^2 - n^{-1}\sum_{i=1}^{n}\hat{\epsilon}_i^2\sigma_{G^*}^2\nonumber\\
=&n^{-1}\sum_{i=1}^{n}\hat{\epsilon}_i^2\Big[(G_i^*-\mu_{G^*})^2-\sigma_{G^*}^2\Big] +2n^{-1}\sum_{i=1}^{n}(\hat{\epsilon}_i^*-\hat{\epsilon}_i)\hat{\epsilon}_i\Big[(G_i^*-\mu_{G^*})^2-\sigma_{G^*}^2\Big]\nonumber\\\label{eqn:sigmadiff}
&+2n^{-1}\sum_{i=1}^{n}(\hat{\epsilon}_i^*-\hat{\epsilon}_i)\hat{\epsilon}_i\sigma_{G^*}^2 +n^{-1}\sum_{i=1}^{n}(\hat{\epsilon}_i^*-\hat{\epsilon}_i)^2(G_i^*-\mu_{G^*})^2,
\end{align}
where under condition (A.1), (A.5)(i) and (A.6)(iii) we have that the order of the last three terms in the expression of $\mu_{G^*}^2\big[\hat{\sigma}_n^{*2}-\hat{\sigma}_n^2\big]$ is $o_{p_*}\big(n^{-1/2}(\log n)^{-1/2}\big)$ on the set $\bm{A}_{1n}$, whereas 
\begin{align*}
\mathbf{P_*}\Big(n^{-1}\sum_{i=1}^{n}\hat{\epsilon}_i^2\Big[(G_i^*-\mu_{G^*})^2-\sigma_{G^*}^2\Big]>K n^{-1/2}(\log n)^{1/2}\Big)=O(p_0n^{-(r-2)/2}),
\end{align*}
by Lemma \ref{lem:concentration}.\\
Therefore, considering Taylor's expansion of $\hat{\sigma}_n^{*-1}$ around $\hat{\sigma}_n^{-1}$, we have
\begin{align}
\bm{R}_n^*&= \bm{T}_{1n}^*-\check{\bm{b}}_n-\big(2\hat{\sigma}_n^2\big)^{-1} \mu_{G^*}^{-3}Z_{1n}^*\breve{\bm{\Sigma}}_n^{-1/2}\bm{D}_n^{(1)}\bm{C}_{11,n}^{-1}\breve{\bm{W}}_n^{*(1)}+Q_{2n}^*\nonumber\\ \label{eqn:Rstarexpand}
&= \bm{R}_{1n}^*-\check{\bm{b}}_n+Q_{2n}^*, \;\;\;\; \text{(say)}
\end{align}
where $Z_{1n}^*=n^{-1}\sum_{i=1}^{n}\hat{\epsilon}_i^2\Big[(G_i^*-\mu_{G^*})^2-\sigma_{G^*}^2\Big]$ and on the set $\bm{A}_{1n}$ we have $||Q_{2n}^*||=o_{p_*}(n^{-1/2})$.

The first three cumulants of $\bm{t}'\bm{R}_{1n}^*$ are given by

$\kappa_1\big(\bm{t}'\bm{R}_{1n}^*\big)=-\dfrac{1}{\sqrt{n}}. \dfrac{1}{2\hat{\sigma}_n^2}\sum_{|\bm{\alpha}|=1}\bm{t}^{\bm{\alpha}}\bar{\bm{\xi}}_n^{*(3)}({\bm{\alpha}}) +o_p(n^{1/2})$

$\kappa_2\big(\bm{t}'\bm{R}_{1n}^*\big)=\mathbf{Var_*}\big(\bm{t}'\bm{R}_{1n}^*\big)=\bm{t}'\bm{t}+o_p(n^{-1/2})$

$\kappa_3\big(\bm{t}'\bm{R}_{1n}^*\big)=\mathbf{E_*}\big(\bm{t}'\bm{R}_{1n}^*\big)^3-3\mathbf{E_*}\big(\bm{t}'\bm{R}_{1n}^*\big)^2. \mathbf{E_*}\big(\bm{t}'\bm{R}_{1n}^*\big)+2\Big(\mathbf{E_*}\big(\bm{t}'\bm{R}_{1n}^*\big)\Big)^3$\\
\hspace*{10mm} $=\dfrac{1}{\sqrt{n}}\bigg[\sum_{|\bm{\alpha}|=3}\bm{t}^{\bm{\alpha}}\bar{\bm{\xi}}_n^{*(1)}({\bm{\alpha}})-\dfrac{3}{\hat{\sigma}_n^2}\sum_{|\bm{\alpha}|=1}\sum_{|\bm{\zeta}|=2}\bm{t}^{\bm{\alpha}+\bm{\zeta}}\bar{\bm{\xi}}_n^{*(3)}({\bm{\alpha}})\bar{\bm{\xi}}_n^{*(1)}({\bm{\zeta}})\Bigg]+o_p(n^{-1/2})$.

Now, using the quadratic form technique of Das and Lahiri (2016), we have on the set $\bm{A}_{1n}$
\begin{align*}
\sup\limits_{B \in  \mathcal{C}_q} \big|\mathbf{P_*}(\bm{R}_{1n}^*\in B) - \int_B\xi^*_{1n}(\bm{x})d\bm{x}\big| = o(n^{-1/2}),
\end{align*}
where 
\begin{align*}
\xi_{1n}^*(\bm{x})=&\phi(\bm{x})\Bigg[1+\dfrac{1}{\sqrt{n}}\bigg[ \dfrac{1}{6}\sum_{|\bm{\alpha}|=3}\bm{t}^{\bm{\alpha}}\bar{\bm{\xi}}_n^{*(1)}({\bm{\alpha}})H_{\bm{\alpha}}(\bm{x})\\ 
&-\dfrac{1}{2\hat{\sigma}_n^2}\Big{\{}\sum_{|\bm{\alpha}|=1}\bm{t}^{\bm{\alpha}}\bar{\bm{\xi}}_n^{*(3)}({\bm{\alpha}})H_{\bm{\alpha}}(\bm{x})
+\sum_{|\bm{\alpha}|=1}\sum_{|\bm{\zeta}|=2}\bm{t}^{\bm{\alpha}+\bm{\zeta}}\bar{\bm{\xi}}_n^{*(3)}({\bm{\alpha}})\bar{\bm{\xi}}_n^{*(1)}({\bm{\zeta}})H_{\bm{\alpha}+\bm{\zeta}}(\bm{x})\Big{\}}\bigg]\Bigg].
\end{align*}

Now, Lemma \ref{lem:Rstarconverge} part (a) follows by Corollary 2.6 of Bhattacharya and Rao (1986) and noting that $\{B+\bm{b}: B\in \mathcal{C}_q\}=\mathcal{C}_q$ and that
\begin{align*}
\mathbf{P}(\bm{R}_{n}^*\in B)&=\mathbf{P}(\bm{R}_{1n}^*\in B+\check{\bm{b}}_n) + o(n^{-1/2})\\
&=\int_{B+\check{\bm{b}}_n}\xi^*_{1n}(\bm{x})d\bm{x} + o(n^{-1/2})\\
&=\int_B\xi^*_{1n}(\bm{x}+\check{\bm{b}}_n)d\bm{x} + o(n^{-1/2})\\
&=\int_{B}\xi^*_{n}(\bm{x})d\bm{x} + o(n^{-1/2}).
\end{align*}
Now for part (b) note that for $n\geq n_0$, on the set $\bm{A}_{1n}$ we have
\begin{align}
\hat{\sigma}_n^*\hat \sigma_n^{-1}\bm{R}_n^*&=\tilde{\bm{\Sigma}}_n^{-1/2}\bm{T}_n^*\nonumber\\
&=\tilde{\bm{\Sigma}}_n^{-1/2}\bm{D}_n^{(1)}\bm{\hat{u}}_n^{*(1)}\nonumber\\
&= \tilde{\bm{\Sigma}}_n^{-1/2}\bm{D}_n^{(1)}\bm{C}_{11,n}^{-1}\Big[\mu_{G^*}^{-1}\breve{\bm{W}}_n^{*(1)}-\dfrac{\lambda_n}{2\sqrt{n}}\tilde{\bm{s}}_n^{*(1)}\Big]\nonumber\\
&=\mu_{G^*}^{-1}\tilde{\bm{\Sigma}}_n^{-1/2}\bm{D}_n^{(1)}\bm{C}_{11,n}^{-1}\breve{\bm{W}}_n^{*(1)}-\check{\bm{b}}_n + Q_{3n}^*\;\;\; \text{(say)}\nonumber\\
&=\mu_{G^*}^{-1}n^{-1/2}\sum_{i=1}^{n}\check{\bm{\xi}}_i^{(0)}\hat{\epsilon}_i(G_i^*-\mu_{G^*})-\check{\bm{b}}_n+Q_{3n}^*\\
&=\bm{T}_{2n}^*-\check{\bm{b}}_n+Q_{3n}^*,\;\;\; \text{(say)}
\end{align}
where $Q_{3n}^*=\tilde{\bm{\Sigma}}_n^{-1/2}\bm{D}_n^{(1)}\bm{C}_{11,n}^{-1}\bm{\Delta}_{n}^{*(1)}+Q_{1n}^*$, where $\bm{\Delta}_n^{*(1)}$ is a $p_0\times 1$ vector with $j$th component
$\lambda_n{n}^{-1/2}\big(\tilde{\beta}_{j,n}^*-\hat{\beta}_{j,n}\big) \gamma \check{s}_{j,n} |\hat{\beta}_{j,n}|^{-1}$ and $Q_{1n}^*$ is as defined in part (a).

Now since $b=0$, by (A.1)(iii)$'$, (A.2)(i), (A.3), Lemma \ref{lem:betahat} and the fact that $||Q_{1n}^*||=o_p(n^{-1/2})$, one can show that on the set $\bm{A}_{1n}$,
\begin{align*}
\mathbf{P_*}\big(||Q_{3n}^*||>K (p_0\lambda_n n^{-1+\delta_2}+o(n^{-1/2})\big)=o(n^{-1/2}).
\end{align*}

Now since by $(A.6)(i)$, $ p_0\lambda_n n^{-1+\delta_2}=o(n^{-1/2})$, similarly to (\ref{eqn:Rstarexpand}), we have
\begin{align}
\bm{R}_n^*&= \bm{T}_{2n}^*-\check{\bm{b}}_n-\big(2\hat{\sigma}_n^2\big)^{-1} \mu_{G^*}^{-3}Z_{1n}^*\check{\bm{\Sigma}}_n^{-1/2}\bm{D}_n^{(1)}\bm{C}_{11,n}^{-1}\breve{\bm{W}}_n^{*(1)}+Q_{4n}^*\nonumber\\
&= \bm{R}_{2n}^*-\check{\bm{b}}_n+Q_{4n}^*\;\;\;\; \text{(say)}
\end{align}
where on the set $\bm{A}_{1n}$ we have $||Q_{4n}^*||=o_{p_*}(n^{-1/2})$. Therefore, two-term Edgeworth expansions of $\bm{R}_{n}^*$ and $\bm{R}_{2n}^*-\check{\bm{b}}_n$ coincide on the set $\bm{A}_{1n}$, by Corollary 2.6 of Bhattachary and Rao (1986). Rest of part (b) of Lemma \ref{lem:Rstarconverge} follows analogously to part (a).  \\

\subsection{Proof of Results}
This section contains the proofs of the proposition and theorems.\\

Note that for any $\bm{t}\in \mathcal{R}^p$ and for each $i \in \{1,\ldots,n\}$, $y_i-\bm{x}'_i\bm{t}=\hat{\epsilon}_i+\bm{x}'_i(\bm{\hat{\beta}}_n-\bm{t})$, and hence 
\begin{align*}
\sum_{i=1}^{n}\big(y_i-\bm{x}'_i\bm{t}^*\big)^2(G_i^*-\mu_{G^*})=&\sum_{i=1}^{n}\big[\bm{x}'_i(\bm{\hat{\beta}}_n-\bm{t})\big]^2(G_i^*-\mu_{G^*})\\
&-2(\bm{t}-\bm{\hat{\beta}}_n)'\breve{\bm{W}}_n^{*}+\sum_{i=1}^{n}\hat{\epsilon}_i^2(G_i^*-\mu_{G^*})^2.
\end{align*}
Therefore,
\begin{align*}
\operatorname*{arg\,min}_{\bm{t}}\bm{L}_1(\bm{t})=\operatorname*{arg\,min}_{\bm{t}}\Big[\sum_{i=1}^{n}\big[\bm{x}'_i(\bm{\hat{\beta}}_n-\bm{t})\big]^2-2\mu_{G^*}^{-1}(\bm{t}-\bm{\hat{\beta}}_n)'\breve{\bm{W}}_n^{*}
+c\lambda_n\sum_{j=1}^{p}c_j|t_{j}|^l\Big].
\end{align*}
Again, since $z_i=\bm{x}'_i\bm{\hat{\beta}}_n+\hat{\epsilon}_i\mu_{G^*}^{-1}(G_i^*-\mu_{G^*})$, we have
\begin{equation*}
\sum_{i=1}^{n}(z_i-\bm{x}'_i\bm{t})^2= \sum_{i=1}^{n}\big[\bm{x}'_i(\bm{\hat{\beta}}_n-\bm{t})\big]^2-2\mu_{G^*}^{-1}(\bm{t}-\bm{\hat{\beta}}_n)'\breve{\bm{W}}_n^{*}+\mu_{G^*}^{-2}\sum_{i=1}^{n}\big[\hat{\epsilon}_i(G_i^*-\mu_{G^*})\big]^2.
\end{equation*}
Therefore, Proposition 2.1 follows.\\

\textbf{Proof of Theorem 4.1}. The KKT condition corresponding to the Alasso criterion function, defined in MTC(11), is
\begin{align*}
2\bm{C}_n^*\bm{w}_n^*-2\bm{W}_n^{*}+\dfrac{\lambda_n^*}{\sqrt{n}}\bm{\Gamma}_n^*\bm{l}_n=\bm{0},
\end{align*}
for some $\bm{l}_n=(l_{1n},\ldots, l_{pn})'$ with $l_{j,n}\in [-1,1]$ for $j=1,\ldots,p$ and $\bm{\Gamma}_n^*=\text{diag}\big(|\tilde{\beta}_{1n}^{*N}|^{-\gamma},$ $\ldots, |\tilde{\beta}_{pn}^{*N}|^{-\gamma}\big)$. This KKT condition can be rewritten through the vector $\bm{w}^*=\big{(}\bm{w}_n^{*(1)\prime}, \bm{w}_n^{*(2)\prime}\big{)}^{\prime}$ as
\begin{align}\label{eqn:NKKT1}
2\bm{C}_{11,n}^*\bm{w}_n^{*(1)}+2\bm{C}_{12,n}^*\bm{w}^{*(2)}_n-2\bm{W}_n^{*(1)}+\dfrac{\lambda_n^*}{\sqrt{n}}\bm{\Gamma}_n^{*(1)}\bm{l}_n^{(1)}=\bm{0}
\end{align}
and for each $j\in \{p_0+1,\dots,p\}$
\begin{align}\label{eqn:NKKT2}
-\dfrac{\lambda_n^*}{2\sqrt{n}}|\tilde{\beta}_{j,n}^{*N}|^{-\gamma}\leq \Big[\big(\bm{C}_{21,n}^*\big)_{j.}\bm{w}_{n}^{*(1)}+\big(\bm{C}_{22,n}^*\big)_{j.}\bm{w}_{n}^{*(2)}-{W}_{j,n}^{*}\Big]\leq \dfrac{\lambda_n^*}{2\sqrt{n}}|\tilde{\beta}_{j,n}^{*N}|^{-\gamma}.
\end{align}
Here, $\bm{W}_n^*=n^{-1/2}\sum_{i=1}^{n}\hat{\epsilon}_i\bm{x}_iG_i^*$, $\bm{W}_n^{*(1)}$ is the vector of the first $p_0$ components of $\bm{W}_n^*$, $W_{j,n}^*$ is the $j$th component of $\bm{W}_n^*$ for $j \in \{1,\dots,p\}$, $\bm{l}_n^{(1)}=(l_{1n},\ldots, l_{p_0n})'$ with $l_{k,n}\in [-1,1]$ for $k=1,\ldots,p_0$ and $\bm{\Gamma}_n^{*(1)}=\text{diag}\big(|\tilde{\beta}_{1n}^{*N}|^{-\gamma},\ldots,$ $ |\tilde{\beta}_{p_0n}^{*N}|^{-\gamma}\big)$ and $\bm{C}_n^*=n^{-1}\sum_{i=1}^{n}\bm{x}_i\bm{x}_i^\prime G_i^*=\begin{bmatrix}
 \bm{C}_{11,n}^* \;\;\;   \bm{C}_{12,n}^*\\
\bm{C}_{21,n}^*\;\;\;  \bm{C}_{22,n}^*
\end{bmatrix}$  
where $\bm{C}_{11,n}^*$ is of dimension $p_0\times p_0$. $\big{(}\bm{C}_{21,n}^*\big{)}_{j\cdot}$ is the $j$th row of $\bm{C}_{21,n}^*$, $j\in \{p_0+1,\dots,p\}$.

Now, to prove part $(a)$ of Theorem 4.1, it is enough to show that $\big{(}\bm{u}_{n2}^{*N\prime},\bm{0}^\prime\big{)}^\prime$ satisfies (\ref{eqn:NKKT1}) and (\ref{eqn:NKKT2}) separately with bootstrap probability $1-o_p(n^{-1/2})$. The vector $\bm{u}_{n2}^{*N}$ is defined as $\bm{u}_{n2}^{*N}=\bm{C}_{11,n}^{*-1}\Big{[}\bm{W}_n^{*(1)}-\dfrac{\lambda_n^*}{\sqrt{n}}\tilde{\bm{s}}_n^{*N(1)}\Big{]}$, where the $j$th component of $\tilde{\bm{s}}_n^{*N(1)}$ is equal to $\text{sgn}(\hat{\beta}_{j,n})|\tilde{\beta}_{jn}^{*N}|^{-\gamma}$, $j\in \{1,\dots,p_0\}$.
 
Note that $\big{(}\bm{u}_{n2}^{*N\prime},\bm{0}^\prime\big{)}^\prime$ exactly satisfies (\ref{eqn:NKKT1}) if $\bm{l}_n^{(1)}=\big{(}\text{sgn}(\hat{\beta}_{1,n}),\dots,\hat{\beta}_{p_0,n})\big{)}$. Thus we can conclude that $\big{(}\bm{u}_{n2}^{*N\prime},\bm{0}^\prime\big{)}^\prime$ satisfies (\ref{eqn:NKKT1}) with bootstrap probability $1-o_p$ $(n^{-1/2})$, if we can show that $\Big{|}\Big{|}\bm{C}_{11,n}^{*-1}\Big{[}\bm{W}_n^{*(1)}-\dfrac{\lambda_n^*}{\sqrt{n}}\bm{\Gamma}_n^{*(1)}\bm{l}_n^{(1)}\Big{]}\Big{|}\Big{|}=o(n^{1/2})$ with bootstrap probability $1-o_p(n^{-1/2})$. Under the assumptions (A.1)(ii) and (A.4)(i) with $r=4$, we have
\begin{align}\label{eqn:Cstar}
\mathbf{P_*}&\big(||\bm{C}_{11,n}^*-\bm{C}_{11,n}\mu_{G^*}||>K.p_0.n^{-1/2}.(\log n)^{1/2}\big) \nonumber\\ &\leq \sum_{j,k=1}^{p_0}\mathbf{P_*}\big(\big|\sum_{i=1}^{n}x_{ij}x_{ik}(G_i^*-\mu_{G^*})\big|>K.n^{-1/2}.(\log n)^{1/2}\big) \nonumber\\
&=o(n^{-1/2})
\end{align}
on the set $\bm{A}_{1n}$ and
\begin{align*}
\mu_{G^*}^{-1}\mathbf{E_*}(\bm{W}_n^{*(1)})=n^{-1/2}\sum_{i=1}^{n}\bm{x}_i^{(1)}\hat{\epsilon}_i=&\dfrac{\lambda_n}{2\sqrt{n}}\Big{(}\text{sgn}({\beta_{1,n}})|\tilde{\beta}_{1,n}|^{-\gamma},\dots,\text{sgn}({\beta_{p_0,n}})|\tilde{\beta}_{p_0,n}|^{-\gamma}\Big{)}^\prime\\ = &\dfrac{\lambda_n}{2\sqrt{n}}\tilde{\bm{s}}_n^{(1)} \text{(say)},
\end{align*}
where on the set $\bm{A}_{1n}$, $|\tilde{\beta}_{j,n}|^{-\gamma}$ is bounded for all $j\in \{1,\dots,p_0\}$ and $n^{-1/2}\lambda_n\rightarrow 0$. These facts along with Proposition \ref{lem:betahatnaive} imply that on the set $\bm{A}_{1n}$
\begin{align*}
\mathbf{P_*}\bigg(\bm{C}_{11,n}^{*-1}\Big{[}\bm{W}_n^{*(1)}-\dfrac{\lambda_n^*}{\sqrt{n}}\bm{\Gamma}_n^{*(1)}\bm{l}_n^{(1)}\Big{]}=o(n^{1/2})\bigg)=1-o(n^{-1/2}).
\end{align*}

Now, note that on the set $\bm{A}_{1n}$
\begin{align*}
\mathbf{P_*}&\Big(\max_{j}\big{\{}||(\bm{C}_{21,n}^*)_{j\cdot}-(\bm{C}_{21,n})_{j\cdot}\mu_{G^*}||: j\in\{p_0+1,\dots,p\}\big{\}}>K.p_0^{1/2}.n^{-1/2}.(\log n)^{1/2}\Big)\\ &\leq \sum_{k=1}^{p_0}\sum_{j=p_0+1}^{p}\mathbf{P_*}\Big(\big|\sum_{i=1}^{n}x_{ij}x_{ik}(G_i^*-\mu_{G^*})\big|>K.n^{-1/2}.(\log n)^{1/2}\Big)\\
&=o(n^{-1/2}),
\end{align*}
and due to Lemma \ref{lem:betahatnaive},
\begin{align*}
\mathbf{P_*}\Big{(}\min_{j}\big{\{}|\tilde{\beta}_{j,n}^{*N}|^{-\gamma}:j\in \{p_0+1,\dots,p\}\big{\}}> K.n^{\gamma/2}(\log n)^{-\gamma/2}\Big{)}=1-o_p(n^{-1/2}).
\end{align*}
Again for $j\in \{p_0+1,\dots,p\}$, 
\begin{align*}
W_{jn}^*=&n^{-1/2}\sum_{i=1}^{n}\hat{\epsilon}_ix_{ij}(G_i^*-\mu_{G^*}) + \mu_{G^*}.n^{-1/2}\sum_{i=1}^{n}x_{ij}\epsilon_i\\
&- \mu_{G^*}\Big(n^{-1}\sum_{i=1}^{n}x_{ij}\bm{x}_i^{(1)}\Big)^\prime\bm{C}_{11,n}^{-1}\Big[n^{-1/2}\sum_{i=1}^{n}\epsilon_i\bm{x}_i^{(1)}-\dfrac{\lambda_n}{2\sqrt{n}}\tilde{\bm{s}}_n^{(1)}\Big]
\end{align*}
Since, $\bm{C}_n\rightarrow \bm{C}$, a pd matrix, and $\max\{\lambda_n, \lambda_n^{*}\}.(\log n/n)^{1/2}\rightarrow 0$, we have 
\begin{align*}
\mathbf{P_*}\Big{(}|W_{jn}^*|> K.(\log n)^{1/2}\Big{)}=1-o_p(n^{-1/2}).
\end{align*}
Hence due to $\min\{\lambda_n, \lambda_n^{*}\}.(\log n)^{-(\gamma+1)/2}.n^{(\gamma-1)/2}\rightarrow \infty$, we have on the set $\bm{A}_{1n}$
\begin{align*}
\mathbf{P_*}\bigg(\big{(}\bm{u}_{n2}^{*N\prime},\bm{0}^\prime\big{)}^\prime\; \text{satisfies}\; (\ref{eqn:NKKT2})\bigg)=1-o_p(n^{-1/2}).
\end{align*}
Therefore part $(a)$ of Theorem 4.1 follows.

Now for part $(b)$, note that since $n^{-1/2}\sum_{i=1}^{n}\bm{x}_i^{(1)}\hat{\epsilon}_i=\dfrac{\lambda_n}{2\sqrt{n}}\Big{(}\text{sgn}({\beta_{1,n}})|\tilde{\beta}_{1,n}|^{-\gamma},\dots,$ $\text{sgn}({\beta_{p_0,n}})$ $|\tilde{\beta}_{p_0,n}|^{-\gamma}\Big{)}^\prime = \dfrac{\lambda_n}{2\sqrt{n}}\tilde{\bm{s}}_n^{(1)}$ (say), so due to (\ref{eqn:Cstar}) and the fact that $n^{-1/2}.$ $(\log n)^{1/2}$ $.\lambda_n\rightarrow 0$, it follows that on the set $\bm{A}_{1n}$,
\begin{align} \label{eq:mtdiff}
\mathbf{P}_*\bigg{(}\sqrt{n}\Big{|}\Big{|}\big{(}\bm{C}_{11,n}^{*-1}-\bm{C}_{11,n}^{-1}\mu_{G^*}^{-1}\big{)}n^{-1/2}\sum_{i=1}^{n}\bm{x}_i^{(1)}\hat{\epsilon}_i\Big{|}\Big{|}_{\infty}= o\big{(}1\big{)}\bigg{)}=1-o(n^{-1/2}).
\end{align}
Again, as $\bm{C}_n\rightarrow \bm{C}$ for some $p\times p$ positive definite matrix $\bm{C}$ and $\mathbf{P}\Big(||\tilde{\bm{\beta}}_n - \bm{\beta}||=O\big{(}n^{-1/2}$ $(\log n)^{1/2}\big{)}\Big) \geq 1 - o\big({n^{-1/2}}\big)$, we have $\mathbf{P}\Big{(}\bm{A}_{1n}\cap \bm{A}_{1n}^{\epsilon}$\Big{)} $\rightarrow 1$ for $A_{1n}^{\epsilon}=\{\lambda_n||\bm{C}_{11,n}^{-1}$ $\tilde{\bm{s}}_n^{(1)}|| > \epsilon^{-1}\}$ for any $\epsilon>0$. Hence, on the set $\bm{A}_{1n}\cap \bm{A}_{1n}^{\epsilon}$ we have
\begin{align*}
\mathbf{P_*}\Big{(}Z_n^*> \epsilon\Big{)}=o_{p}\big(n^{-1/2}\big).
\end{align*}
Therefore part $(b)$ follows.

Now to prove part (c), It is enough to show
\begin{align}\label{eq:r}
\sup\limits_{\bm{x}\in \mathcal{R}^{p_0}}\Big{|}\mathbf{P_*}\big{(}\bm{F}_n^{*(1)}\leq \bm{x}\big{)}-\mathbf{P}\big{(}\bm{F}_n^{(1)} \leq \bm{x}\big{)}\Big{|} \geq K. \dfrac{\lambda_n}{\sqrt{n}}\; \text{for some}\; K>0. 
\end{align}
where $\bm{F}_n^{*(1)}$ and $\bm{F}_n^{(1)}$ are sub vectors of $\bm{F}_n^{*}$ and $\bm{F}_n$ respectively, comprising of first $p_0$ components. Note that
\begin{align*}
\bm{F}_n^{*(1)}&=\bm{C}_{11,n}^{*-1}\Big{[}\bm{W}_n^{*(1)}-\dfrac{\lambda_n^*}{\sqrt{n}}\tilde{\bm{s}}_n^{*N(1)}\Big{]}\\
&= \bm{C}_{11,n}^{-1}\mu_{G^*}^{-1}\Big{[}\bm{W}_n^{*(1)}-\dfrac{\lambda_n^*}{\sqrt{n}}\tilde{\bm{s}}_n^{*N(1)}\Big{]} + \big{(}\bm{C}_{11,n}^{*-1}-\bm{C}_{11,n}^{-1}\mu_{G^*}^{-1}\big{)}\Big{[}\bm{W}_n^{*(1)}-\dfrac{\lambda_n^*}{\sqrt{n}}\tilde{\bm{s}}_n^{*N(1)}\Big{]}\\
&= \breve{F}_n^{*(1)} + \breve{\bm{R}}_{1n}^*\;\; \text{(say)}
\end{align*}
where $\bm{W}_n^{*(1)}= n^{-1/2}\sum_{i=1}^{n}\hat{\epsilon}_i\bm{x}_i(G_i^*-\mu_{G^*})+n^{-1/2}\sum_{i=1}^{n}\hat{\epsilon}_i\bm{x}_i\mu_{G^*}$ with $n^{-1/2}\sum_{i=1}^{n}\bm{x}_i^{(1)}\hat{\epsilon}_i=\dfrac{\lambda_n}{2\sqrt{n}}\tilde{\bm{s}}_n^{(1)}$. Hence due to the fact that $\max\{\lambda_n,\lambda_n^*\}.n^{-1/2}\rightarrow 0$ and $\tilde{\bm{s}}_n^{*N(1)}$ \& $\tilde{\bm{s}}_n^{(1)}$ are bounded in respective probabilities, it follows from Lemma \ref{lem:concentration} that 
\begin{align*}
\mathbf{P}_*\bigg{(}\big{|}\big{|}\breve{\bm{R}}_{1n}^*\big{|}\big{|}\leq c_n .n^{-1/2}\bigg{)}=1-o_p(1)
\end{align*}
where $\{c_n\}$ is a sequence of positive constants increasing to $\infty$ with $c_n = o(\sqrt{\log n})$.

Now write $\breve{\bm{F}}_n^{*(1)}=\tilde{\bm{F}}_n^{*(1)}+\tilde{\bm{Ad}}_n^{(1)}$, where $\tilde{\bm{Ad}}_n^{(1)}=\bm{C}_{11,n}^{-1}n^{-1/2}\sum_{i=1}^{n}\hat{\epsilon}_i\bm{x}_i^{(1)}$. Now similar to (\ref{eq:rrs}), it can be shown that for sufficiently large $n$,
\begin{align*}
&\bm{F}_n^{(1)} = n^{-1/2}\sum_{i=1}^{n}\big(\tilde{\bm{\xi}}_i^{(0)}+\tilde{\bm{\eta}}_i^{(0)}\big)\epsilon_i +\tilde{\bm{R}}_{2n}\\
&\tilde{\bm{F}}_n^{*(1)} = \mu_{G^*}^{-1}n^{-1/2}\sum_{i=1}^{n}\big(\tilde{\bm{\xi}}_i^{(0)}\hat{\epsilon}_i+\tilde{\bm{\eta}}_i^{(0)}\bar{\epsilon}_i\big)(G_i^*-\mu_{G^*})  +\tilde{\bm{R}}_{2n}^*
\end{align*}
where $\mathbf{P}\bigg{(}\big{|}\big{|}\tilde{\bm{R}}_{2n}\big{|}\big{|}= o(n^{-1/2})\bigg{)}=1-o(1)$ and $\mathbf{P}_*\bigg{(}\big{|}\big{|}\tilde{\bm{R}}_{2n}^*\big{|}\big{|} = o(n^{-1/2})\bigg{)}=1-o_p(1)$. Here, $\tilde{\bm{\xi}}_i^{(0)} = \bm{C}_{11,n}^{-1}\bm{x}_i^{(1)}$, $\tilde{\bm{\eta}}_i^{(0)} = \bm{C}_{11,n}^{-1}\tilde{\eta}_i$ with $j$ th component \big{[}$j\in\mathcal{A} = \{k:\beta_j\neq 0\}$\big{]} of $\tilde{\eta}_i$ is $\Big(\dfrac{\lambda_n}{2n}\tilde{x}_{i,j}\dfrac{\gamma}{|\tilde{\beta}_{j,n}|^{\gamma+1}}sgn(\hat{\beta}_{j,n})\Big)$. Here we have assumed without loss of generality that $\mathcal{A}=\{1,\dots,p_0\}$. and $\hat{\epsilon}_i$ and $\bar{\epsilon}_i$ are respectively Alasso and OLS residuals.
Then by Berry-Essen Theorem and Lemma 3.1 of Bhattacharya and Rao (1986), we have
\begin{align}\label{eq:rrp}
&\sup\limits_{\bm{x}\in \mathcal{R}^{p_0}}\Big{|}\mathbf{P}\big{(}\bm{F}_n^{(1)}\leq \bm{x}\big{)}-\bm{\Phi}_{\bm{V}_n}(\bm{x})\Big{|}= O(n^{-1/2})\nonumber\\
\text{and}\;\;&\sup\limits_{\bm{x}\in \mathcal{R}^{p_0}}\Big{|}\mathbf{P_*}\big{(}\tilde{\bm{F}}_n^{*(1)}+\breve{\bm{R}}_{1n}^*\leq \bm{x}\big{)}-\bm{\Phi}_{\tilde{\bm{V}}_n}(\bm{x})\Big{|}= O_p(c_n. n^{-1/2})
\end{align}
where $\bm{V}_n = n^{-1}\sum_{i=1}^{n}\big(\tilde{\bm{\xi}}_i^{(0)}+\tilde{\bm{\eta}}_i^{(0)}\big)^\prime\big(\tilde{\bm{\xi}}_i^{(0)}+\tilde{\bm{\eta}}_i^{(0)}\big)\sigma^2$ and $\tilde{\bm{V}}_n = n^{-1}\sum_{i=1}^{n}\big(\tilde{\bm{\xi}}_i^{(0)}\hat{\epsilon}_i+\tilde{\bm{\eta}}_i^{(0)}\bar{\epsilon}_i\big)^\prime$ $\big(\tilde{\bm{\xi}}_i^{(0)}\hat{\epsilon}_i+\tilde{\bm{\eta}}_i^{(0)}\bar{\epsilon}_i\big)$. Now similar to Lemma \ref{lem:Sigma}, it can be shown that $\Big{|}\Big{|}\tilde{\bm{V}}_n - \bm{V}_n\Big{|}\Big{|} = o_p(c_n.n^{-1/2})$ with $c_n$, as defined earlier.
Hence by Turnbull (1930) and noting (14.66) of Lemma 14.6 of Bhattacharya and Rao (1986) and the facts that $\tilde{\bm{V}}_n=O_p(1)$ \& $\bm{V}_n=O(1)$, we have
\begin{align}\label{eq:rr}
\sup\limits_{\bm{x}\in \mathcal{R}^{p_0}}\Big{|}\bm{\Phi}_{\tilde{\bm{V}}_n}(\bm{x})-\bm{\Phi}_{\bm{V}_n}(\bm{x})\big{)}\Big{|}\leq \Big{|}\Big{|}\tilde{\bm{V}}_n - \bm{V}_n\Big{|}\Big{|} = o_p(c_n.n^{-1/2})
\end{align}
Therefore by (\ref{eq:rrp}) and (\ref{eq:rr}) and noting that $c_n = o(\sqrt{\log n})$, we have
\begin{align}\label{eq:rrq}
&\sup\limits_{\bm{x}\in \mathcal{R}^{p_0}}\Big{|}\mathbf{P_*}\big{(}\tilde{\bm{F}}_n^{*(1)}+\breve{\bm{R}}_{1n}^*\leq \bm{x}\big{)} - \mathbf{P}\big{(}\bm{F}_n^{(1)}\leq \bm{x}\big{)}\Big{|}= o_p(\lambda_n. n^{-1/2})
\end{align}

Now defining $\bm{Ad}_n^{(1)}=\bm{C}_{11,n}^{-1}\dfrac{\lambda_n}{2\sqrt{n}}\bm{s}_n^{(1)}$, by (\ref{eq:rrp}), (\ref{eq:rrq}) and Taylor expansion, we have for any $\bm{x}\in \mathcal{R}^{p_0}$,
\begin{align*}
\mathbf{P_*}\big{(}\bm{F}_n^{*(1)}\leq \bm{x}\big{)}&=\mathbf{P_*}\Big{(}\tilde{\bm{F}}_n^{*(1)}+\breve{\bm{R}}_{1n}^*+\tilde{\bm{Ad}}_n^{(1)}\leq \bm{x}\Big{)}\\
&=\mathbf{P}\Big{(}\bm{F}_n^{(1)}\leq \bm{x}-\bm{Ad}_n^{(1)}+O(n^{-1/2})\Big{)} + o_p(\lambda_n.n^{-1/2})\\
&=\bm{\Phi}_{\bm{V}_n}\Big{(}\bm{x}-\bm{Ad}_n^{(1)}+O(n^{-1/2})\Big{)} +O(n^{-1/2}) + o_p(\lambda_n.n^{-1/2})\\
&=\bm{\Phi}_{\bm{V}_n}(\bm{x}) - \dfrac{\lambda_n}{2\sqrt{n}}\Big{[}\tilde{\bm{s}}_n^{(1)\prime}\bm{C}_{11,n}^{-1}(D_1,\dots,D_{p})^{\prime}\Phi_{\bm{V}_n}(\tilde{\bm{x}})\Big{]}+ o_p(\lambda_n.n^{-1/2})\\
&=\mathbf{P}\Big{(}\bm{F}_n^{(1)}\leq \bm{x}\Big{)} - \dfrac{\lambda_n}{2\sqrt{n}}\Big{[}\tilde{\bm{s}}_n^{(1)\prime}\bm{C}_{11,n}^{-1}(D_1,\dots,D_{p})^{\prime}\Phi_{\bm{V}_n}(\tilde{\bm{x}})\Big{]}+ o_p(\lambda_n/\sqrt{n})
\end{align*}
for some $\tilde{\bm{x}}$ with $||\tilde{\bm{x}}-\bm{x}||\leq ||\bm{Ad}_n^{(1)}||$.

Therefore (\ref{eq:r}) follows from the triangle inequality and  the fact that $\sup\limits_{\bm{x}\in \mathcal{R}^{p_0}}[f(\bm{x})+g(\bm{x})]\leq \sup\limits_{\bm{x}\in \mathcal{R}^{p_0}}f(\bm{x})+\sup\limits_{\bm{x}\in \mathcal{R}^{p_0}}g(\bm{x})$.\\

\textbf{Proof of Theorem 5.1}. By Lemma \ref{lem:Rstarconverge} we have 
\begin{equation}
\sup\limits_{B \in  \mathcal{C}_q} \big|\mathbf{P_*}(\bm{R}_n^*\in B) - \int_B\xi^*_n(\bm{x})d\bm{x}\big| = o_p(n^{-1/2}).
\end{equation} 

Now, retracting the steps of Lemma \ref{lem:Rstarconverge} and using the fact that $||\hat{\bm{\Sigma}}_n-\bm{\Sigma}_n||=o_p(n^{-(1+\delta_1)/2})$ [cf. Lemma \ref{lem:Sigma}], it can be shown that 
\begin{equation}
\sup\limits_{B \in  \mathcal{C}_q} \big|\mathbf{P}(\bm{R}_n\in B) - \int_B\xi_n(\bm{x})d\bm{x}\big| = o(n^{-1/2}),
\end{equation} 
where
\begin{align*}
\xi_n(\bm{x})=&\phi(\bm{x})\Bigg[1+\sum_{k=1}^{r}\dfrac{1}{k!}\big{\{}\sum_{\bm{\alpha}=k}\tilde{\bm{b}}_n^{\alpha}H_{\bm{\alpha}}(\bm{x})\big{\}}+\dfrac{1}{\sqrt{n}}\bigg[ -\dfrac{\mu_3}{2\sigma^3}\sum_{|\bm{\alpha}|=1}\bm{t}^{\bm{\alpha}}\bar{\bm{\xi}}_n({\bm{\alpha}})H_{\bm{\alpha}}(\bm{x})\\
&+\dfrac{\mu_3}{6\sigma^3}\Big{\{}\sum_{|\bm{\alpha}|=3}\bm{t}^{\bm{\alpha}}\bar{\bm{\xi}}_n({\bm{\alpha}})H_{\bm{\alpha}}(\bm{x})-3\sum_{|\bm{\alpha}|=3}\sum_{|\bm{\zeta}|=1}\bm{t}^{\bm{\alpha}+\bm{\zeta}}\bar{\bm{\xi}}_n({\bm{\alpha}})\bar{\bm{\xi}}_n({\bm{\zeta}})H_{\bm{\alpha}+\bm{\zeta}}(\bm{x})\Big{\}}\bigg]\Bigg],
\end{align*}
where $x\in \mathcal{R}^q$, $\bar{\bm{\xi}}_{n}(\bm{\alpha})=n^{-1}\sum_{i=1}^{n}\Big(\bm{\Sigma}_n^{-1/2}\bm{\xi}_i^{(0)}\Big)^{\bm{\alpha}}$. For details see the proof of Theorem 8.2 of Chatterjee and Lahiri (2013).
Now due to assumption (A.6)(i), Lemma \ref{lem:betahat} and Lemma \ref{lem:Sigma} and the facts that $||\bm{b}_n||=O(n^{-\delta_1})$ and $||\check{\bm{b}}_n||=O_p(n^{-\delta_1})$, the coefficients of $n^{-1/2}$ in $\xi^*_n(\bm{x})$ converge to those of $\xi_n(\bm{x})$ in probability and $||\tilde{\bm{b}}_n^{\bm{\alpha}}-\check{\bm{b}}_n^{\bm{\alpha}}||=o(n^{-1/2})$, for all $\bm{\alpha}$ such that $|\bm{\alpha}|\leq r_1$. Therefore Theorem 5.1 follows.\\

\textbf{Proof of Theorem 5.2}. By Lemma \ref{lem:Rstarconverge}, on the set $\bm{A}_{1n}$, we have for $n>n_1$, $\hat{\mathcal{A}}_n=\mathcal{A}_n^*$ and
\begin{align}
\bm{T}_n^*+\breve{\bm{b}}_n^*
&= \bm{D}_n^{(1)}\bm{C}_{11,n}^{-1}\Big[\mu_{G^*}^{-1}\breve{\bm{W}}_n^{*(1)}-\dfrac{\lambda_n}{2\sqrt{n}}\tilde{\bm{s}}_n^{*(1)}\Big]+\bm{D}_n^{(1)}\bm{C}_{11,n}^{-1}\hat{s}_n^{*(1)}\dfrac{\lambda_n}{2\sqrt{n}}\nonumber\\
&=\mu_{G^*}^{-1}\bm{D}_n^{(1)}\bm{C}_{11,n}^{-1}\breve{\bm{W}}_n^{*(1)}+\dfrac{\lambda_n}{2\sqrt{n}}\bm{D}_n^{(1)}\bm{C}_{11,n}^{-1}\big(\tilde{\bm{s}}_n^{*(1)}-\hat{\bm{s}}_n^{*(1)}\big)\nonumber\\\label{eqn:Tstarplusbstar}
&=\mu_{G^*}^{-1}\bm{D}_n^{(1)}\bm{C}_{11,n}^{-1}\breve{\bm{W}}_n^{*(1)}+Q_{4n}^*,\;\;\;\; \text(say)
\end{align}
where the $j$th element of $\hat{s}_n^{*(1)}$ is $sgn\big(\hat{\beta}_{j,n}^*\big)|\tilde{\beta}_{j,n}^*|^{-\gamma}$. Now since $||\hat{\bm{\beta}}_n^*-\hat{\bm{\beta}}_n||_{\infty}=O_{p_*}(n^{-1/2})$ on the set $\bm{A}_{1n}$, one can conclude that on the set $\bm{A}_{1n}$, $\mathbf{P_*}\big(\tilde{\bm{s}}_n^{*(1)} = \hat{\bm{s}}_n^{*(1)}\big)=1$ for sufficiently large $n$. Hence we can conclude that $\mathbf{P_*}\big(||Q_{4n}^*||\neq 0\big)=o(n^{-1})$.

Now expansion and error bounds of the quantity $\big[\check{\sigma}_n^{*2}-\check{\sigma}_n^2\big]$, similar to (\ref{eqn:sigmadiff}), hold. Thus by Taylor's expansion of $\check{\sigma}_n^*$ around $\check{\sigma}_n$ and by (\ref{eqn:Tstarplusbstar}), one has
\begin{align}
\check{\bm{R}}_n^* &=\mu_{G^*}^{-1}\tilde{\bm{\Sigma}}_n^{-1/2}\bm{D}_n^{(1)}\bm{C}_{11,n}^{-1}\breve{\bm{W}}_n^{*(1)}\Big[1-\dfrac{1}{2\hat{\sigma}_n^2}(\hat{\sigma}_n^*-\hat{\sigma}_n)+\dfrac{3}{4\hat{\sigma}_n^4}\dfrac{(\hat{\sigma}_n^*-\hat{\sigma}_n)^2}{2}\Big] +Q_{5,n}^*\nonumber\\
&=\bm{R}_{3n}^*+Q_{5n}^*,\;\;\; \text{(say)}
\end{align}
where on the set $\bm{A}_{1n}$,
\begin{align*}
\mathbf{P_*}\big(||Q_{5n}^*||=o(n^{-1}) \big)=o(n^{-1}).
\end{align*}
Thus by Corollary 2.6 of Bhattacharya and Rao (1986), the Edgeworth expansions of $\bm{R}_{3n}^*$ and $\check{\bm{R}}_{n}^*$ agree up to order $o(n^{-1})$. Now, similarly to Lemma \ref{lem:Rstarconverge}, using the transformation technique of Bhattacharya and Ghosh (1978), one can obtain the three-term Edgeworth expansion of $\bm{R}_{3n}^*$, say $\pi^*_n(\bm{x})$, which will contain terms involving $n^{-1}$ as well as $n^{-1/2}$. 
The coefficients in $\pi^*_n(\bm{x})$ will involve $\check{\sigma}_n^2$, $\mu_{G^*}$, $\mathbf{E_*}(G_1^*-\mu_{G^*})^4$, $\bar{\bm{\xi}}_{n}^{*(j)}(\bm{\alpha})=n^{-1}\sum_{i=1}^{n}\Big(\check{\bm{\xi}}_i^{(0)}\hat{\epsilon}_i^j\Big)^{\bm{\alpha}}$ (for $j=1,3$) and $\bar{\bm{\eta}}_{n}^{*(j)}(\bm{\alpha})=n^{-1}\sum_{i=1}^{n}\Big(\check{\bm{\eta}}_i^{(0)}\hat{\epsilon}_i^j\Big)^{\bm{\alpha}}$ (for $j=1,3$), where $\bm{\alpha}\in\mathcal{N}^q$ such that $|\bm{\alpha}|=1,\ldots,4$. Similarly, one can construct a three-term Edgeworth expansion of $\breve{\bm{R}}_n$, say $\pi_n(\bm{x})$, which will involve $\sigma^2$, $\mu_3$, $\mu_4$, $\tilde{\bm{\xi}}_{n}(\bm{\alpha})=n^{-1}\sum_{i=1}^{n}\Big(\bar{\bm{\Sigma}}_n^{-1/2}\bm{\xi}_i^{(0)}\Big)^{\bm{\alpha}}$, and $\tilde{\bm{\eta}}_n(\bm{\alpha})=n^{-1}\sum_{i=1}^{n}\Big(\bar{\bm{\Sigma}}_n^{-1/2}\bm{\eta}_i^{(0)}\Big)^{\bm{\alpha}}$, $j=1,3$ and $\bm{\alpha}\in\mathcal{N}^q$ such that $|\bm{\alpha}|=1,\ldots,4$, in the coefficients of $n^{-l/2}$, $l=1,2$. It is easy to see that the coefficient of $n^{-1/2}$ in $\pi_n(\bm{x})$ and $\pi^*_n(\bm{x})$ match with that in $\xi_{1n}$ and $\xi_{1n}^*$ respectively, where $\xi_{1n}^*$ is as defined in the proof of Lemma \ref{lem:Rstarconverge} and $\xi_{1n}(\bm{x})=\xi_{n}(\bm{x})-\sum_{k=1}^{r}\dfrac{1}{k!}\big{\{}\sum_{\bm{\alpha}=k}\tilde{\bm{b}}_n^{\alpha}H_{\bm{\alpha}}(\bm{x})\big{\}}\phi(\bm{x})$ with $\xi_{n}(\bm{x})$ being defined as in Theorem 5.1 after replacing $\bm{\Sigma}_n$ with $\bar{\bm{\Sigma}}_n$. Now due to the conditions (A.1)--(A.6) with $r=8$, $(\check{\sigma}_n^2-\sigma_n^2)=O_p(n^{-1/2})$ and $||\tilde{\Sigma}_n-\sigma^2\Sigma_n||=O_p(n^{-1/2})$ [Lemma \ref{lem:Sigma}] and the fact that $||\tilde{\Sigma}_n^{-1/2}-\sigma^{-1}\Sigma_n^{-1/2}||\leq K.||\tilde{\Sigma}_n-\sigma^2\Sigma_n||$ [cf. Turnbull (1930)], the coefficient of $n^{-1/2}$ in $\xi^*_{1n}(\bm{x})$ converges to that of $\xi_{1n}(\bm{x})$ [Similarly as in the proof of Theorem 5.1], whereas the coefficients of $n^{-1}$ in $\pi^*_n(\bm{x})$ and $\pi_n(\bm{x})$ are bounded in respective probabilities. Therefore, theorem 5.2 follows.\\

\textbf{Proof of Theorem 5.3}. The first part follows by Lemma \ref{lem:Rstarconverge} (b) and retracing the proof of Theorem 5.1. And the second part follows analogously to the proof of Theorem 5.2.\\

\textbf{Proof of Theorem 5.4 and 5.5}. The first part follows by Lemma \ref{lem:Rstarconverge} (b) with the use of Hoeffding's and Bernstein's inequality in place of Lemma \ref{lem:concentration} and retracing the proof of Theorem 5.1. And the second part follows analogously to the proof of Theorem 5.2.

\section{Conclusion} \label{sec:1.777}
Second order results of Perturbation Bootstrap method in Alasso are established. It is shown that the naive perturbation bootstrap of Minnier et al. (2011) is not sufficient for correcting the distribution of the Alasso estimator upto second order. Novel modification is proposed in bootstrap objective function to achieve second order correctness even in high dimension. The modification is also shown to be computationally efficient. Thus, in a way the results in this paper establish perturbation bootstrap method as a significant refinement of the approximation of the exact distribution of the Alasso estimator over oracle normal approximation. This is an important finding from the perspective of valid inferences regarding the regression parameters based on adaptive lasso estimator.

%

%
%

\end{document}